\pdfoutput=1
\documentclass[11pt,twoside,a4paper,cmspaper,final,collab]{cms-tdr}
\def\svnVersion{58a52a7-D}\def\svnDate{2021/04/09}\def\cmsCernNoTag{CERN-EP-2020-222}\def\cmsCernDate{\today}\def\cmsMessage{"Published in the Journal of High Energy Physics as \href{http://dx.doi.org/10.1007/JHEP03(2021)257}{\doi{10.1007/JHEP03(2021)257}.}"}
\begin{document}\cmsNoteHeader{HIG-19-018}

\providecommand{\cmsResize}[1]{\resizebox{\textwidth}{!}{#1}}
\newcommand{\HH}{\ensuremath{{\PH\PH}}\xspace}
\newcommand{\ppHHbbgg}{\ensuremath{\Pp\Pp\to\PH\PH\to\gamma\gamma\bbbar}\xspace}
\newcommand{\Zee}{\ensuremath{\PZ \to \Pe\Pe }\xspace}            
\newcommand{\Zmumu}{\ensuremath{\PZ \to \PGm\PGm}\xspace}
\newcommand{\HHbbgg}{\ensuremath{\PH\PH\to\gamma\gamma\bbbar}\xspace}
\newcommand{\mHH}{\ensuremath{m_{\PH\PH}}\xspace}
\newcommand{\Hgg}{\ensuremath{\PH\to\gamma\gamma}\xspace}
\newcommand{\Hbb}{\ensuremath{\PH\to \bbbar}\xspace}

\newcommand{\ggH}{\ensuremath{\Pg\Pg\PH}\xspace}
\newcommand{\VH}{\ensuremath{\PV\PH}\xspace}
\newcommand{\VBFH}{\ensuremath{\mathrm{VBF}\ \PH}\xspace}
\newcommand{\ttH}{\ensuremath{\ttbar\PH}\xspace}

\newcommand{\sigmaHH}{\ensuremath{\sigma_{\PH\PH}\xspace }}
\newcommand{\sigmaVBF}{\ensuremath{\sigma_{\text{VBF}~\PH\PH}\xspace}}
\newcommand{\BR}{\ensuremath{\mathcal{B}(\HH\to\bbgg)}}

\newcommand{\bbgg}{\ensuremath{\gamma\gamma\bbbar}\xspace}
\newcommand{\bb}{\ensuremath{\bbbar}\xspace}
\newcommand{\bbbb}{\ensuremath{\bbbar\bbbar}\xspace}
\newcommand{\DIPHO}{\ensuremath{\gamma\gamma+\text{jets}}\xspace}
\newcommand{\GAMJETS}{\ensuremath{\gamma+\text{jets}}\xspace}
\newcommand{\mt}{\ensuremath{m_\cPqt}\xspace}
\newcommand{\Mtilde}{\ensuremath{\widetilde{M}_{\mathrm{X}}}\xspace}
\newcommand{\Mgg}{\ensuremath{m_{\gamma\gamma}}\xspace}
\newcommand{\Mjj}{\ensuremath{m_\text{jj}\xspace}}
\newcommand{\Mjjvbf}{\ensuremath{m^{\mathrm{VBF}}_\text{jj}\xspace}}
\newcommand{\dEtavbf}{\ensuremath{\abs{\Delta\eta^{\mathrm{VBF}}_\text{jj}}\xspace}}
\newcommand{\etavbf}{\ensuremath{\eta^{\mathrm{VBF}}\xspace}}
\newcommand{\ptvbf}{\ensuremath{\pt^{\mathrm{VBF}}\xspace}}
\newcommand{\acosthetastar}{\ensuremath{\abs{\cos {\theta^\text{CS}_{\HH}}}\xspace}}
\newcommand{\acosthetabb}{\ensuremath{\abs{\cos {\theta_\text{jj}}}}\xspace}
\newcommand{\acosthetagg}{\ensuremath{\abs{\cos {\theta_{\gamma \gamma}}}\xspace}}
\newcommand{\Mggjj}{\ensuremath{m_{\gamma\gamma \text{jj}}}}
\newcommand{\ptgg}{\ensuremath{\pt^{\gamma\gamma}}}
\newcommand{\ptjj}{\ensuremath{\pt^{\text{jj}}}}
\newcommand{\ptg}{\ensuremath{\pt^{\gamma}}}
\newcommand{\ptj}{\ensuremath{\pt^{\text{j}}}}

\newcommand{\kapt}{\ensuremath{\kappa_{\PQt}}\xspace}
\newcommand{\kapl}{\ensuremath{\kappa_{\lambda}}\xspace}
\newcommand{\ctwov}{\ensuremath{c_{2\PV}}\xspace}
\newcommand{\conev}{\ensuremath{c_{\PV}}\xspace}
\newcommand{\HVV}{\ensuremath{\PH\PV\PV}\xspace}
\newcommand{\HHVV}{\ensuremath{\PH\PH\PV\PV}\xspace}
\newcommand{\ctwo}{\ensuremath{c_2}\xspace}
\newcommand{\cg}{\ensuremath{c_{\Pg}}\xspace}
\newcommand{\cgg}{\ensuremath{c_{2\Pg}}\xspace}
\newcommand{\ptgone}{\ensuremath{\pt^{\gamma 1}}\xspace}
\newcommand{\ptgtwo}{\ensuremath{\pt^{\gamma 2}}\xspace}
\newcommand{\mH}{\ensuremath{m_{\PH}}\xspace}
\newcommand{\lbdSM}{\ensuremath{\lambda^\mathrm{SM}_{\PH\PH\PH}}\xspace}
\newcommand{\lbdHHH}{\ensuremath{\lambda_{\PH\PH\PH}}\xspace}
\newcommand{\DRgj}{\ensuremath{\Delta R_{\gamma \text{j}}}\xspace}
\newcommand{\DRbj}{\ensuremath{\Delta R_{\PQb \text{j}}}\xspace}
\newcommand{\minDRgj}{\ensuremath{\Delta R^{\text{min}}_{\gamma \text{j}}}\xspace}
\newcommand{\Dphigj}{\ensuremath{\Delta \phi_{\gamma \text{j}}}\xspace}
\newcommand{\Detagj}{\ensuremath{\Delta \eta_{\gamma \text{j}}}\xspace}
\newcommand{\yt}{\ensuremath{y_{\PQt}}\xspace}
\newcommand{\ytSM}{\ensuremath{y_{\PQt}^\mathrm{SM}\xspace}}

\cmsNoteHeader{HIG-19-018}
\title{Search for nonresonant Higgs boson pair production in final states with two bottom quarks and two photons in proton-proton collisions at \texorpdfstring{$\sqrt{s} = 13\TeV$}{sqrt(s) = 13 TeV}}

\date{\today}

\abstract{A search for nonresonant production of Higgs boson pairs via gluon-gluon and vector boson fusion processes in final states with two bottom quarks and two photons is presented. The search uses data from proton-proton collisions at a center-of-mass energy of $\sqrt{s}=13\TeV$ recorded with the CMS detector at the LHC, corresponding to an integrated luminosity of 137\fbinv. No significant deviation from the background-only hypothesis is observed. An upper limit at 95\% confidence level is set on the product of the Higgs boson pair production cross section and branching fraction into $\bbgg$. The observed (expected) upper limit is determined to be 0.67 ($0.45$)\unit{fb}, which corresponds to 7.7 (5.2) times the standard model prediction. This search has the highest sensitivity to Higgs boson pair production to date. Assuming all other Higgs boson couplings are equal to their values in the standard model, the observed coupling modifiers of the trilinear Higgs boson self-coupling $\kapl$ and the coupling between a pair of Higgs bosons and a pair of vector bosons $\ctwov$ are constrained within the ranges $-3.3<\kapl<8.5$ and $-1.3<\ctwov<3.5$ at 95$\%$ confidence level. Constraints on $\kapl$ are also set by combining this analysis with a search for single Higgs bosons decaying to two photons, produced in association with top quark-antiquark pairs, and by performing a simultaneous fit of $\kapl$ and the top quark Yukawa coupling modifier $\kapt$.}

\hypersetup{
pdfauthor={CMS Collaboration},
pdftitle={Search for nonresonant Higgs boson pair production in final states with two bottom quarks and two photons in proton-proton collisions at sqrt(s) = 13 TeV},
pdfsubject={CMS},
pdfkeywords={CMS, Higgs boson, trilinear self-coupling, photons, b-jets}}

\maketitle

\section{Introduction}\label{sec:intro}
Following the discovery of the Higgs boson (H) by the ATLAS and CMS Collaborations \cite{HiggsdiscoveryAtlas, Chatrchyan:2012ufa, Chatrchyan:2013lba}, there has been significant interest in thoroughly understanding the Brout--Englert--Higgs mechanism \cite{Higgs1, Higgs:1964pj}. With the last remaining free parameter, the mass of the Higgs boson (\mH), now measured to be around 125\GeV, the Higgs boson self-coupling and the structure of the scalar Higgs field potential are precisely predicted in the standard model (SM). Therefore, measuring the Higgs boson's trilinear self-coupling \lbdHHH is of particular importance because it provides valuable information for reconstructing the shape of the scalar potential.

At the CERN LHC, the trilinear self-coupling of the Higgs boson is only directly accessible via Higgs boson pair (\HH) production. 
This rare process dominantly occurs via gluon-gluon fusion (ggF). Vector boson fusion (VBF) is the second largest production mode. 
In the SM, the ggF production cross section in proton-proton ($\Pp\Pp$) collisions at $\sqrt{s} = 13\TeV$ is $31.1^{+1.4}_{-2.0}$\unit{fb}~\cite{Grazzini:2018bsd,Dawson:1998py,Borowka:2016ehy,Baglio:2018lrj,deFlorian:2013jea,Shao:2013bz,deFlorian:2015moa}, calculated at
next-to-next-to-leading order (NNLO) with the resummation at next-to-next-to-leading-logarithm
accuracy and including top-quark mass effects at next-to-leading order (NLO). For VBF, the production cross section is calculated to be $1.73\pm 0.04$\unit{fb}~\cite{Dreyer:2018qbw,Baglio:2012np,Liu-Sheng:2014gxa} at next-to-NNLO in quantum chromodynamics (QCD). The
uncertainties in the values of the cross sections include variations of the factorisation and renormalisation scales, parton distribution function (PDF), and the value of the strong force coupling constant ($\alpS$). The cross sections are calculated for $\mH = 125\GeV$.

\begin{figure*}[hbt]
\centering
\includegraphics[width=0.3\textwidth]{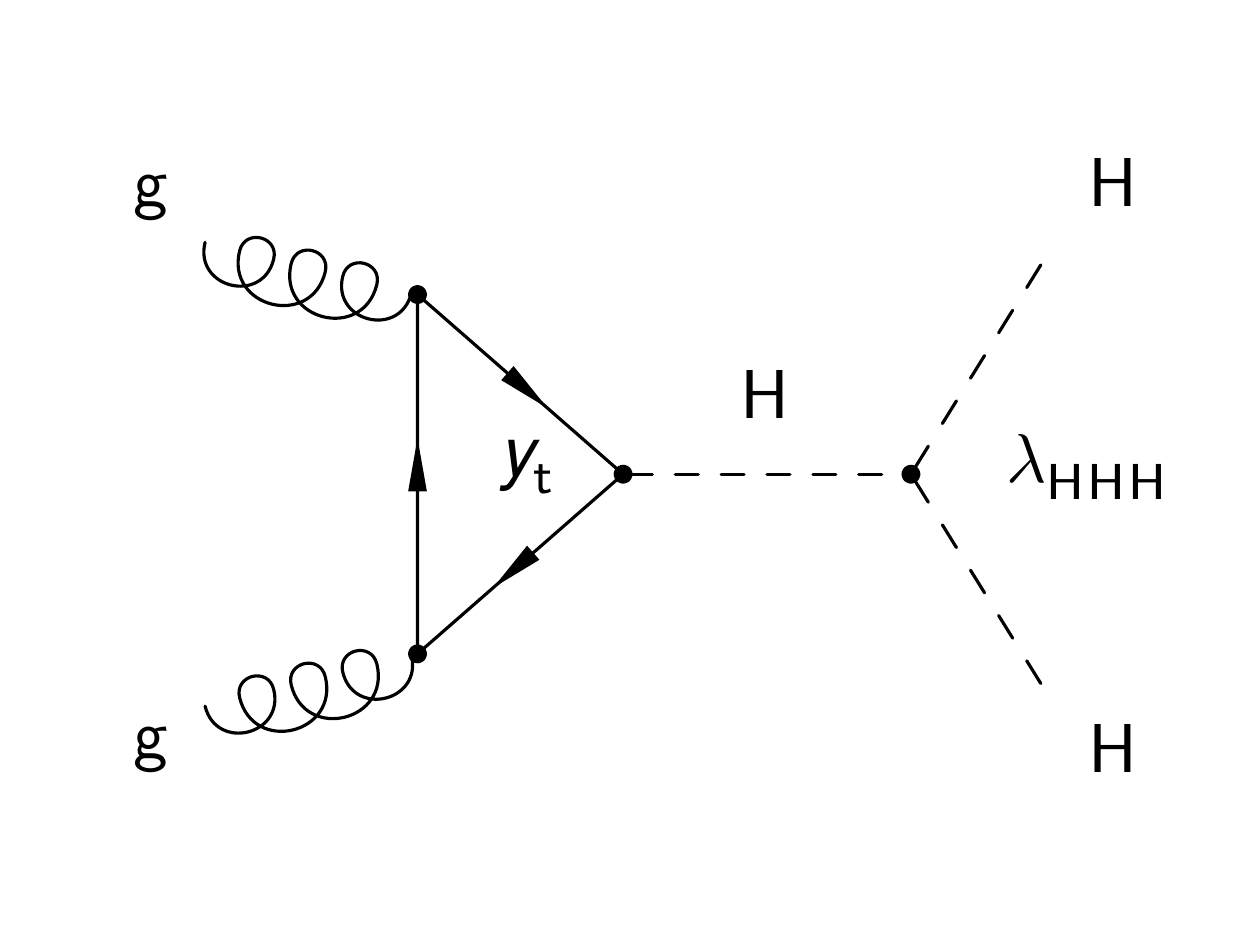}
\includegraphics[width=0.3\textwidth]{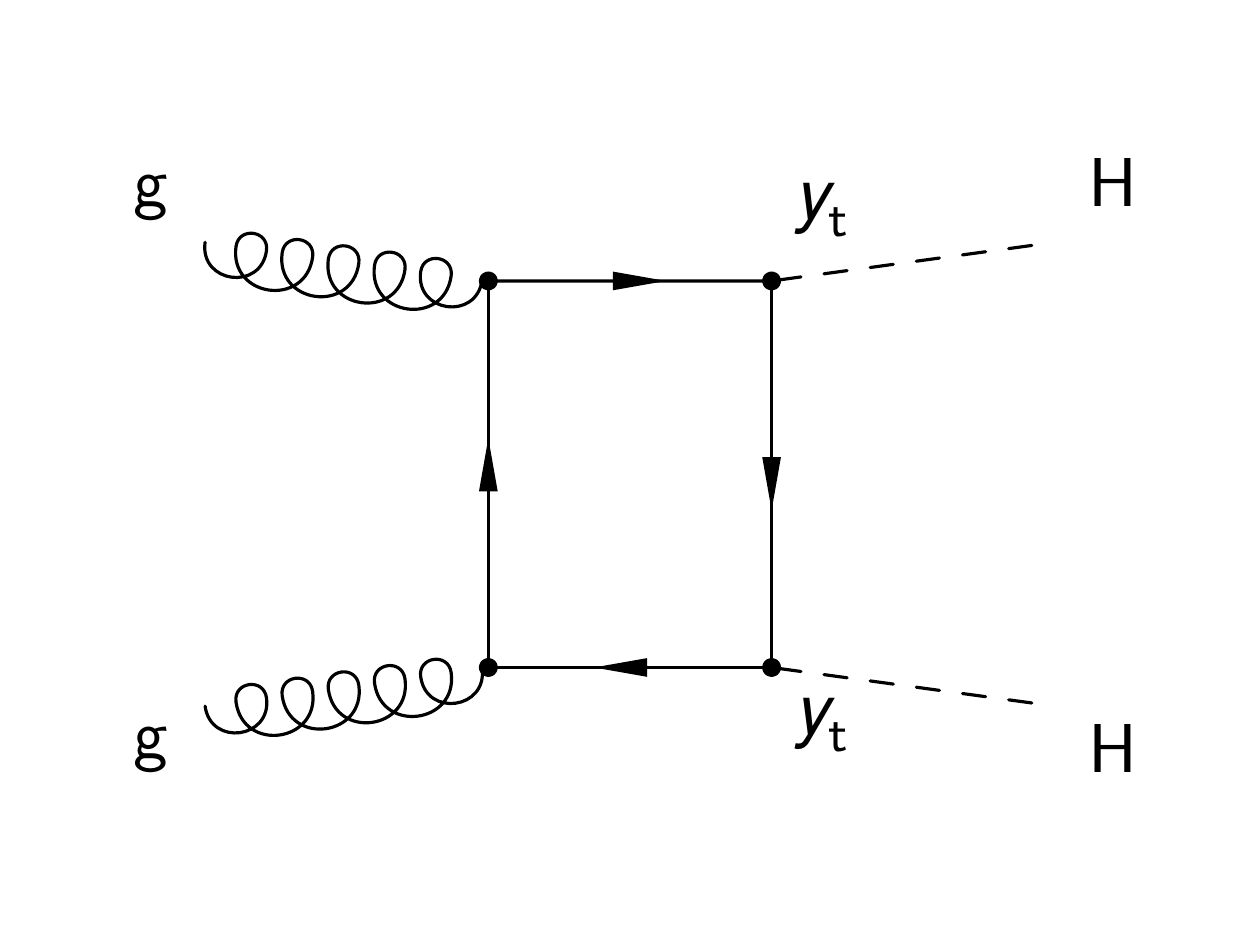}\\
\includegraphics[width=0.3\textwidth]{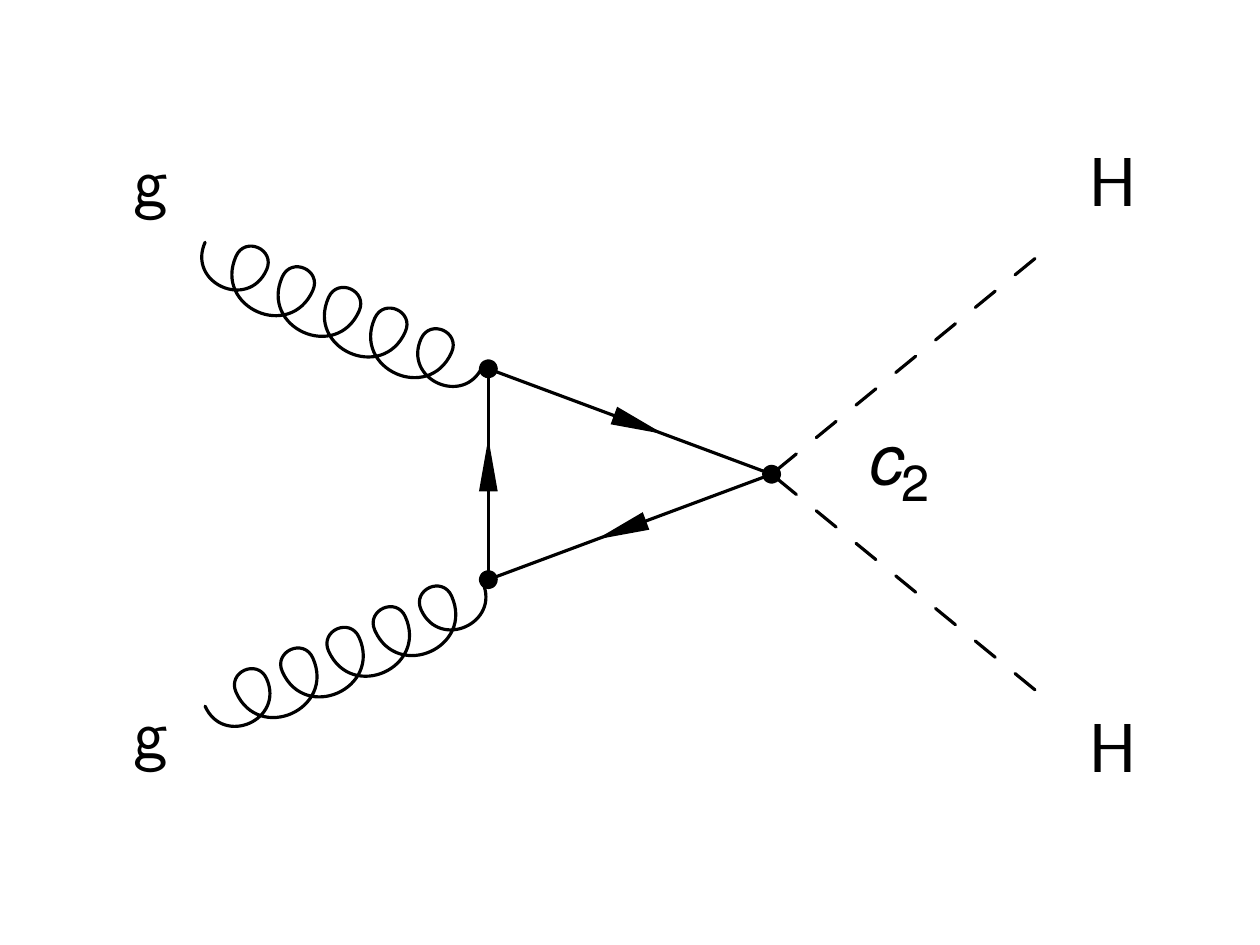}
\includegraphics[width=0.3\textwidth]{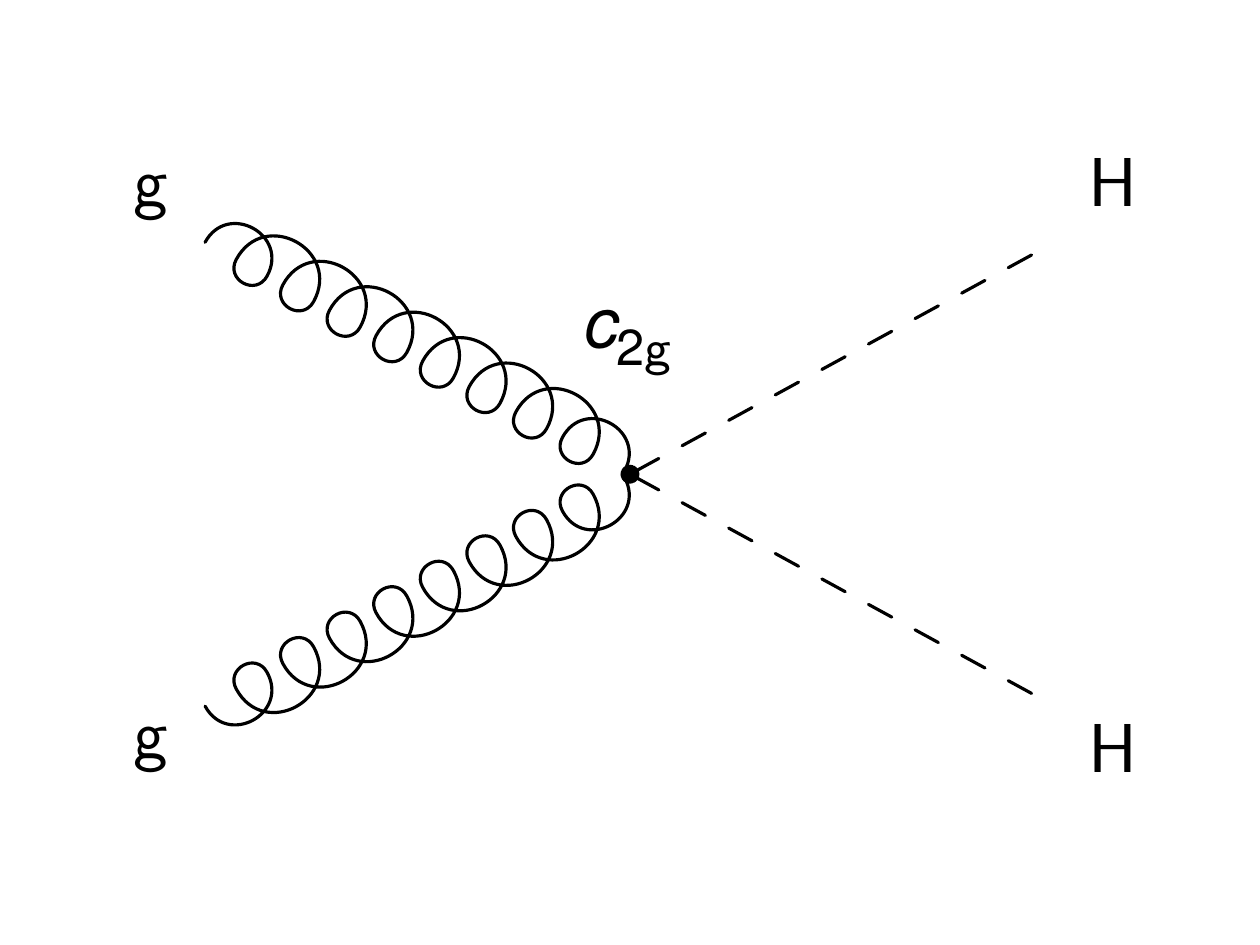}
\includegraphics[width=0.3\textwidth]{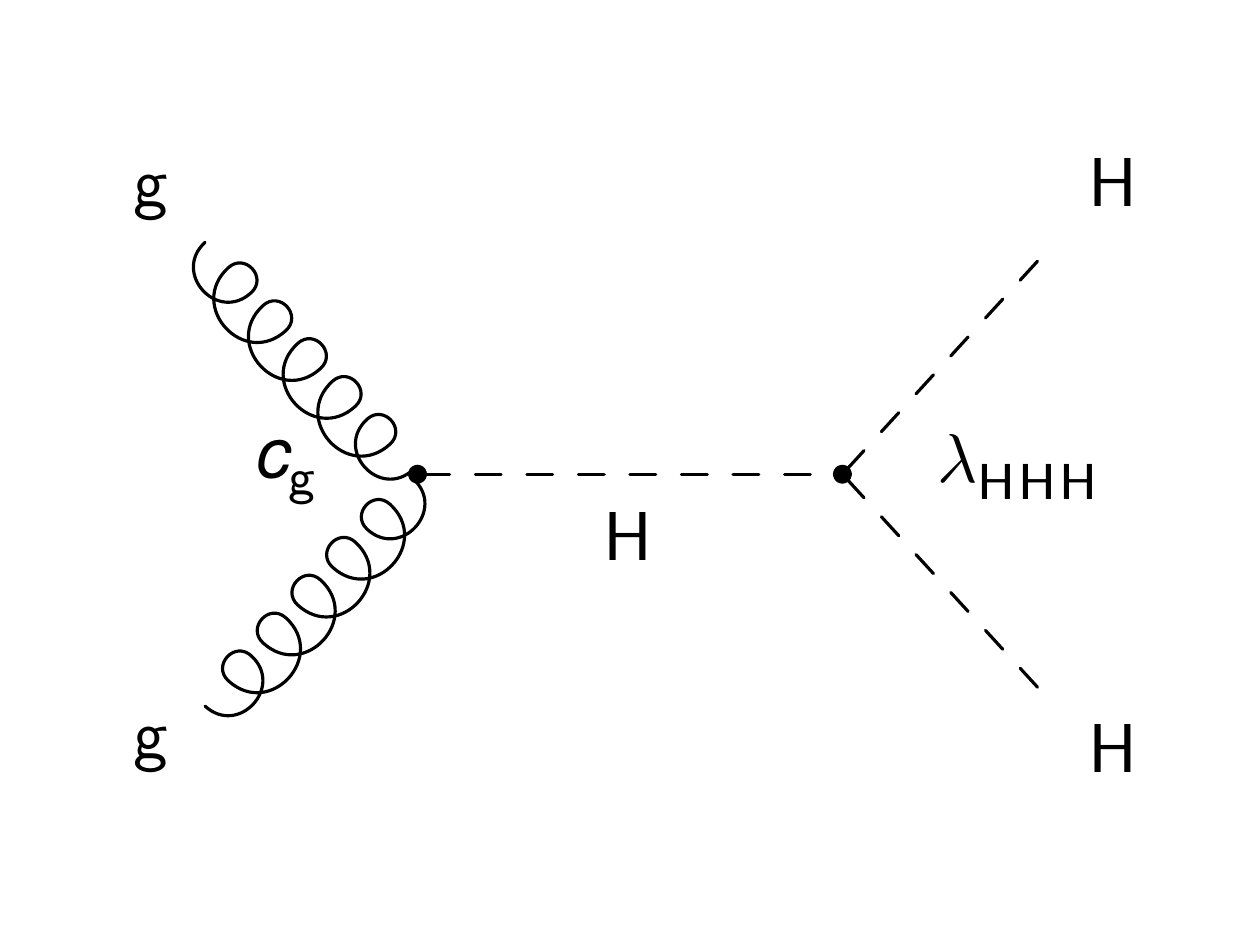}\\
\caption{Feynman diagrams of the processes contributing to the production of Higgs boson pairs via ggF at LO. The upper diagrams correspond to SM processes, involving the top Yukawa coupling \yt and the trilinear Higgs boson self-coupling \lbdHHH, respectively. The lower diagrams correspond to BSM processes: the diagram on the left involves the contact interaction of two Higgs bosons with two top quarks (\ctwo), the middle diagram shows the quartic coupling between the Higgs bosons and two gluons (\cgg), and the diagram on the right describes the contact interactions between the Higgs boson and gluons (\cg).\label{fig:dia}}
\end{figure*}

Contributions from physics beyond the SM (BSM) can 
significantly enhance the \HH production cross section, as well as change the kinematical properties 
of the produced Higgs boson pair, and consequently those of the decay products.
 The modification of the properties of nonresonant \HH production via ggF from BSM effects can be parametrized through an effective Lagrangian that extends the SM one with dimension-6 operators~\cite{deFlorian:2016spz,Goertz:2014qta}. This parametrization results in five couplings: \lbdHHH, the coupling between the Higgs boson and the top quark (\yt), and three additional couplings not present in the SM. Those three couplings represent contact interactions between two Higgs bosons and two gluons ($\cgg$), between one Higgs boson and two gluons ($\cg$), and between two Higgs bosons and two top quarks ($\ctwo$). The Feynman diagrams contributing to ggF \HH production at leading order (LO) are shown in Fig.~\ref{fig:dia}. All five of these couplings are investigated in this analysis.

\begin{figure*}[hbt]
\centering
\includegraphics[width=0.3\textwidth]{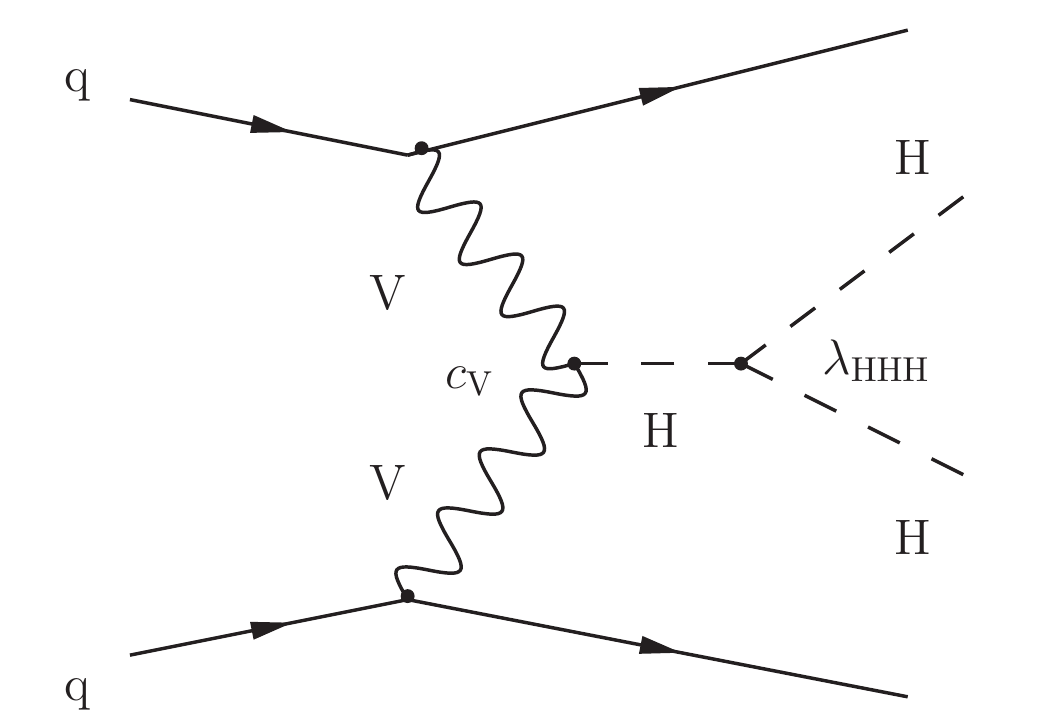}      
\includegraphics[width=0.3\textwidth]{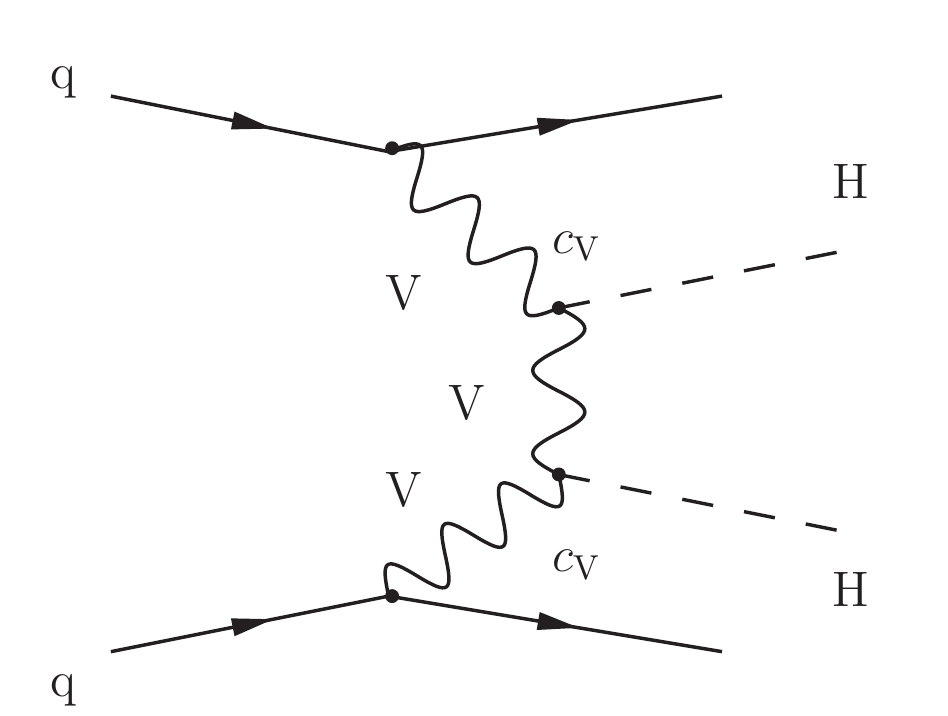}
\includegraphics[width=0.29\textwidth]{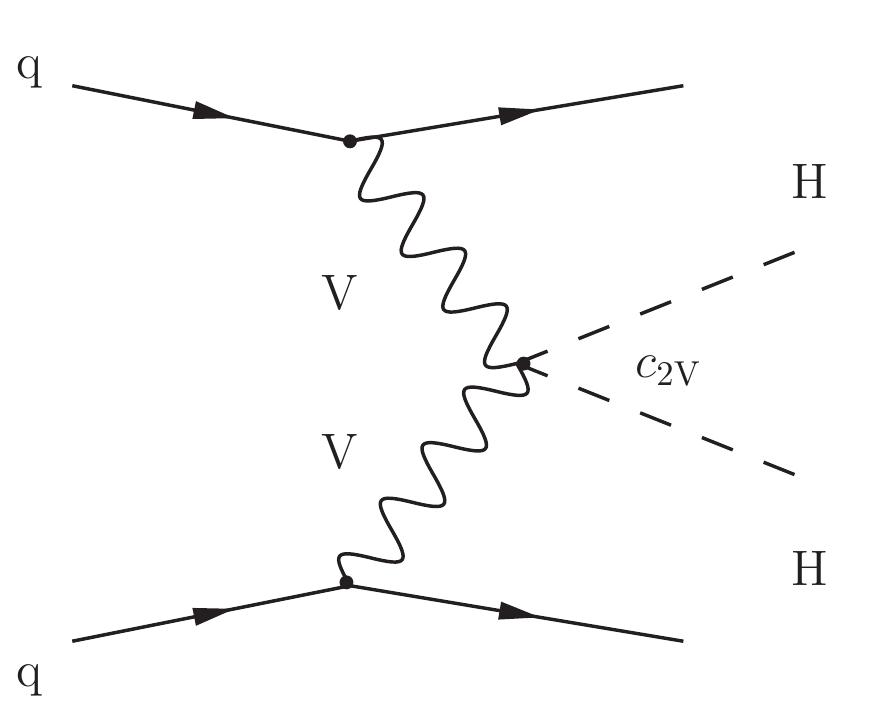}
\caption{Feynman diagrams that contribute to the production of Higgs boson pairs via VBF at LO. On the left the diagram involving the $\PH\PH\PH$ vertex (\lbdHHH), in the middle the diagram with two $\HVV$ vertices ($\conev$), and on the right the diagram with the $\HHVV$ vertex ($\ctwov$).\label{fig:diavbf}}
\end{figure*}

The VBF \HH production mode gives access to \lbdHHH, as well as to the coupling between two vector bosons and the Higgs boson ($\HVV$) and the coupling between a pair of Higgs bosons and a pair of vector bosons ($\HHVV$). The Feynman diagrams contributing to this production mode at LO are shown in Fig.~\ref{fig:diavbf}. While \lbdHHH is mainly constrained from measurements of \HH production via ggF, and the $\HVV$ coupling modifier ($\conev$) is constrained by measurements of vector boson associated production of a single Higgs boson and the decay of the Higgs boson to a pair of bosons~\cite{Sirunyan:2018koj}, the $\HHVV$ coupling modifier ($\ctwov$) is only directly measurable via VBF \HH production. Anomalous values of $\ctwov$ are investigated to establish the presence of the $\HHVV$-mediated process as a probe of BSM physics.

Previous searches for nonresonant production of a Higgs boson pair via ggF were performed by both the ATLAS and CMS Collaborations using the LHC data collected at $\sqrt{s}=8$ and 13\TeV \cite{Aad:2014yja, Aad:2015uka, Aad:2015xja, Aaboud:2016xco,Aad:2019uzh,Khachatryan:2016sey,Aaboud:2018ftw,Aad:2019uzh,Sirunyan:2017tqo, Sirunyan:2017djm, Sirunyan:2017guj,Sirunyan:2018iwt}. Searches in the $\bbgg$ channel performed by the ATLAS~\cite{Aaboud:2018ftw} and CMS~\cite{Sirunyan:2018iwt} Collaborations using up to 36.1\fbinv of $\Pp\Pp$ collision data at $\sqrt{s}=13\TeV$ set upper limits at 95\% confidence level (\CL) on the product of the \HH cross section and the branching fraction into $\bbgg$. The observed upper limits are found to be 24 (30 expected) and 26 (20 expected) times the SM expectation for the ATLAS and CMS searches, respectively.
Statistical combinations of search results in various decay
channels were also performed by the two experiments~\cite{Aad:2019uzh,Sirunyan:2018ayu}. 
Recently, the first search for \HH production via VBF was carried out by the ATLAS Collaboration in the $\bbbb$ channel~\cite{Aad:2020kub}. 

This paper describes a search for nonresonant
production of pairs of Higgs bosons decaying to $\bbgg$
using a data sample of 137\fbinv collected by the CMS experiment from 2016 to 2018.
The $\bbgg$ final state has a combined branching fraction of $2.63 \pm 0.06\times 10^{-3}$~\cite{deFlorian:2016spz} for a Higgs boson mass of 125\GeV.
This channel is one of the most sensitive to \HH production because of the large SM branching fraction of Higgs boson decays to bottom quarks, the good mass resolution of the $\PH \to \gamma \gamma$ channel, and relatively low background rates.

The analysis targets the main \HH production modes: ggF and VBF. Both modes are analyzed following similar strategies.
After reducing the nonresonant $\bbgg$ background and the background coming from single Higgs boson production in association with a top quark-antiquark pair (\ttH), the events are categorized into ggF- and VBF-enriched signal regions using a multivariate technique. The signal is extracted from a fit to the invariant masses of the 
Higgs boson candidates in the $\bb$ and $\gamma\gamma$ final states.
The analysis described in this paper advances the previous \ppHHbbgg search~\cite{Sirunyan:2018iwt} by a factor of four, benefiting equally from the larger collected data sets, and the innovative analysis techniques. The enhanced sensitivity of the present analysis was achieved by improving the \PQb jet energy resolution with a dedicated energy regression, introducing new multivariate methods for background rejection, optimizing the event categorization, and adding dedicated VBF categories.

Finally, the search for Higgs boson pair production is combined with an independent analysis that targets \ttH production, where the Higgs boson decays to a diphoton pair~\cite{Sirunyan:2020sum}. The \ttH production cross section depends on \yt, and also includes a trilinear Higgs boson self-coupling contribution from NLO electroweak corrections~\cite{Maltoni:2017ims,Degrassi:2016wml}. The combination enables \lbdHHH and \yt to be measured simultaneously and provides constraints applicable to a wide range of theoretical models, where both couplings have anomalous values.

This paper is organized as follows: after a brief description of the CMS detector in Section~\ref{sec:detector}, the production of Higgs boson pairs is described in Section~\ref{sec:BSM}. The data samples and simulation, event reconstruction, and analysis strategy are discussed in Sections~\ref{sec:samples},~\ref{sec:objects}, and~\ref{sec:hhsystem}, respectively. Sections~\ref{sec:ttHDiscSec} and~\ref{sec:bkg_rej} are dedicated to the description of the background rejection methods. The event categorization is described in Section~\ref{sec:categor}. Sections~\ref{sec:signal} and~\ref{sec:background} describe the modeling of the signal and background, respectively. The systematic uncertainties are discussed in Section~\ref{sec:syst}. Finally, the results are presented in Section~\ref{sec:res}. The analysis and its results are then summarized in Section~\ref{sec:sum}.
 
\section{The CMS detector}
\label{sec:detector}
The central feature of the CMS apparatus is a superconducting solenoid of 6\unit{m} internal diameter, providing a magnetic field of 3.8\unit{T}. Within the solenoid volume are a silicon pixel and strip tracker, a lead tungstate crystal electromagnetic calorimeter (ECAL), and a brass and scintillator hadron calorimeter (HCAL), each composed of a barrel and two endcap sections. Forward calorimeters extend the pseudorapidity ($\eta$) coverage provided by the barrel and endcap detectors. Muons are detected in gas-ionization chambers embedded in the steel flux-return yoke outside the solenoid.

A more detailed description of the CMS detector, 
together with a definition of the coordinate system used and the relevant kinematic variables, 
can be found in Ref.~\cite{CMSdetector}.

Events of interest are selected using a two-tiered trigger system~\cite{Khachatryan:2016bia}. 
The first level (L1), composed of custom hardware processors, 
uses information from the calorimeters and muon detectors to select events 
at a rate of around 100\unit{kHz} within a time interval of less than 4\mus. 
The second level, known as the high-level trigger, 
consists of a farm of processors running a version of the full event reconstruction software 
optimised for fast processing, and reduces the event rate to around 1\unit{kHz} before data storage~\cite{CMS_HLT}.

The particle-flow algorithm~\cite{ParticleFlow} (PF) aims to reconstruct 
and identify each individual particle in an event (PF candidate), 
with an optimised combination of information from the various elements of the CMS detector. 
The energy of photons is obtained from the ECAL measurement.
The energy of electrons is determined from a combination of the track momentum at the main interaction vertex, the corresponding ECAL cluster energy, and the energy sum of all bremsstrahlung photons attached to the track.
The momentum of muons is obtained from the curvature of the corresponding track. 
The energy of charged hadrons is determined from a combination of their momentum 
measured in the tracker and the matching ECAL and HCAL energy deposits, 
corrected for zero-suppression effects 
and for the response function of the calorimeters to hadronic showers. 
Finally, the energy of neutral hadrons is obtained 
from the corresponding corrected ECAL and HCAL energies.

For each event, hadronic jets are clustered from these reconstructed particles using the infrared and collinear safe anti-\kt algorithm~\cite{Cacciari:2008gp, Cacciari:2011ma} with a distance parameter of 0.4. Jet momentum is determined as the vectorial sum of all particle momenta in the jet, and is found from simulation to be, on average, within 5 to 10\% of the true momentum over the whole \pt spectrum and detector acceptance. Additional proton-proton interactions within the same or nearby bunch crossings can contribute additional tracks and calorimetric energy depositions, increasing the apparent jet momentum. To mitigate this effect, tracks identified to be originating from pileup vertices are discarded and an offset correction is applied to correct for remaining contributions. Jet energy corrections are derived from simulation studies so that the average measured energy of jets becomes identical to that of particle level jets. In situ measurements of the momentum balance in dijet, photon+jet, $\PZ$+jet, and multijet events are used to determine any residual differences between the jet energy scale in data and in simulation, and appropriate corrections are made~\cite{Khachatryan:2016kdb}. Additional selection criteria are applied to each jet to remove jets potentially dominated by instrumental effects or reconstruction failures. The jet energy resolution amounts typically to 15--20\% at 30\GeV, 10\% at 100\GeV, and 5\% at 1\TeV~\cite{Khachatryan:2016kdb}.

The missing transverse momentum vector \ptvecmiss is computed as the negative vector sum 
of the transverse momenta of all the PF candidates in an event, 
and its magnitude is denoted as \ptmiss~\cite{METperformance}. 
The \ptvecmiss is modified to account for corrections to the energy scale of the reconstructed jets in the event.

\section{Higgs boson pair production}
\label{sec:BSM}
Nonresonant ggF \HH production at the LHC can be described using an effective field theory (EFT) approach~\cite{deFlorian:2016spz}. Considering operators up to dimension 6~\cite{Goertz:2014qta}, the tree-level interactions of the Higgs boson are modeled by five parameters. Deviations from the SM values of \lbdHHH and \yt are parametrized as $\kapl \equiv \lbdHHH/\lbdSM$ and $\kapt \equiv \yt/\ytSM$, where the SM values of the couplings are defined as $\lbdSM \equiv \mH^2 /(2 v^2) = 0.129$, $\ytSM = \mt/v \approx 0.7$. Here, $v = 246\GeV$ is the vacuum expectation value of the Higgs field, and $\mt \approx 173\GeV$ is the top quark mass. The anomalous couplings \cgg, \ctwo, and \cg are not present in the SM. The corresponding part of the Lagrangian can be written as ~\cite{Giudice:2007fh}:
\ifthenelse{\boolean{cms@external}}{
\begin{multline}
\mathcal{L}_{\HH} =  {\kapl}\,  \lbdSM v\, \PH^3
  - \frac{ \mt }{v}\bigl({\kapt} \,   \PH  +  \frac{\ctwo}{v}   \, \PH^2 \bigr) \,\bigl( \overline{\PQt}_\text{L}\PQt_\text{R} + \text{h.c.}\bigr)  \\+ \frac{1}{4} \frac{\alpS}{3 \pi v} \bigl(   \cg \, \PH -  \frac{\cgg}{2 v} \, \PH^2  \bigr) \,  G^{\mu \nu}G_{\mu\nu},
\label{eq:lag}
\end{multline}
}{
\begin{multline}
\mathcal{L}_{\HH} =  {\kapl}\,  \lbdSM v\, \PH^3
  - \frac{ \mt }{v}\bigl({\kapt} \,   \PH  +  \frac{\ctwo}{v}   \, \PH^2 \bigr) \,\bigl( \overline{\PQt}_\text{L}\PQt_\text{R} + \text{h.c.}\bigr)  + \frac{1}{4} \frac{\alpS}{3 \pi v} \bigl(   \cg \, \PH -  \frac{\cgg}{2 v} \, \PH^2  \bigr) \,  G^{\mu \nu}G_{\mu\nu},
\label{eq:lag}
\end{multline}
}

where  $\PQt_\text{L}$ and $\PQt_\text{R}$ are the top quark fields with left and right chiralities, respectively. The Higgs boson field is denoted as $\PH$, $G^{\mu \nu}$ is the gluon field strength tensor, and $h.c.$ denotes the Hermitian conjugate.

At LO the full cross section of ggF Higgs boson pair production can be expressed by a polynomial with 15 terms corresponding to five individual diagrams, shown in Fig.~\ref{fig:dia}, and their interference.
It has been observed in Ref.~\cite{Dall'Osso:2015aia} that twelve benchmark hypotheses, described by various combinations of the five parameters ($\kapl$, $\kapt$, $\ctwo$, $\cg$, $\cgg$), are able to represent the distributions of the main kinematic observables of the \HH processes over the full phase space. The parameter values for these benchmark hypotheses are summarized in Table \ref{tab:bench}. The simulated samples generated with the EFT parameters that describe the twelve benchmark hypotheses are combined to cover all possible kinematic configurations of the EFT parameter space.
 The specific kinematic configurations at any point in the full 5D parameter space are obtained through a corresponding reweighting procedure~\cite{Dall'Osso:2015aia,Carvalho:2017vnu} that parametrizes the changes in the differential ggF \HH cross section. 

The reweighting procedure described in Ref.~\cite{Dall'Osso:2015aia} to obtain the distributions of the kinematic observables is implemented for LO only, and cannot be applied to the higher-order simulation because of the presence of additional partons at the matrix element level. Therefore, the 12 BSM signal benchmark hypotheses summarized in Table \ref{tab:bench} are investigated using an LO Monte Carlo (MC) simulation, and only anomalous values of \kapl and \kapt are studied with the NLO simulation, as described in Section~\ref{sec:samples}.

\begin{table*}[h]
  \centering
\topcaption{Coupling parameter values in the SM and in twelve BSM benchmark
hypotheses identified using the method described in Ref.~\cite{Dall'Osso:2015aia}. \label{tab:bench}}
\newcolumntype{R}{>{$}{r}<{$}}
\cmsResize{
  \begin{tabular}{l  R R R   R R R   R R R  R R R  R R }
&     1 &   2 &    3 &    4 &   5 &   6 &   7 &    8 &   9 &   10 &  11 &   12 & $\text{SM}$\\
\hline
$\kapl$   &  7.5 &  1.0 &  1.0 & -3.5 &  1.0 &  2.4 &  5.0 & 15.0 &  1.0 & 10.0 &  2.4 & 15.0 & 1.0\\
$\kapt$  &  1.0 &  1.0 &  1.0 &  1.5 &  1.0 &  1.0 &  1.0 &  1.0 &  1.0 &  1.5 &  1.0 &  1.0 & 1.0 \\
$\ctwo$   & -1.0 &  0.5 & -1.5 & -3.0 &  0.0 &  0.0 &  0.0 &  0.0 &  1.0 & -1.0 &  0.0 &  1.0 &0.0 \\
$\cg$    &  0.0 & -0.8 &  0.0 &  0.0 &  0.8 &  0.2 &  0.2 & -1.0 & -0.6 &  0.0 &  1.0 &  0.0&0.0\\
$\cgg$  &  0.0 &  0.6 & -0.8 &  0.0 & -1.0 & -0.2 & -0.2 &  1.0 &  0.6 &  0.0 & -1.0 &  0.0&0.0\\
\end{tabular}
}
\end{table*}

In the SM, three different couplings are
involved in \HH production via VBF: \lbdHHH, $\HVV$, and $\HHVV$.
The Lagrangians corresponding to the left, middle, and right diagrams in Fig.~\ref{fig:diavbf}~scale with $\conev\kapl$, $\conev^2$, and $\ctwov$, respectively, where $\ctwov$ and $\conev$ are the $\HHVV$ and $\HVV$ coupling modifiers, normalized to the SM values.

\section{Data sample and simulated events} \label{sec:samples}
The analyzed data correspond to a total integrated luminosity of 137\fbinv and were collected over a data-taking period spanning three years: 35.9\fbinv in 2016, 41.5\fbinv in 2017, and 59.4\fbinv in 2018.
Events are selected using double-photon triggers with asymmetric thresholds on the photon transverse momenta of $\ptgone > 30\GeV$ and $\ptgtwo > 18 (22)$\GeV for the data collected during 2016 (2017 and 2018). In addition, loose calorimetric identification requirements~\cite{Sirunyan:2018ouh}, based on the shape of the electromagnetic shower, the isolation of the photon candidate, and the ratio between the hadronic and electromagnetic energy deposit of the shower, are imposed on the photon candidates at the trigger level. 

The ggF \HH signal samples are simulated at NLO~\cite{Bagnaschi:2011tu,Heinrich:2017kxx,Heinrich:2019bkc,Jones:2017giv,Heinrich:2020ckp} including
the full top quark mass dependence~\cite{Buchalla:2018yce} using \POWHEG 2.0. 
The samples are generated for different values of \kapl. As shown in Ref.~\cite{Heinrich:2019bkc} the dependence of the ggF \HH cross section on \kapl and \kapt can be reconstructed from three terms corresponding to the diagrams involving \kapl, \kapt and the interference. Therefore, samples corresponding to any point in the (\kapl, \kapt) parameter space can be obtained from the linear combination of any three of the generated MC samples with different values of \kapl.

In addition, LO signal samples are generated for the BSM benchmark hypotheses described in Section~\ref{sec:BSM} using \MGvATNLO v2.2.2 (2016) or v2.4.2 (2017 and 2018) \cite{Alwall:2014hca, Hespel:2014sla, Frederix:2014hta}.
The simulated LO signal samples, corresponding to the 12 BSM benchmark hypotheses, are added together to increase the number of events, and then reweighted to any coupling configuration ($\kapl$, $\kapt$, $\ctwo$, $\cg$, $\cgg$) using generator-level information on the \HH system. 

The VBF \HH signal samples are generated at LO~\cite{Alwall:2014hca} using \MGvATNLO~v2.4.2. The simulated samples are generated for different combinations of the coupling modifier values ($\kapl$, $\conev$, $\ctwov$). Similarly to what is done for the ggF \HH samples generated at NLO, samples corresponding to any point in the ($\kapl$, $\conev$, $\ctwov$) parameter space can be obtained from the linear combination of any six of the generated samples. 

We apply a global k-factor to the generated ggF \HH and VBF \HH signal samples to scale the cross section to NNLO and next-to-NNLO accuracy respectively. The k-factor is obtained for the cross section prediction in the SM and applied to all considered scenarios. The k-factor for the ggF \HH cross section depends on the invariant mass of the two Higgs bosons, however, within the region of sensitivity of this analysis, this effect is covered by the total scale uncertainty.

The dominant backgrounds in this search are irreducible prompt diphoton production ($\DIPHO$) and the reducible background from $\GAMJETS$ events, where the jets are misidentified as isolated photons and \PQb jets. Although these backgrounds are estimated using data-driven methods, simulated samples are used for the training of multivariate discriminants and the optimization of the analysis categories. The $\DIPHO$ background is modeled with \SHERPA
v.2.2.1~\cite{Gleisberg:2008ta} at LO and includes up to three additional
partons at the matrix element level. In addition, a b-enriched diphoton background is generated with \SHERPA at LO requiring up to two \PQb jets to increase the number of simulated events in the analysis region of interest.
The $\GAMJETS$ background is modeled with \PYTHIA~8.212 \cite{Sjostrand:2014zea} at LO.

Single Higgs boson production, where the Higgs boson decays to a pair of photons, is considered as a resonant background. These production processes are simulated at NLO in QCD precision using \POWHEG 2.0~\cite{Nason:2004rx, POWHEG_Frixione:2007vw, Alioli:2010xd, Bagnaschi:2011tu} for ggF \PH (\ggH) and VBF \PH, and \MGvATNLO v2.2.2 (2016) / v2.4.2 (2017 and 2018) for \ttH, vector boson associated production (\VH), and production associated with a single top quark. The cross sections and decay branching fractions are
taken from Ref.~\cite{deFlorian:2016spz}. The contribution from the other single \PH decay modes is negligible.  

All simulated samples are interfaced with \PYTHIA for parton showering and fragmentation with the standard \pt-ordered parton shower (PS) scheme. The underlying event is modeled with \PYTHIA, using the CUETP8M1 tune for 2016 and the CP5 tune for 2017--2018~\cite{Khachatryan:2015pea, Sirunyan:2019dfx}. PDFs are taken from the NNPDF3.0~\cite{Ball:2014uwa} NLO (2016) or NNPDF3.1~\cite{Ball:2017} NNLO (2017 and 2018) set for all simulated samples except for the signal simulated at LO, for which the {PDF4LHC15\_NLO\_MC} set at NLO~\cite{Carrazza:2015hva, Butterworth:2015oua, Dulat:2015mca, Harland-Lang:2014zoa, Ball:2014uwa} is used. The response of the CMS detector is modeled using the \GEANTfour~\cite{Agostinelli:2002hh} package. The simulated events include additional $\Pp\Pp$ interactions within the same or nearby bunch crossings (pileup), as observed in the data.

Additionally, the simulated VBF \HH signal events are also interfaced with the \PYTHIA dipole shower scheme to model initial-state radiation (ISR) and final-state radiation (FSR)~\cite{Cabouat_2018}. The dipole shower scheme correctly takes into account the structure of the color flow between incoming and outgoing quark lines, and its predictions are found to be in good agreement with the NNLO QCD calculations, as reported in Ref.~\cite{Jager:2020hkz}. These simulated samples are used to derive the uncertainties associated with the \PYTHIA PS ISR and FSR parameters.

\section{Event reconstruction and selection}\label{sec:objects}

The photon candidates are reconstructed from energy clusters in the ECAL not linked to charged-particle tracks (with the exception of converted photons). The photon energies measured by the ECAL are corrected with a multivariate regression technique based on simulation that accounts for radiation lost in material upstream of the ECAL and imperfect shower containment~\cite{Sirunyan:2018ouh}. The ECAL energy scale in data is corrected using simulated $\Zee$ events, while the photon energy in simulated events is smeared to reproduce the resolution measured in data. 

Photons are identified using a boosted decision tree (BDT)-based multivariate analysis (MVA) technique trained to separate photons from jets (photon ID)~\cite{Sirunyan:2018ouh}. The photon ID is trained using variables that describe the shape of the photon electromagnetic shower and the isolation criteria, defined using sums of the transverse momenta of photons, and of charged 
hadrons, inside a cone of radius $\Delta R = \sqrt{\smash[b]{(\Delta \eta)^2 +(\Delta \phi)^2}} = 0.3$ around the photon candidate direction, where $\phi$ is the azimuthal angle in radians. The imperfect MC simulation modeling of the input variables is corrected to match the data using a chained quantile regression method~\cite{cqr} based on studies of $\Zee$ events. In this method, a set of BDTs is trained to predict the cumulative distribution function for a given input. Its prediction is conditional upon the three kinematic variables ($\PT$, $\abs{\eta}$ , $\phi$) and the global event energy density~\cite{Sirunyan:2018ouh}, which are the input variables to the BDTs. The corrections are then applied to the simulated photons such that the predicted cumulative distribution function of the simulated variables is morphed onto the one observed in data.

Events are required to have at least two identified photon candidates that are within the ECAL and tracker fiducial region ($\abs{\eta} < 2.5$), excluding the ECAL barrel-endcap transition region ($1.44 < \abs{\eta} < 1.57$) because the reconstruction of a photon object in this region is not optimal. The photon candidates are required to pass the following criteria: $100 < \Mgg < 180 \GeV$, $\ptgone/\Mgg > 1/3$ and $\ptgtwo/\Mgg > 1/4$, where $\Mgg$ is the invariant mass of the photon candidates. When more than two photon candidates are found, the photon pair with the highest transverse momentum $\ptgg$ is chosen to construct the Higgs boson candidate.

The primary $\Pp\Pp$ interaction vertex in the event is identified using a multivariate technique based on a BDT following the same approach described in Ref.~\cite{Khachatryan:2014ira}. The BDT is trained on simulated \ggH events and has observables related to tracks recoiling against the identified diphoton system as inputs. The efficiency of the correct vertex assignment is greater than 99.9\%, thanks to the requirement of at least two jets in the \bbgg final state.

Jet candidates are required to have $\PT > 25 \GeV$ and $\abs{\eta} < 2.4$ (2.5) for 2016 (2017--2018) and to be separated from the identified photons by a distance of $\DRgj \equiv \sqrt{\smash[b]{(\Detagj)^2 +(\Dphigj)^2}}  > 0.4$. The jet $\eta$ range is extended for the 2017 and 2018 data-taking years because of the new CMS pixel detector installed during the Phase-1 upgrade~\cite{Vormwald_2019}. In addition, identification criteria
are applied to remove spurious jets associated with calorimeter noise~\cite{PUJID}. Jets from the hadronization of \PQb quarks are tagged by a secondary vertex algorithm, \textsc{DeepJet}, 
based on the score from a deep neural network (DNN)~\cite{CMS-DP-2018-058,Bols:2020bkb}. We will refer to the
output of this DNN as the \PQb tagging score.

In addition to standard CMS jet energy corrections~\cite{JINST6}, a \PQb jet energy regression~\cite{Sirunyan:2019wwa} is used to improve the energy resolution of \PQb jets and, therefore, the $\Mjj$ resolution. The energy correction and resolution estimator are computed for each of the Higgs boson candidate jets through a regression implemented in a DNN and trained on jet properties. The regression simultaneously provides a \PQb jet energy correction and a resolution estimator.

In events with more than two jets, the Higgs boson candidate is reconstructed from the two jets with the highest \PQb tagging scores. The dijet invariant mass is required to be~$70 < \Mjj < 190\GeV$.

An additional regression was developed specifically for the $\bbgg$ final states to further improve the dijet invariant mass resolution. This regression exploits the fact that there is no genuine missing transverse momentum from the hard-scattering process in the $\bbgg$ final state, and follows a similar approach as used in Ref.~\cite{Sirunyan:2018iwt}. The regression targets the dijet invariant mass at the generator level, and is trained using the kinematic properties of the event and \ptmiss. The regression is trained on a simulated sample of b-enriched $\DIPHO$ events.

The two regression techniques were validated on data collected by the CMS experiment.
The two-step regression technique improves the dijet invariant mass resolution of the SM \HH signal by about 20\%, and the $\Mjj$ peak position is shifted by 5.5\GeV (5\%) closer to the expected Higgs boson mass.

To select events corresponding to \HH production via VBF, additional requirements are imposed. The VBF
process is characterized by the presence of two additional energetic jets, corresponding to two quarks from each of the colliding protons scattered away from the beam line. These ``VBF-tagged'' jets are expected to have a large pseudorapidity separation, $\dEtavbf$, and a large dijet invariant mass, $\Mjjvbf$.  VBF-tagged jets are required to have $\pt > 40~(30)$\GeV for the leading (subleading) jet, $\abs{\eta} < 4.7$, and be separated from the selected photon and \PQb jet candidates by $\DRgj > 0.4$ and $\DRbj > 0.4$. Jets must also pass an identification criterion designed to reduce the number of selected jets originating from pileup~\cite{PUJID}. The dijet pair with the highest dijet invariant mass $\Mjjvbf$ is selected as the two VBF-tagged jets. We will refer to these requirements as ``VBF selection criteria''.

\section{Analysis strategy}
\label{sec:hhsystem}
To improve the sensitivity of the search, MVA techniques are used to distinguish the ggF and VBF \HH signal from the dominant nonresonant background. The output of the MVA classifiers is then used to define mutually exclusive analysis categories targeting VBF and ggF \HH production. 
 The \HH signal is extracted from a fit to the 
invariant masses of the two Higgs boson candidates in the ($\Mgg$, $\Mjj$) plane simultaneously in all categories.
 
\begin{figure*}[hbt!]
\centering
\includegraphics[width=0.4\textwidth]{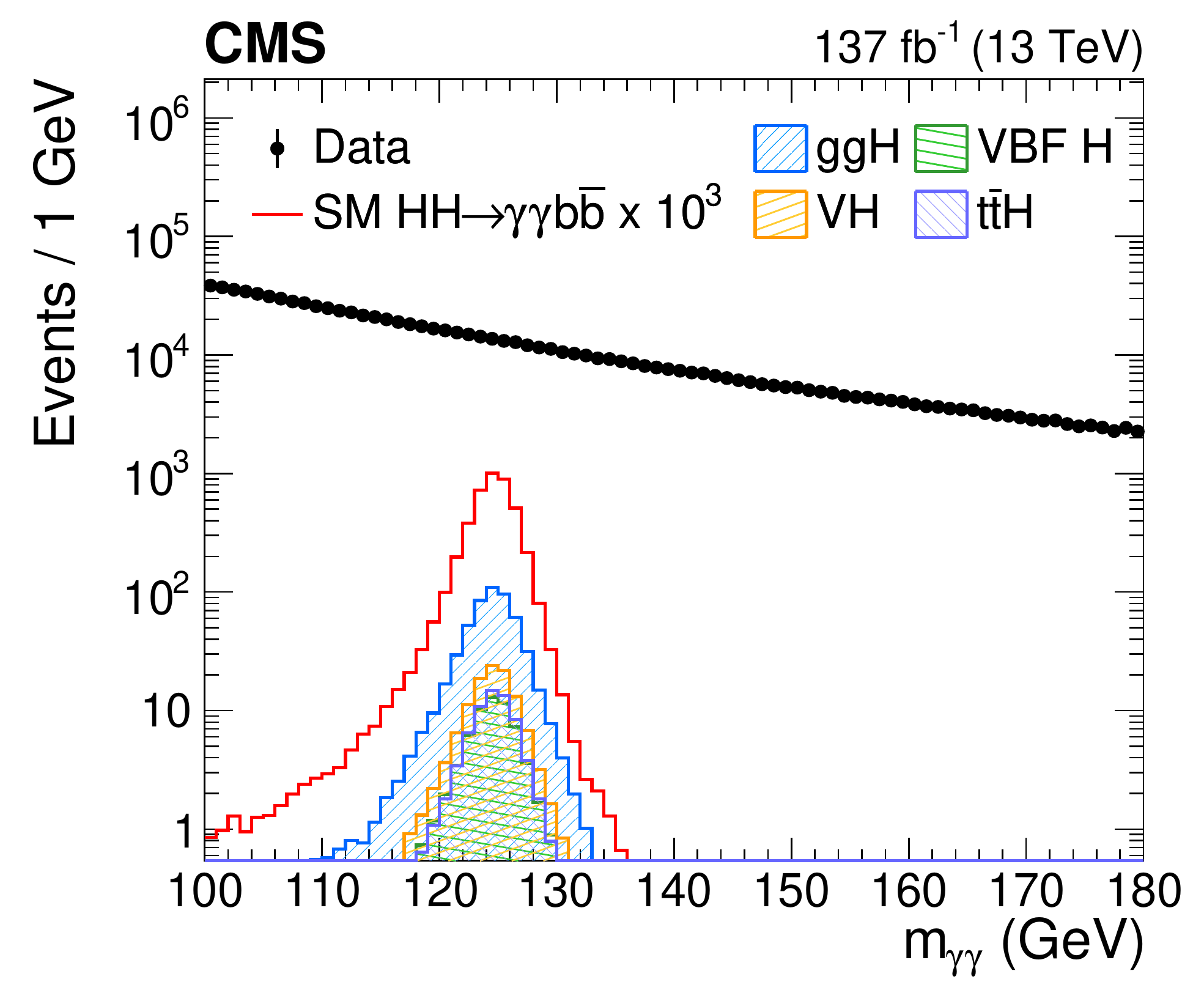}
\includegraphics[width=0.4\textwidth]{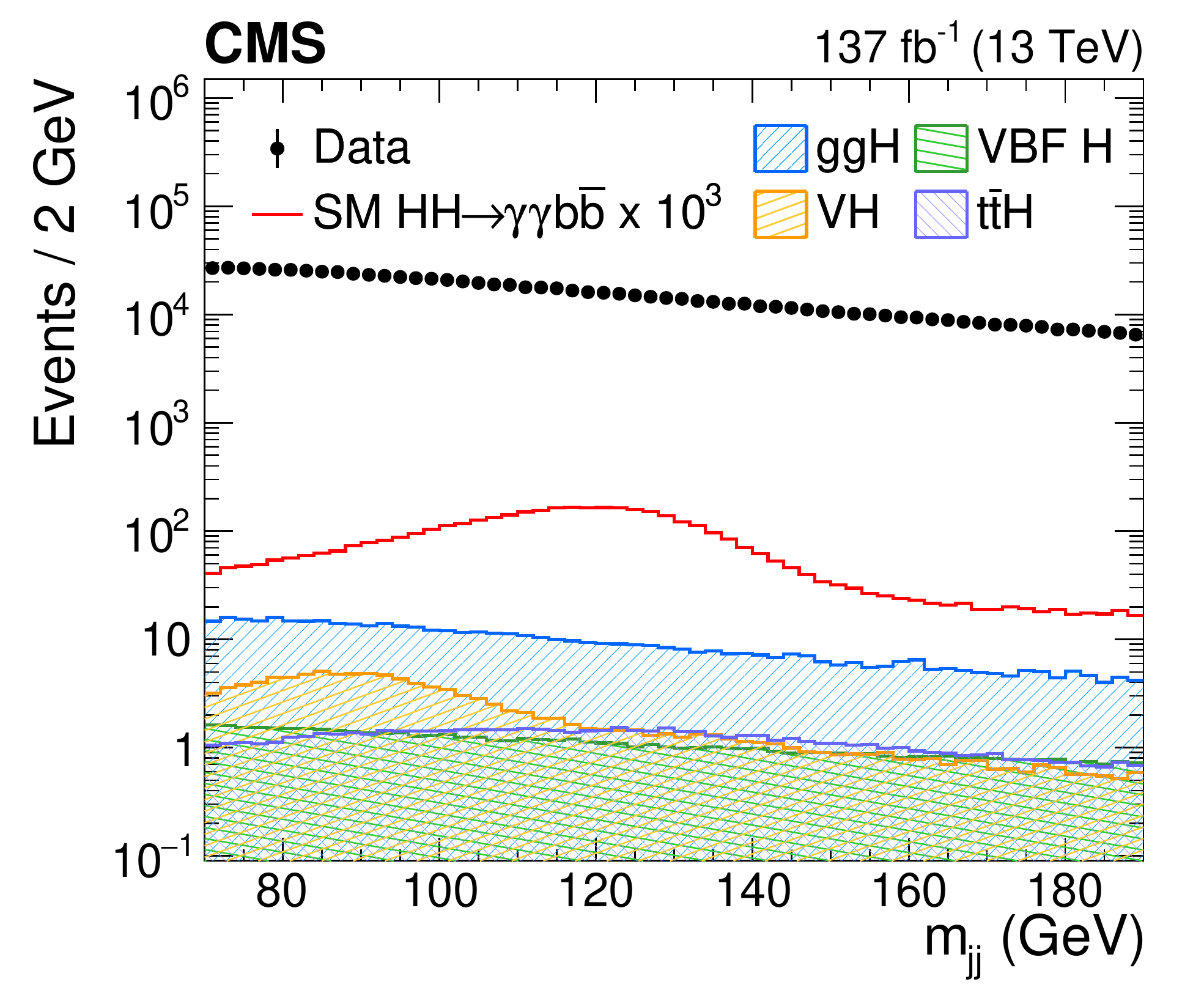}
\caption{ 
The invariant mass distributions of the reconstructed Higgs boson candidates $\Mgg$ (left) and $\Mjj$ (right)
 in data and simulated events. Data, dominated by the $\DIPHO$ and $\GAMJETS$ backgrounds, are compared to the SM ggF $\HH$ signal samples and single \PH samples (\ttH, \ggH, \VBFH, \VH) after imposing the selection criteria described in Section~\ref{sec:objects}. The error bars on the data points indicate statistical uncertainties. The \HH signal has been scaled by a factor of $10^{3}$ for display purposes. }
\label{fig:masses}\end{figure*}

We study the properties of the \HH system, built from the reconstructed diphoton and dijet candidates, to identify observables that can help us distinguish between the signal and background. The invariant mass distributions are shown in Fig.~\ref{fig:masses} for diphoton and dijet pairs in data and in signal and background simulation after imposing the selection criteria described in Section~\ref{sec:objects}. 
The signal has a peaking distribution in $\Mgg$ and $\Mjj$. The data distribution, dominated by the $\DIPHO$ and $\GAMJETS$ backgrounds, exhibits a falling spectrum because of the nonresonant nature of these processes.
In this analysis, these characteristics are used to extract the signal via a fit to $\Mgg$ and $\Mjj$.

\begin{figure*}[hbt!]
\centering
\includegraphics[width=0.4\textwidth]{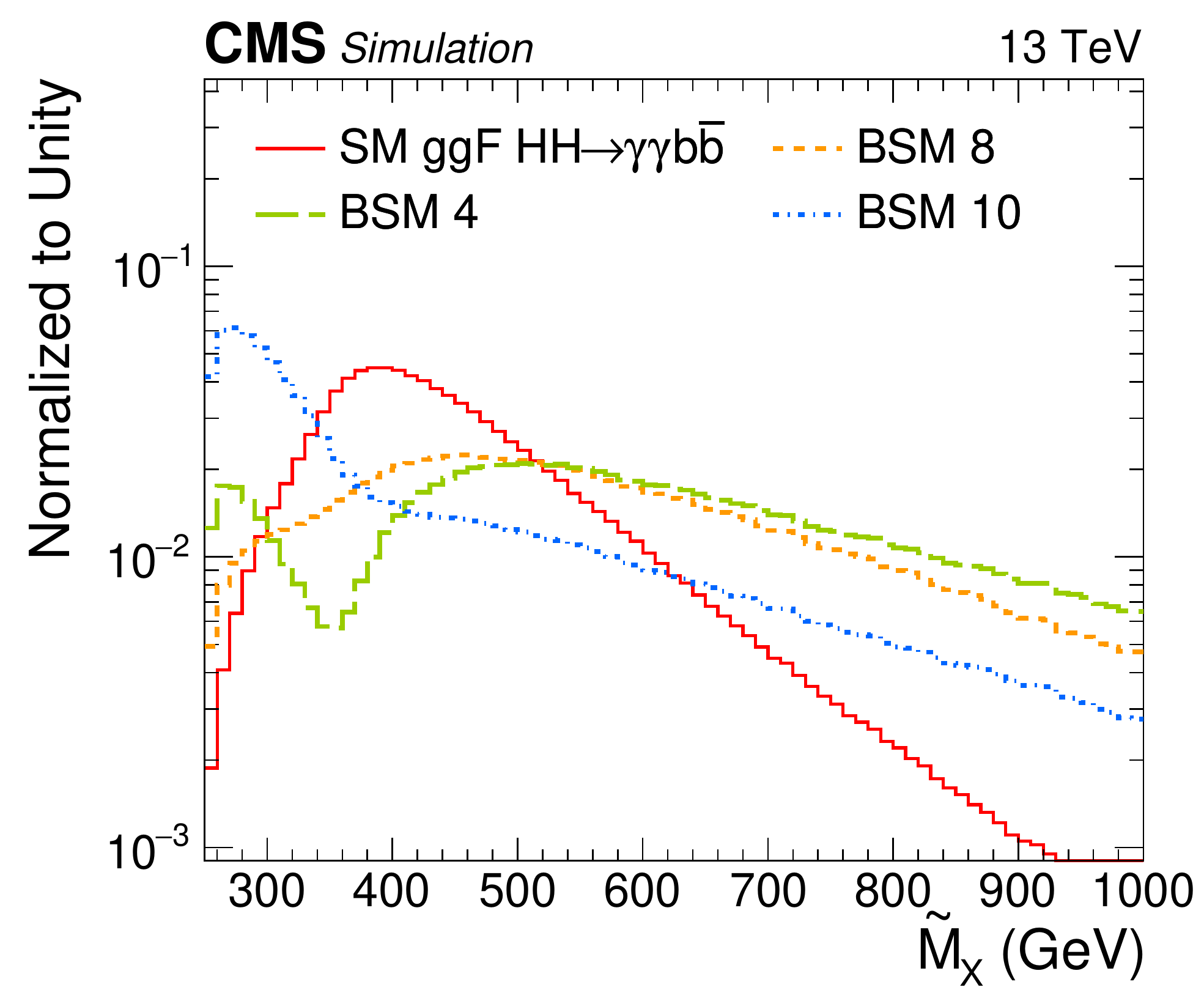}
\includegraphics[width=0.4\textwidth]{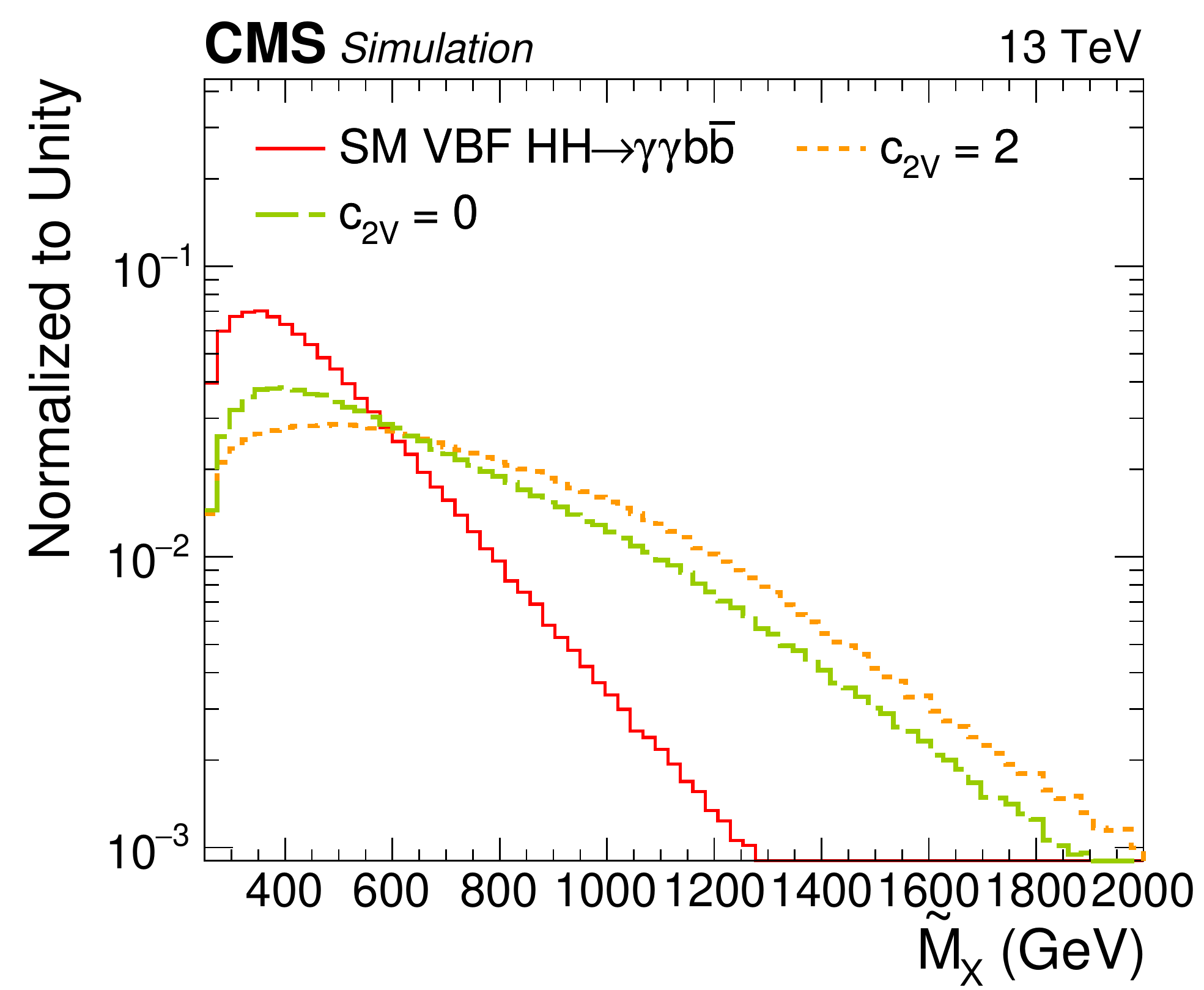}
\caption{Distributions of $\Mtilde$. The SM ggF \HH
signal is compared with several BSM hypotheses listed in Table~\ref{tab:bench} (left),
and the SM VBF \HH signal is compared with two different anomalous values of $\ctwov$ (right). All distributions are normalized to unity.}
\label{fig:mx}\end{figure*}

The distribution of \Mtilde, defined as:
\begin{equation}
\Mtilde = \Mggjj - (\Mjj - \mH) - (\Mgg - \mH),
\end{equation}

where $\Mggjj$ is the invariant mass of the two Higgs boson candidates, is particularly sensitive to different values of the couplings described in Section~\ref{sec:BSM}.
The \Mtilde distribution is less dependent on the dijet and diphoton energy resolutions than $\Mggjj$ if the dijet and diphoton pairs originate from a Higgs boson decay~\cite{Kumar:2014bca}. 
In Fig.~\ref{fig:mx}, the distribution of \Mtilde is shown for several BSM benchmark hypotheses affecting ggF \HH production (described in Table~\ref{tab:bench}) and for different values of $\ctwov$ affecting the VBF \HH production mode. The SM \HH process exhibits a broad structure in \Mtilde, induced by the interference between different processes contributing to $\HH$ production and shaped by the analysis selection. The signals with $\ctwov=0$ and $\ctwov=2$ have a much harder spectrum than the SM VBF \HH signal.

\section{The \texorpdfstring{$\ttH$}{ttH} background rejection}
\label{sec:ttHDiscSec}
Single Higgs boson production is an important resonant background in the \bbgg final state, with \ttH production being dominant in high purity signal regions. To reduce \ttH background contamination, a dedicated classifier (ttHScore) was developed. The classifier is trained on a mixture of SM \HH events and events generated for the twelve BSM benchmark hypotheses (described in Table~\ref{tab:bench}) as signal, and \ttH events as background. The discriminant uses a combination of low-level information from the individual PF candidates and high-level features describing kinematic properties of the event. The kinematic variables used in the training can be classified in three groups: angular variables, variables to distinguish semileptonic decays of \PW bosons produced in the top quark decay, and variables to distinguish hadronic decays of \PW bosons. The ttHScore discriminant is implemented with a DNN 
combining feed-forward and long short-term memory neural networks~\cite{Schmidhuber}, based on the topology-classifier architecture introduced in Ref.~\cite{Nguyen:2018ugw}. The network is implemented in \textsc{Keras}~\cite{chollet2015keras} using the \textsc{TensorFlow}~\cite{tensorflow2015-whitepaper} backend, and the hyperparameters are chosen through Bayesian optimization. 
The ttHScore output is shown in Fig.~\ref{fig:cumulative} (left) for data and simulated events. 
The events entering the analysis are required to pass a selection based on this classifier, which is optimized as described in Section~\ref{sec:categor}.

\section{Nonresonant background rejection}\label{sec:bkg_rej}
\subsection{Background reduction in the ggF \texorpdfstring{\HH}{HH} signal region}
\label{sec:background_ggf}
An MVA discriminant implemented with a BDT is used to separate the ggF \HH signal and the dominant nonresonant $\DIPHO$ and $\GAMJETS$ backgrounds. We select several discriminating observables to be used in the training. They can be classified in three groups: kinematic variables, object identification variables, and object resolution variables. The first group exploits the kinematic properties of the \HH system, the second helps to separate the signal from the reducible $\GAMJETS$ background, and the third takes into account the
resonant nature of the $\gamma\gamma$ and $\bbbar$ final states for signal. The following discriminating variables were chosen: 

\begin{itemize}
\item The $\PH$ candidate kinematic variables: $\ptg/\Mgg$, $\ptj/\Mjj$ for leading and subleading photons and jets, where $\ptg$ and $\ptj$ are the transverse momenta of the selected photon and jet candidates.
\item The \HH transverse balance: $\ptgg/\Mggjj$ and $\ptjj/\Mggjj$, where $\ptgg$ and $\ptjj$ are the transverse momenta of the diphoton and dijet candidates.
\item Helicity angles: \acosthetastar, \acosthetabb, \acosthetagg, where $\acosthetastar$ is the Collins-Soper angle~\cite{PhysRevD.16.2219} between the direction of the \Hgg candidate and the average beam direction in the \HH center-of-mass frame, while $\acosthetabb$ and $\acosthetagg$ are the angles between one of the Higgs boson decay products and the direction defined by the Higgs boson candidate.
\item Angular distance: minimum \DRgj between a photon and a jet, $\minDRgj$, considering all combinations between objects passing the selection criteria, and \DRgj between the other photon-jet pair not used in the $\minDRgj$ calculation. 
\item \PQb tagging: the \PQb tagging score of each jet in the dijet candidate.
\item photon ID: photon identification variables for leading and subleading photons.
\item Object resolution: energy resolution for the leading and subleading photons and jets obtained from the photon~\cite{Sirunyan:2018ouh} and \PQb jet~\cite{Sirunyan:2019wwa} energy regressions, the mass resolution estimators for the diphoton and dijet candidates.
\end{itemize}

\begin{figure*}[thb]
  \centering
  \includegraphics[width=0.45\textwidth]{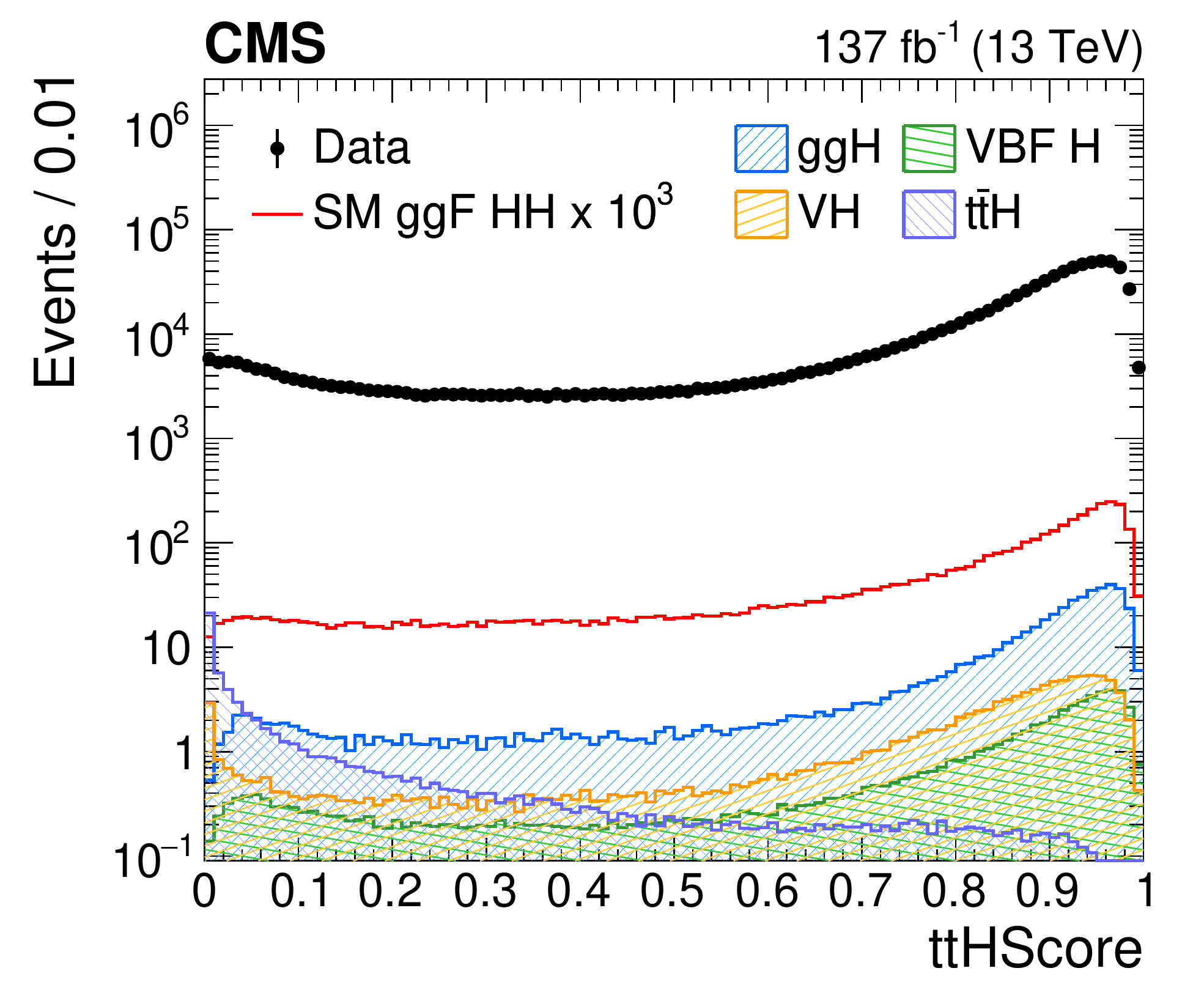}\hfil 
  \includegraphics[width=0.45\textwidth]{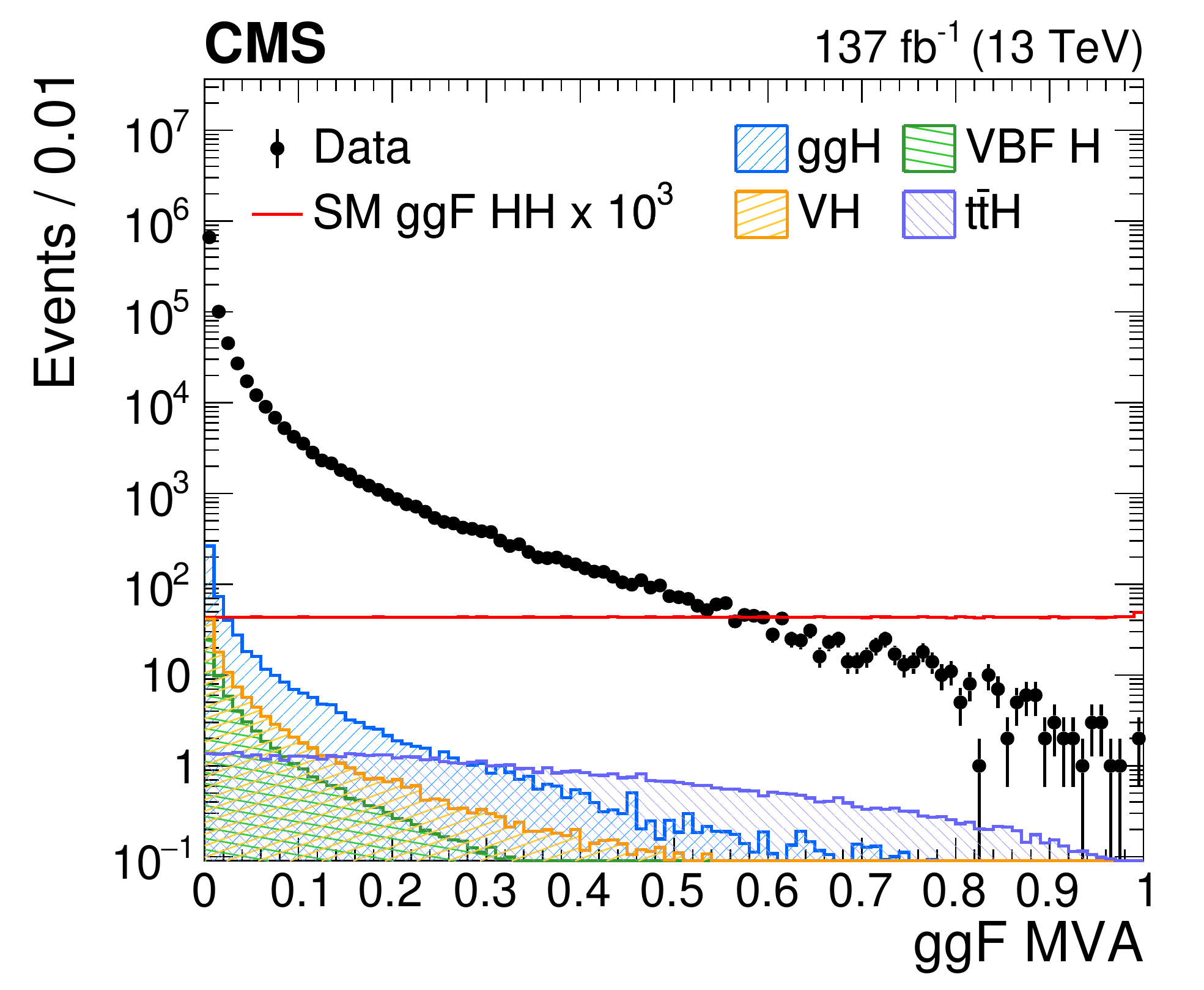}\hfil 
  \caption{The distribution of the ttHScore (left) and MVA output (right) in data and simulated events. Data, dominated by $\DIPHO$ and $\GAMJETS$ background, are compared to the SM ggF $\HH$ signal samples and single \PH samples (\ttH, \ggH, \VBFH, \VH) after imposing the selection criteria described in Section~\ref{sec:objects}. The error bars on the data points indicate statistical uncertainties. The \HH signal has been scaled by a factor of $10^{3}$ for display purposes. }
  \label{fig:cumulative}
\end{figure*}

The BDT is trained using the \textsc{xgboost}~\cite{xgboost} software package using a gradient boosting algorithm.
The \DIPHO and $\GAMJETS$ MC samples are used as background, while an ensemble of SM \HH and the 12 BSM \HH benchmark hypotheses listed in Table~\ref{tab:bench} is used as signal. Training on an ensemble of BSM and SM \HH signals makes the BDT sensitive to a broad spectrum of theoretical scenarios. During the training, signal
events are weighted with the product of the inverse mass resolution of the diphoton and dijet systems. These resolutions are obtained using the per-object
resolution estimators provided by the energy regressions developed for photons and \PQb jets. 
In the training, the mass dependence of the classifier is removed by using only dimensionless kinematic variables.
The inverse resolution weighting at
training time improves the performance by bringing back the information about the resonant nature of the signal.
Independent training and testing samples are created
by splitting the signal and background samples. The classifier hyperparameters are optimized using a randomized grid search and a 5-fold cross-validation technique~\cite{Hastie}. The BDT is trained separately for the 2016, 2017, and 2018 data-taking years.
 The BDT output distribution is very similar among the three years, leading to the same definitions of optimal signal regions based on the BDT output. Therefore, during the event categorization, a single set of analysis categories is defined using data from 2016--2018.
The distributions of the BDT output for signal and background are very well separated. In order to avoid problems of numerical precision when defining optimal signal-enriched regions, the BDT output is transformed such that the signal distribution is uniform. This transformation is applied to all events, both in simulation
and data. The distribution of the MVA output for data and simulated events is shown in Fig.~\ref{fig:cumulative} (right). 

\subsection{Background reduction in the VBF \texorpdfstring{\HH}{HH} signal region}
\label{sec:background_vbf}

Similarly to the ggF \HH analysis strategy, an MVA discriminant is employed to separate the VBF \HH signal from the background. As for the ggF case, the \DIPHO and $\GAMJETS$ processes are the dominant sources of background. For the VBF production mode, the ggF \HH events are considered as background.
About a third of the ggF \HH events passing the selection requirements described in Section~\ref{sec:objects} also pass the dedicated VBF selection criteria. The distinctive topology of the VBF \HH process is used to separate the VBF \HH signal from the various sources of background. In addition to the discriminating features of the \HH signal described in Sections~\ref{sec:hhsystem} and~\ref{sec:background_ggf}, the following set of VBF-discriminating features were identified: 

\begin{itemize}
\item VBF-tagged jet kinematic variables: $\ptvbf/\Mjjvbf$, $\etavbf$ for VBF-tagged jets.
\item VBF-tagged jet invariant mass: invariant mass $\Mjjvbf$ of the VBF-tagged jets.
\item Rapidity gap: product of and difference in the pseudorapidity of the two VBF-tagged jets.
\item Quark-gluon likelihood~\cite{CMS:2013kfa,JetsInRun2} of the two VBF-tagged jets. A likelihood discriminator used to distinguish between jets originating from quarks and from gluons.
\item Kinematic variables related to the \HH system: $\Mtilde$ and the transverse momentum of the pair of reconstructed Higgs bosons. 
\item Angular distance: minimum $\DR$ between a photon and a VBF-tagged jet, and between a \PQb jet and a VBF-tagged jet.
\item Centrality variables for the reconstructed Higgs boson candidates: 
\begin{equation}
C_{\PH} = \exp\left[-\frac{4}{(\eta^{\mathrm{VBF}}_1-\eta^{\mathrm{VBF}}_2)^2}\left(\eta^{\PH}-\frac{\eta^{\mathrm{VBF}}_{1}+\eta^{\mathrm{VBF}}_{2}}{2} \right)^2  \right],
\end{equation}
where $\PH$ is the Higgs boson candidate reconstructed either from diphoton or dijet pairs, and $\eta^{\mathrm{VBF}}_{1}$ and $\eta^{\mathrm{VBF}}_{2}$ are the pseudorapidities of the two VBF-tagged jets.

\end{itemize}

We split events into two regions: $\Mtilde<500$\GeV and $\Mtilde>500$\GeV. While the region of $\Mtilde>500$\GeV is sensitive to anomalous values of $\ctwov$, the $\Mtilde<500$\GeV region retains the sensitivity to SM VBF \HH production.

\begin{figure*}[thb]
  \centering
  \includegraphics[width=0.45\textwidth]{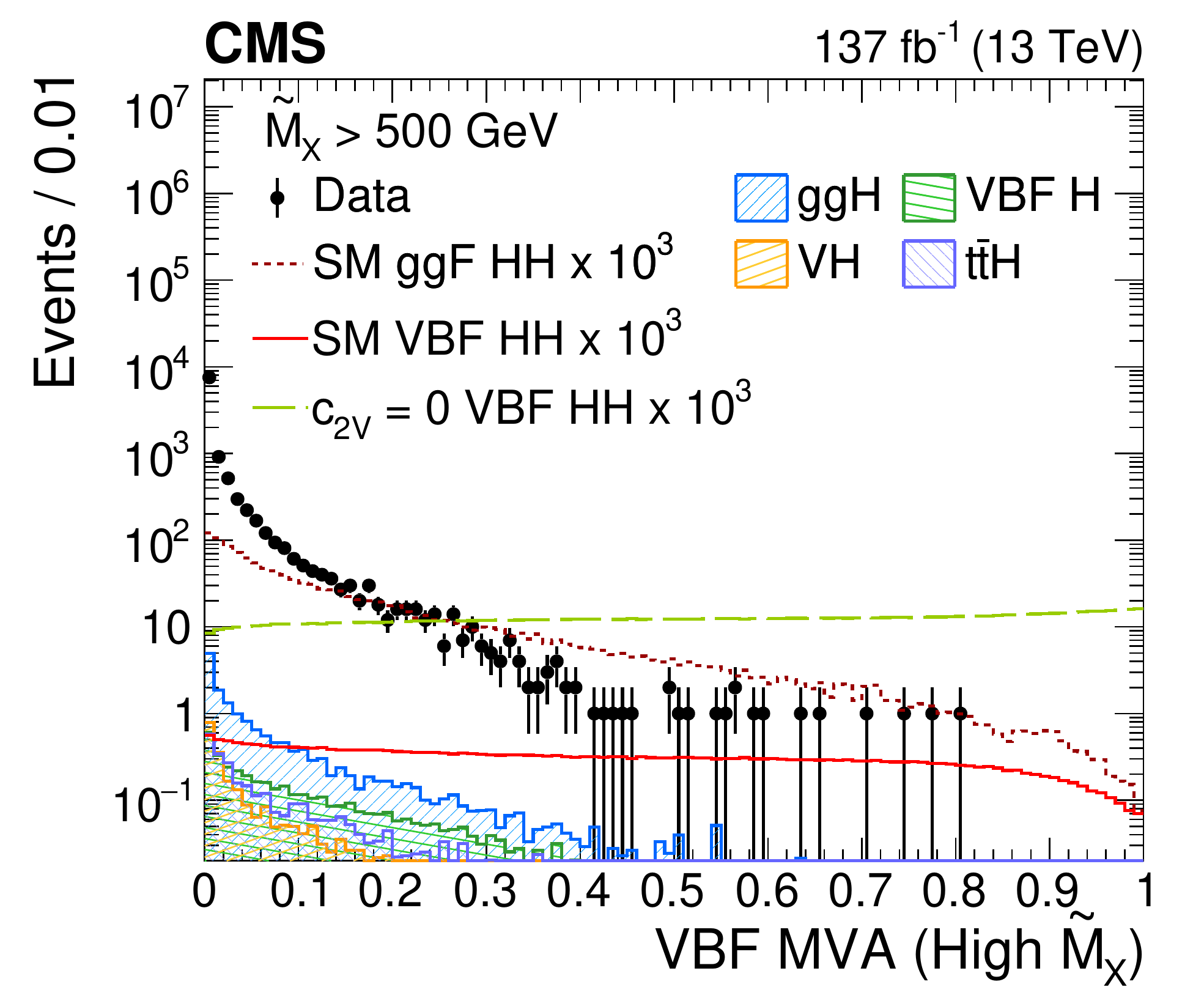}\hfil 
  \includegraphics[width=0.45\textwidth]{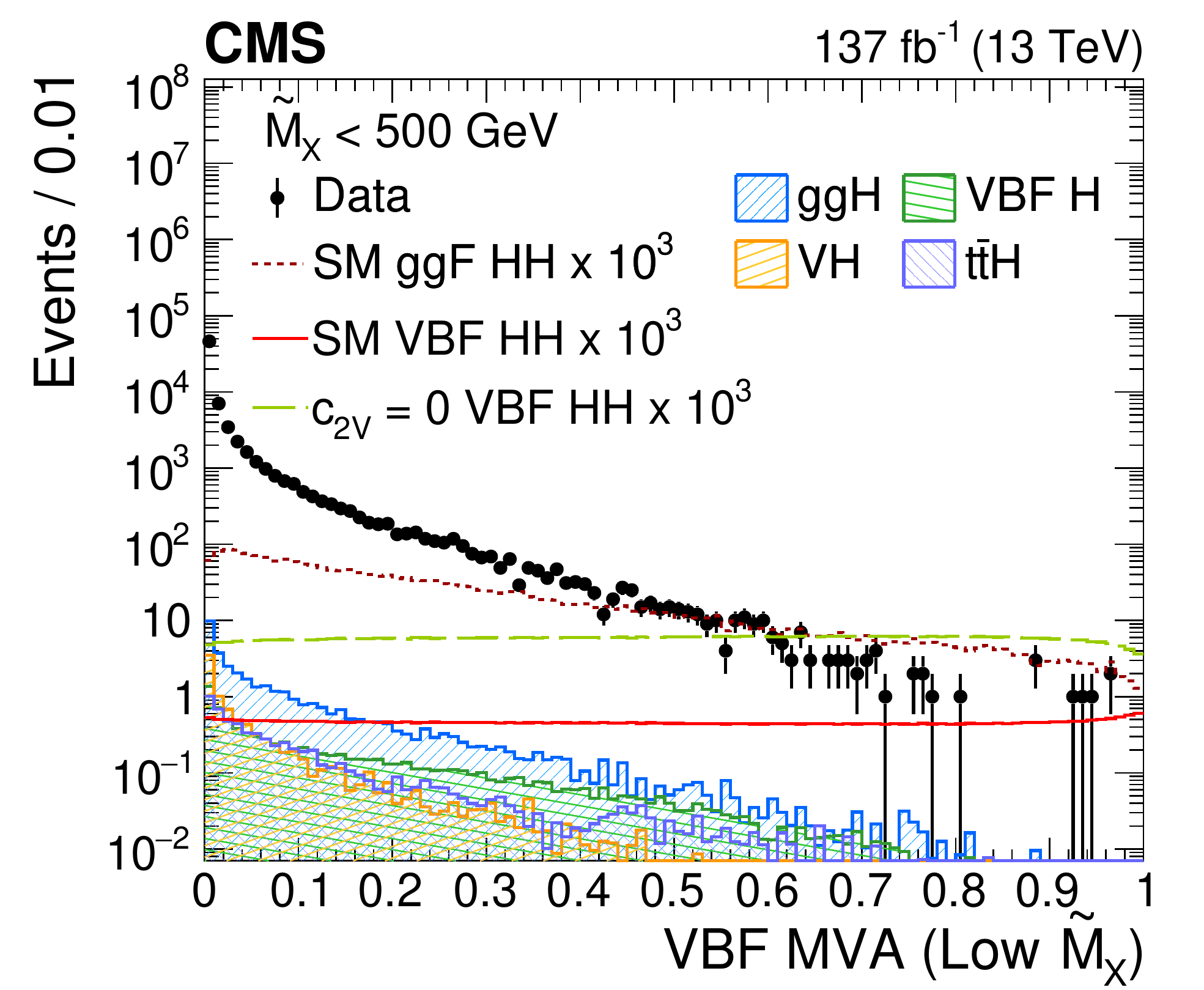}\hfil 
  \caption{The distribution of the two MVA outputs is shown in data and simulated events in the two VBF $\Mtilde$ regions: $\Mtilde>500$\GeV (left) and $\Mtilde<500$\GeV (right). Data, dominated by the $\DIPHO$ and $\GAMJETS$ backgrounds, are compared to the VBF $\HH$ signal samples with SM couplings and $\ctwov=0$, SM ggF $\HH$ and single \PH samples (\ttH, \ggH, \VBFH, \VH) after imposing the VBF selection criteria described in Section~\ref{sec:objects}. The error bars on the data points indicate statistical uncertainties. The \HH signal has been scaled by a factor of $10^{3}$ for display purposes. }
  \label{fig:cumulativeVBF}
\end{figure*}

A multi-class BDT, using a gradient boosting algorithm and implemented in the \textsc{xgboost} \cite{xgboost} framework, is trained to separate the VBF \HH signal from the \DIPHO, $\GAMJETS$, and SM ggF \HH background. A mix of VBF \HH samples with the SM couplings and quartic coupling $\ctwov=0$ is used as signal.
Training on the mix of samples makes the BDT sensitive to both SM and BSM scenarios.
Although the kinematic properties of different BSM signals with anomalous values of $\ctwov$ are similar, the cross section of the signal with $\ctwov=0$ is significantly enhanced with respect to that predicted by the SM. Therefore, the signal samples used for the training were chosen to maximize sensitivity of the analysis to a range of potential signals. Signal events are weighted with the inverse of the mass resolution of the diphoton and dijet systems during the training, as it is done for the ggF MVA.
The BDT is trained separately for each of the three data-taking years in the two $\Mtilde$ regions. As it is done for the ggF MVA output, data from 2016--2018 are merged to create a single set of analysis categories based on the BDT output. The BDT output is transformed such that the distribution of the mix of the VBF \HH signals with SM couplings and quartic coupling $\ctwov=0$ is uniform. The transformation is applied to all events in the two $\Mtilde$ regions. The distribution of the MVA outputs for data and simulated events is shown in Fig.~\ref{fig:cumulativeVBF}.

\section{Event categorization}
\label{sec:categor}
In order to maximize the sensitivity of the search, events are split into different categories
according to the output of the MVA classifier 
and the mass of the Higgs boson pair system $\Mtilde$. The $\Mtilde$ distribution changes significantly for different BSM hypotheses, as shown in Fig.~\ref{fig:mx}. Therefore, a categorization of \HH events in $\Mtilde$ creates signal regions sensitive to multiple theoretical scenarios. In the search for VBF \HH production, the categories in $\Mtilde$ are defined before the MVA is trained, as described in Section~\ref{sec:background_vbf}. 
For the categories that target ggF \HH production, categories in $\Mtilde$ are defined after the MVA is trained.

The categorization is optimized by maximizing the expected significance estimated as the sum in quadrature of $\mathrm{S}/\sqrt{\mathrm{B}}$ over all categories in a window centered on \mH: $115<\Mgg<135$\GeV. Here, S and B are
the numbers of expected signal and background events, respectively.
 Simulated events are used for this optimization. The SM \HH process is considered as signal, while the background consists of the $\DIPHO$, $\GAMJETS$, and $\ttH$ processes. The MVA categories are optimized simultaneously with a threshold on the value of ttHScore. Two VBF and three ggF categories are optimized based on the MVA output. For ggF \HH in each MVA category a set of $\Mtilde$ categories is then optimized. The optimization procedure leads to 12 ggF analysis categories: four categories in $\Mtilde$ in each of the three categories in the MVA score. The optimized selection on $\text{ttHScore} > 0.26$ corresponds to 80 (85)\% \ttH background rejection at 95 (90)\% signal efficiency for the 12 ggF (2 VBF) categories. 
The categorization is summarized in Table~\ref{tab:categories}. 
The VBF and ggF categories are mutually exclusive, as we only consider events that do not enter the VBF categories for the ggF categories. 
Events with VBF MVA scores below 0.52 (0.86) for $\Mtilde>500$ ($\Mtilde<500$)\GeV are not considered in the VBF signal region. Because of the overwhelming background contamination such events do not
improve the expected sensitivity of the analysis. Similarly, events with ggF MVA scores below 0.37 are not considered in the ggF signal region. 

\begin{table*}
    \centering
    \topcaption{Summary of the analysis categories. Two VBF- and twelve ggF-enriched categories are defined based on the output of the MVA classifiers 
and the mass of the Higgs boson pair system $\Mtilde$. The VBF and ggF categories are mutually exclusive.}
        \begin{tabular}{ l  c  c  c   c   c   c }
          Category   & MVA &  \Mtilde (\GeVns{})\\  \hline
      VBF CAT 0 & 0.52--1.00 & $>$500 \\
      VBF CAT 1 & 0.86--1.00 & 250--500  \\ 
      ggF CAT 0 &  0.78--1.00 & $>$600 \\
      ggF CAT 1 & & 510--600  \\ 
      ggF CAT 2&  & 385--510   \\
      ggF CAT 3&  & 250--385  \\ 
      ggF CAT 4& 0.62--0.78  & $>$540  \\ 
      ggF CAT 5 & & 360--540  \\
      ggF CAT 6 &  & 330--360  \\ 
      ggF CAT 7 & & 250--330  \\
      ggF CAT 8 & 0.37--0.62& $>$585 & \\ 
      ggF CAT 9 & & 375--585  \\ 
      ggF CAT 10& & 330--375  \\ 
      ggF CAT 11& & 250--330  \\ 
    \end{tabular}
    \label{tab:categories}
\end{table*}

\subsection{Combination of the \texorpdfstring{\HH}{HH} and \texorpdfstring{\ttH}{ttH} signals to constrain \texorpdfstring{$\kapl$}{kappa lambda} and \texorpdfstring{$\kapt$}{kappa top}}
\label{sec:klkaptfit}

As discussed in Section~\ref{sec:BSM}, the \HH production cross section depends on \kapl and \kapt. The production cross section of the single \PH processes also depends on \kapl, as a result of NLO electroweak corrections~\cite{Maltoni:2017ims}. The \ggH and \ttH production cross sections additionally depend on \kapt. Therefore, the \HHbbgg signal can be combined with the single \PH production modes to provide an improved constraint on the \kapl and \kapt parameters. In the case of anomalous values of \kapl, the single \PH process with the largest modification of the cross section is \ttH. For this reason, additional orthogonal categories targeting the \ttH process are included in the analysis: the ``$\ttH$ leptonic'' and the ``$\ttH$ hadronic'' categories, developed and optimized for the measurement of the \ttH production cross section in the diphoton decay channel~\cite{Sirunyan:2020sum}.
The events that do not pass the selections for the \HH categories defined in Table~\ref{tab:categories} are tested for the \ttH categories. This ensures the orthogonality between the events selected by the \HH and \ttH categories. 

The \Hgg candidate selection is the same as described in Section~\ref{sec:objects}. 
The \ttH leptonic categories target \ttH events where at least one \PW boson, originating from the top or antitop quark, decays leptonically. At least one isolated electron (muon) with $\abs{\eta} <2.4$ and $\PT>10$ (5)\GeV, and at least one jet with $\PT>25$\GeV are required. The \ttH hadronic categories target hadronic decays of \PW bosons. In these categories at least three jets are required, one of which must be \PQb tagged, and a lepton veto is imposed. In order to maximize the sensitivity, an MVA approach is used to separate the \ttH events from the background, dominated by $\DIPHO$, $\GAMJETS$, $\ttbar$ + jets, $\ttbar$ + $\gamma$, and $\ttbar$ + $\gamma\gamma$ events. A BDT classifier is trained for each of the two channels using simulated events.
The variables used for the training include kinematic properties of the reconstructed objects, object identification variables, and global event properties such as jet and lepton multiplicities. The BDT input variables also include the outputs of other machine learning algorithms trained specifically to target different backgrounds.
These include DNN classifiers trained to reduce the $\ttbar$ + $\gamma\gamma$ and $\DIPHO$ background, and a top quark tagger based on a BDT~\cite{Sirunyan:2017wif}.
 The output scores of the BDTs are used to reject background-like events and to classify the remaining  events in four subcategories for each of the two channels. The boundaries of the categories are optimized by maximizing the expected significance of the \ttH signal.

\section{Signal model}
\label{sec:signal}

In each of the \HH categories, a parametric fit in the ($\Mgg,\Mjj$) plane is performed. In the \ttH categories, the \Mgg distribution is fitted to extract the signal. When the \HH and \ttH categories are combined, both the \HH and \ttH production modes are considered as signals.

\begin{figure} [hbtp]
	\centering
		{\includegraphics[width=0.45\textwidth]{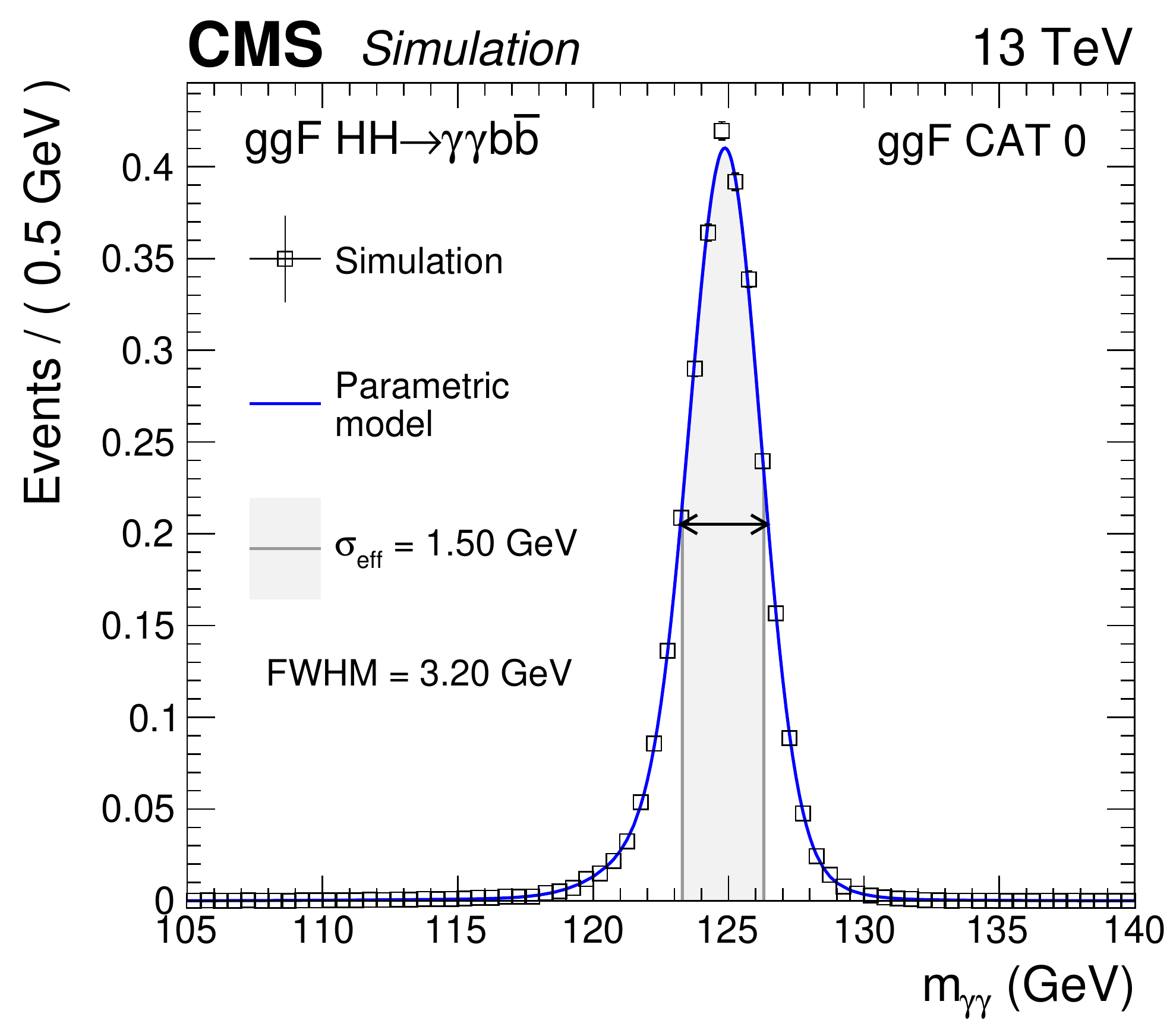}}
		{\includegraphics[width=0.41\textwidth]{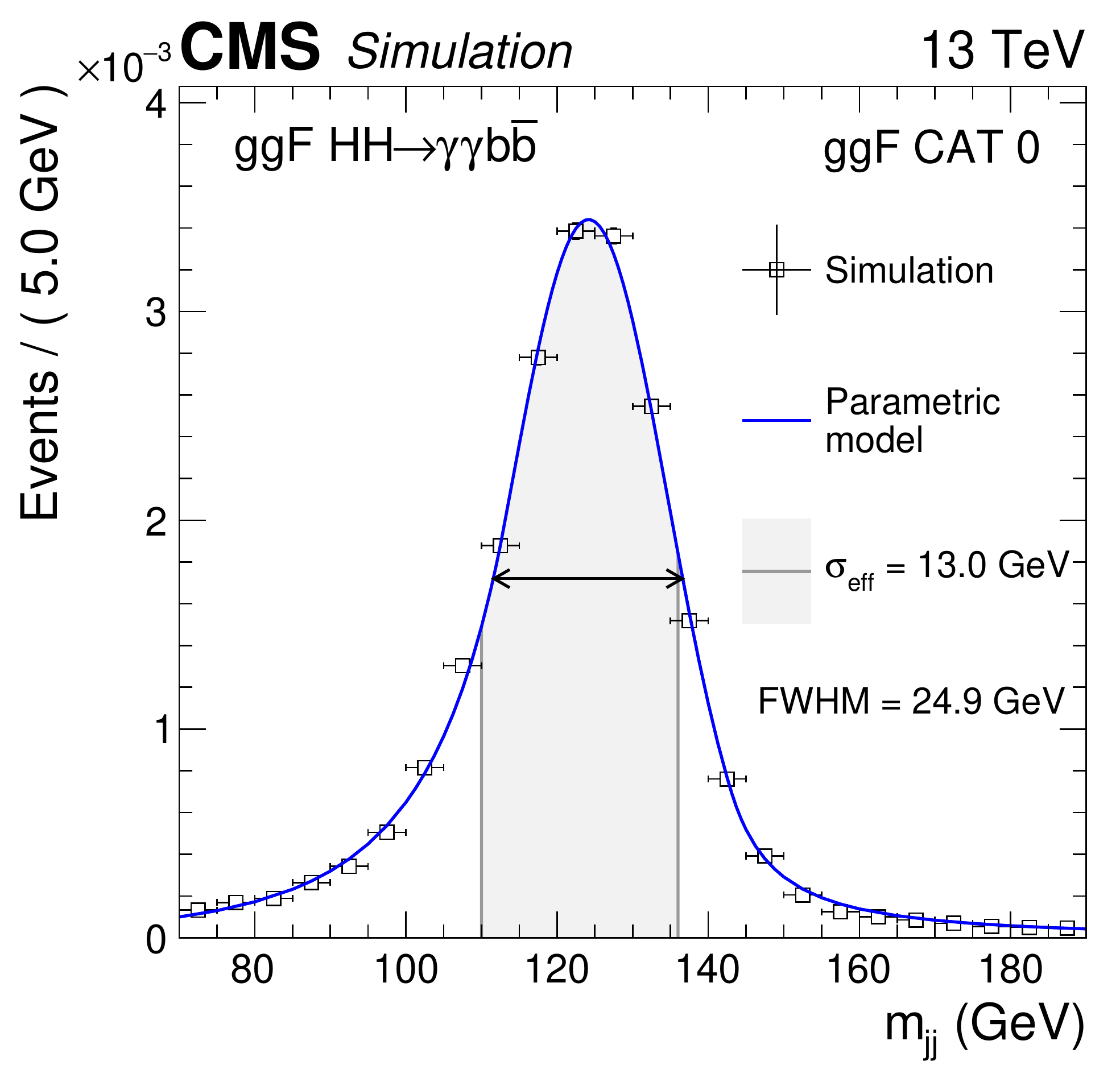}}\\
		{\includegraphics[width=0.45\textwidth]{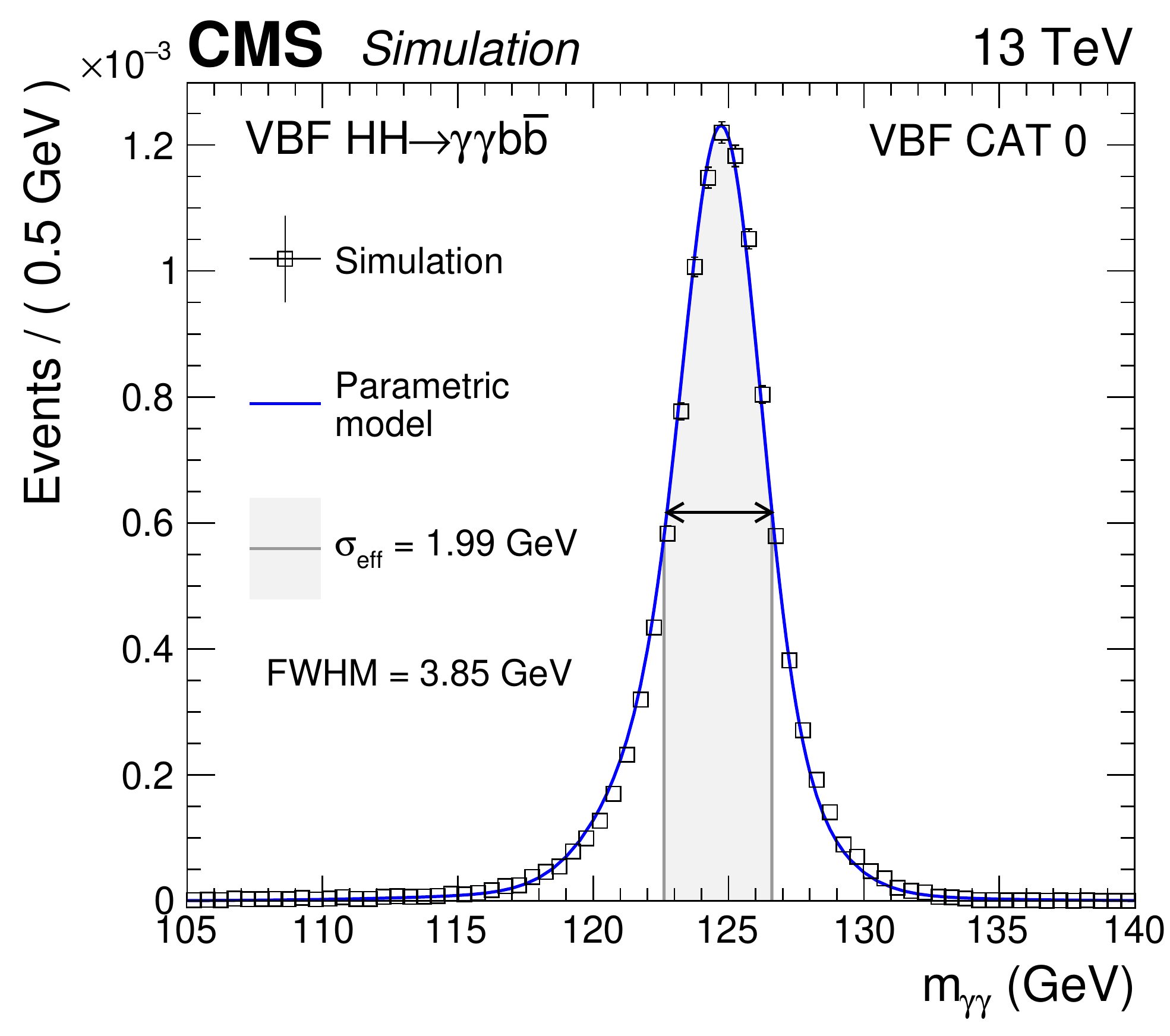}}
		{\includegraphics[width=0.41\textwidth]{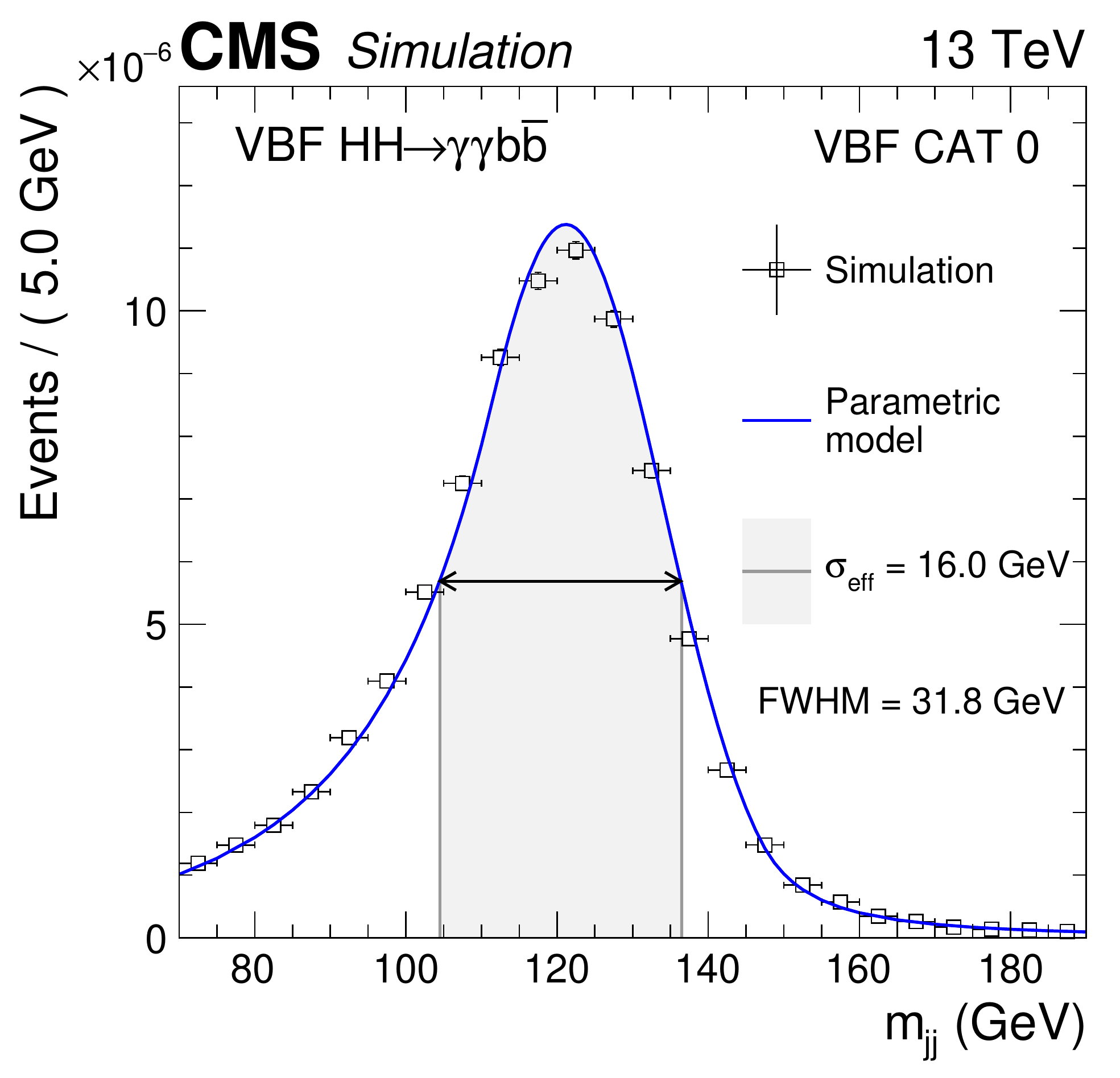}}
  \caption{Parametrized signal shape for $\Mgg$ (left) and \Mjj (right) in the best resolution ggF (upper) and VBF (lower) categories. The open
  squares represent simulated events and the blue lines are the
  corresponding models. Also shown are the $\sigma_{\text{eff}}$ value
  (half the width of the narrowest interval containing 68.3\% of the
  invariant mass distribution) and the corresponding interval as a gray
  band, and the full width at half the maximum (FWHM) and the corresponding
  interval as a double arrow.}
		\label{fig:statAnalysisSigPlots}
\end{figure}

The shape templates of the diphoton and dijet invariant mass distributions are constructed from simulation. 
In each \HH and \ttH analysis category, the $\Mgg$ distribution is fitted using a sum
of, at most, five Gaussian functions.
Figure~\ref{fig:statAnalysisSigPlots} (left) shows the signal model for $\Mgg$ in the VBF and ggF CAT0 categories, which are the categories with the best resolution.

For the \HH categories, the $\Mjj$ distributions are modeled with a double-sided Crystal Ball (CB) function, a modified version of the standard CB function~\cite{CrystalBallRef}
with two independent exponential tails. Figure~\ref{fig:statAnalysisSigPlots} (right) shows the signal model for $\Mjj$ in the VBF and ggF categories with the best resolution.

For the \HH signal, the final two-dimensional (2D) signal probability distribution function is a product of the independent $\Mgg$ and $\Mjj$ models.
The possible correlations are investigated by comparing the 2D $\Mgg$-$\Mjj$ distributions in the simulated signal samples
with the 2D probability distributions built as a product of the one-dimensional (1D) ones. With the statistical precision available in this analysis, the correlations have been found to be negligible.

\section{Background model}
\label{sec:background}
\subsection{Single Higgs background model}
\label{sec:singleHiggsbackground}
The SM single \PH background shape is constructed from
the simulation following the same methodology as used for the signal model described in Section~\ref{sec:signal}. For each analysis category and single \PH production mode,
 the $\Mgg$ distributions are fitted using a sum
of, at most, five Gaussian functions. The $\Mjj$ modeling in the \HH categories depends on the production mechanism, and a parametrisation is obtained from the 
simulated distributions: for the \ggH and \VBFH processes, the $\Mjj$ distribution is modeled with a Bernstein polynomial; for \VH production, a CB function is used to model the distribution of the hadronic decays of vector bosons; for \ttH, where the two \PQb jets are produced 
from a top quark decay, a Gaussian function with a mean around 120\GeV is used.
 Like for the signal modeling, the final 2D SM single \PH model is a product of the independent models of the $\Mgg$ and $\Mjj$ distributions.

\subsection{Nonresonant background model}
\label{sec:nonresbackground}
The model used to describe the nonresonant background is extracted from data using the discrete profiling method~\cite{DiscreteProfilingMethod}
as described in Ref.~\cite{Khachatryan:2014ira}. This
technique was designed as a way to estimate the systematic uncertainty
associated with choosing a particular analytic function to fit the
background $\Mgg$ and $\Mjj$ distributions. The method treats the choice of the
background function as a discrete nuisance parameter in the likelihood 
fit to the data. This method is used to model $\Mgg$ distribution of the nonresonant background in the $\ttH$ categories. For the \HH categories, the method is generalized to the 2D model case as a product of two 1D models for $\Mgg$ and $\Mjj$.

A set of MC pseudo-experiments was generated with positive and negative correlations between $\Mgg$ and $\Mjj$ injected and then fitted with the factorized 2D model. A negligible bias has been observed, and the correlations have been found to be within the statistical precision of the analysis.

\section{Systematic uncertainties}
\label{sec:syst}
The systematic uncertainties only affect the signal model
 and the resonant single \PH background, since the nonresonant background model is constructed in a data-driven way with the
 uncertainties associated with the choice of a background fit function taken into account by the discrete profiling method described
 in Section~\ref{sec:nonresbackground}. The systematic uncertainties can affect the overall normalization, or a variation in category yields, representing event migration between the categories. Theoretical uncertainties have been applied to the \HH and single \PH normalizations. The following sources of theoretical uncertainty are considered: the uncertainty in the signal cross section arising from scale variations, uncertainties on $\alpS$, PDFs and in the prediction of the branching fraction $\mathcal{B}(\HH\to\bbgg)$. 
 The dominant theoretical uncertainties arise from the prediction of the SM \HH and \ttH production cross sections. 
In addition, a conservative PS uncertainty is assigned to the VBF \HH signal, defined as the full symmetrized difference in yields in each category obtained with simulated samples of VBF \HH events interfaced with the standard \pt-ordered and dipole shower PS schemes. 

The dominant experimental uncertainties are:

\begin{itemize}
\item \textit{Photon identification BDT score}:
  the uncertainty arising from the imperfect MC simulation of the input variables to the photon ID 
  is estimated by rederiving the corrections with equally sized subsets 
of the \Zee events used to train the quantile regression BDTs. Its magnitude corresponds to the standard deviation of the event-by-event differences in the photon ID evaluated on the two different sets of corrected input variables.
This uncertainty reflects the limited capacity of the BDTs 
arising from the finite size of the training set. It is seen to cover the residual discrepancies between data and simulation. 
  The uncertainty in the signal yields is
  estimated by propagating this uncertainty through the full category selection procedure.
\item \textit{Photon energy scale and resolution}: 
  the uncertainties associated with the corrections applied to the photon energy scale in data
  and the resolution in simulation are evaluated using \Zee events~\cite{Khachatryan:2015iwa}.
\item \textit{Per-photon energy resolution estimate}: 
  the uncertainty in the per-photon resolution is
  parametrized as a rescaling of the resolution by
  $\pm 5\%$ around its nominal value. 
  This is designed to cover all differences between data and simulation 
  in the distribution, which is an output of the energy regression.
\item \textit{Jet energy scale and resolution corrections}: 
  the energy scale of jets is measured using the \pt balance of jets with \PZ bosons and photons in
  \Zee, \Zmumu, and $\GAMJETS$ events, as well as using the \pt balance between jets 
  in dijet and multijet events \cite{JetsInRun2,Khachatryan:2016kdb}. The uncertainty in the jet energy scale and resolution
  is a few percent and depends on \pt and $\eta$. The impact of uncertainties on the event yields is evaluated by varying the jet energy corrections within their uncertainties and 
  propagating the effect to the final result. Some sources of the jet energy scale uncertainty are fully (anti-)correlated, while others are considered uncorrelated.
\item \textit{Jet \PQb tagging}: 
  uncertainties in the \PQb tagging efficiency are evaluated 
  by comparing data and simulated distributions for the \PQb tagging
  discriminator~\cite{CMSbtagging}. These include the statistical uncertainty in the
estimate of the fraction of heavy- and light-flavor jets in data and
simulation.
\item \textit{Trigger efficiency}: 
  the efficiency of the trigger selection is measured with 
  $\Zee$ events using a tag-and-probe technique~\cite{CMS:2011aa}. An additional uncertainty is introduced to account for a 
  gradual shift in the timing of the inputs of the ECAL L1 trigger in the region $\abs{\eta} > 2.0$, 
  which caused a specific trigger inefficiency during 2016 and 2017 data taking. 
  Both photons and, to a greater extent, jets can be affected by this inefficiency, which has a small impact. 
\item \textit{Photon preselection}: 
  the uncertainty in the preselection efficiency
  is computed as the ratio between the efficiency measured in data and in simulation. The preselection efficiency in data is measured with the tag-and-probe technique in \Zee events~\cite{CMS:2011aa}.
\item \textit{Integrated luminosity}: 
  uncertainties are determined by the CMS luminosity monitoring 
  for the 2016--2018 data-taking years~\cite{CMSlumi2016,CMSlumi2017,CMSlumi2018} and are in the range of 2.3--2.5\%. To account for common sources of uncertainty in the luminosity measurement schemes, some sources are fully (anti-)correlated across the different data-taking years, while others are considered uncorrelated. The total 2016--2018 integrated luminosity has an uncertainty of 1.8\%.
\item \textit{Pileup jet identification}: 
  the uncertainty in the pileup jet classification output score is estimated by
  comparing the score of jets in events with a \PZ boson and one balanced jet
  in data and simulation. The assigned uncertainty depends on $\pt$ and $\eta$, and is designed to cover all differences between data and simulation in the distribution.
\end{itemize}

Most of the experimental uncertainties are uncorrelated among the three data-taking years. Some sources of uncertainty in the measured luminosity and jet energy corrections are fully (anti-)correlated, while others are considered uncorrelated.
This search is statistically limited, and the total impact of systematic uncertainties on the result is about 2\%.

\section{Results}
\label{sec:res}

An unbinned
maximum likelihood fit to the $\Mgg$ and $\Mjj$ distributions is performed simultaneously in the 14 \HH categories to extract the \HH signal. A likelihood function is defined for each analysis category using analytic models to describe the $\Mgg$ and $\Mjj$ distributions of signal and background events, with nuisance parameters to account for the experimental and theoretical systematic uncertainties described in Section~\ref{sec:syst}. The fit is performed in the mass ranges $100 < \Mgg < 180\GeV$ and $70 < \Mjj < 190 \GeV$ for all categories apart from ggF CAT10 and CAT11. In those two categories, a small but nonnegligible shoulder was observed in the $\Mjj$ distribution. Therefore, the $\Mjj$ fit range is reduced to $90 < \Mjj < 190 \GeV$ to avoid a possible bias with minimal
impact on the analysis sensitivity. 

In order to determine \kapl and \kapt, the \HH and \ttH categories are used together in a simultaneous maximum likelihood fit. 
In the \ttH categories, a binned maximum likelihood fit is performed to $\Mgg$ in the mass range $100 < \Mgg < 180\GeV$. 

\begin{figure}[htbp]
 \centering
\includegraphics[width=0.40\textwidth]{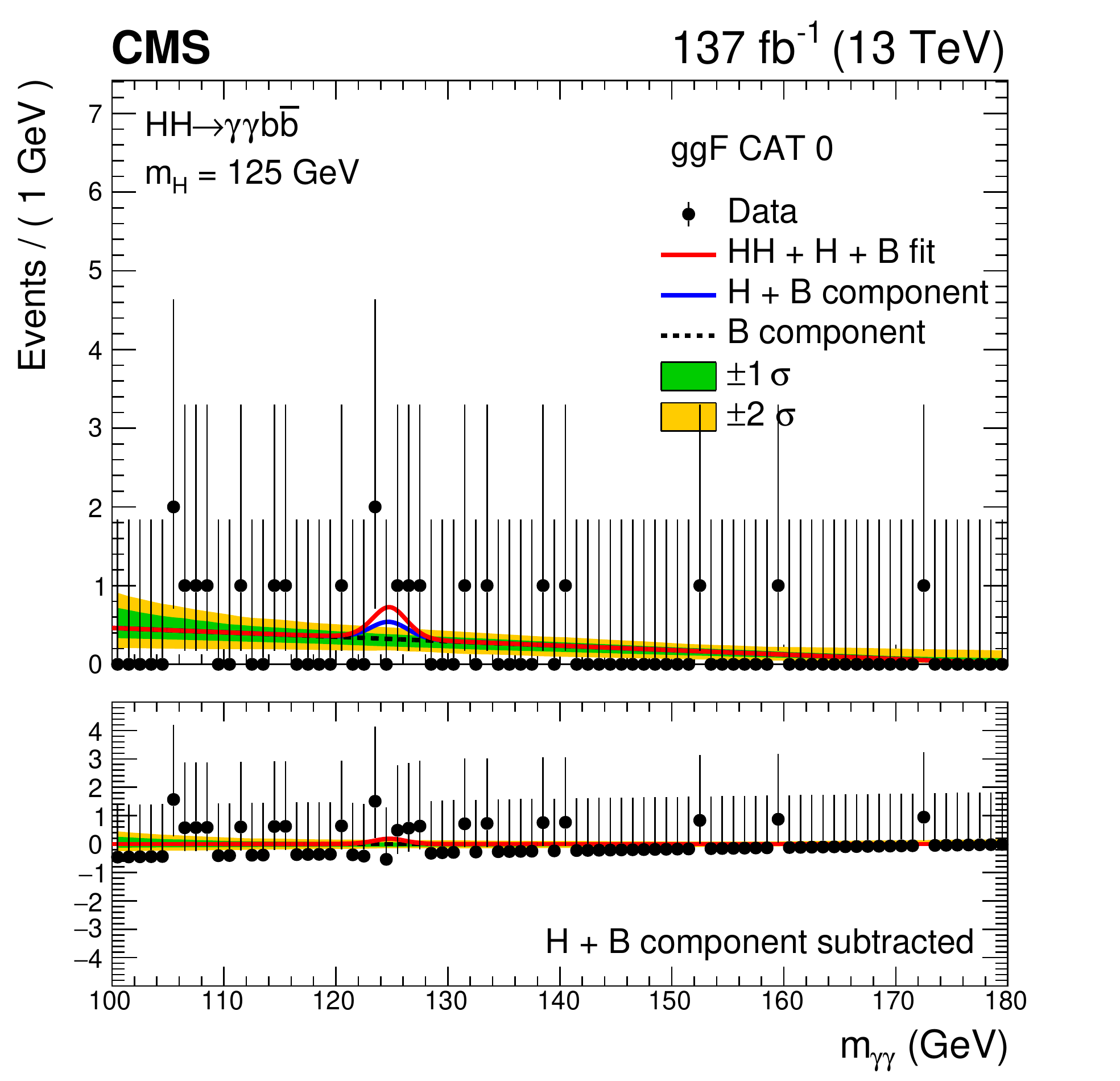}
\includegraphics[width=0.40\textwidth]{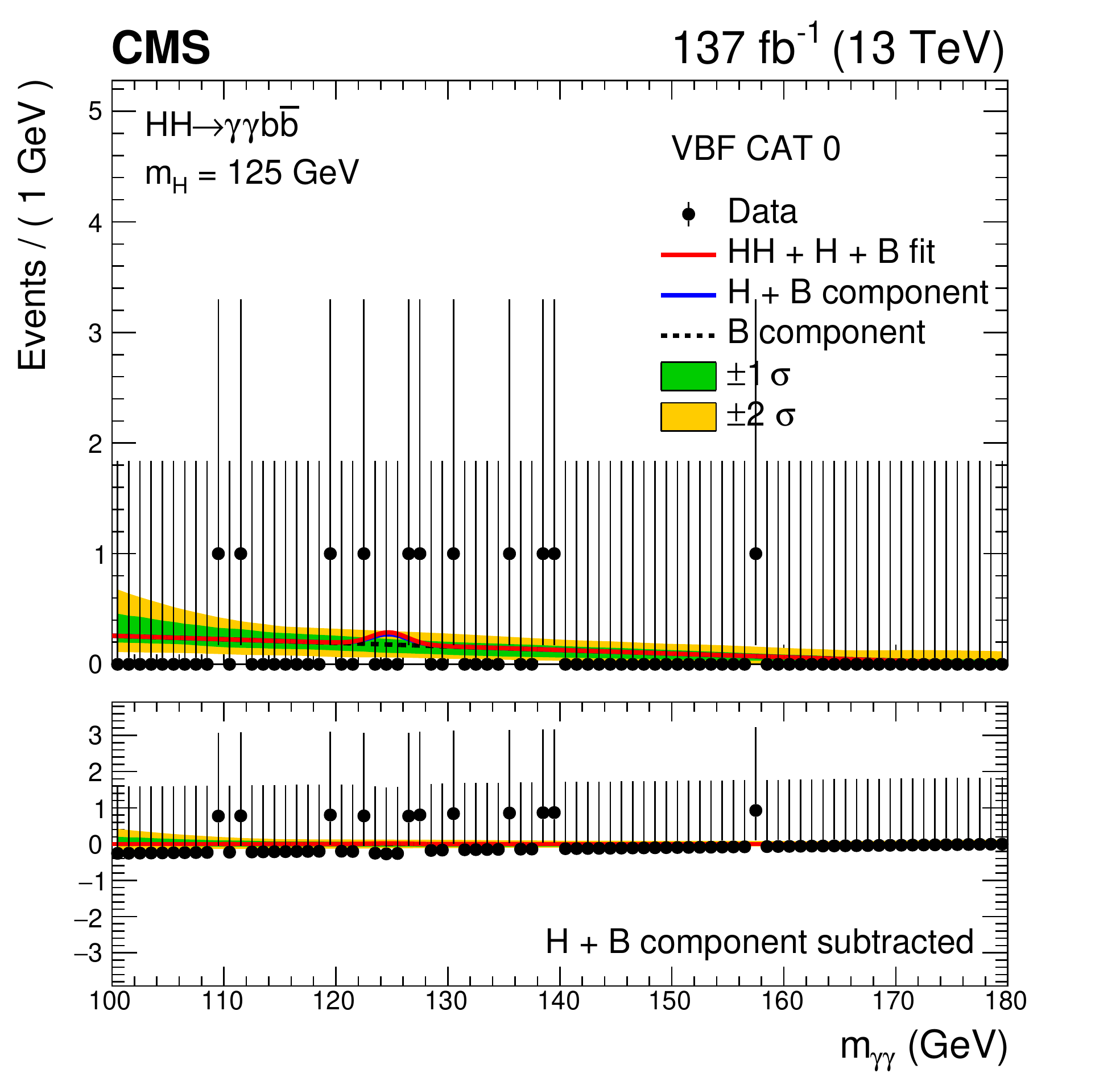}\\
\includegraphics[width=0.40\textwidth]{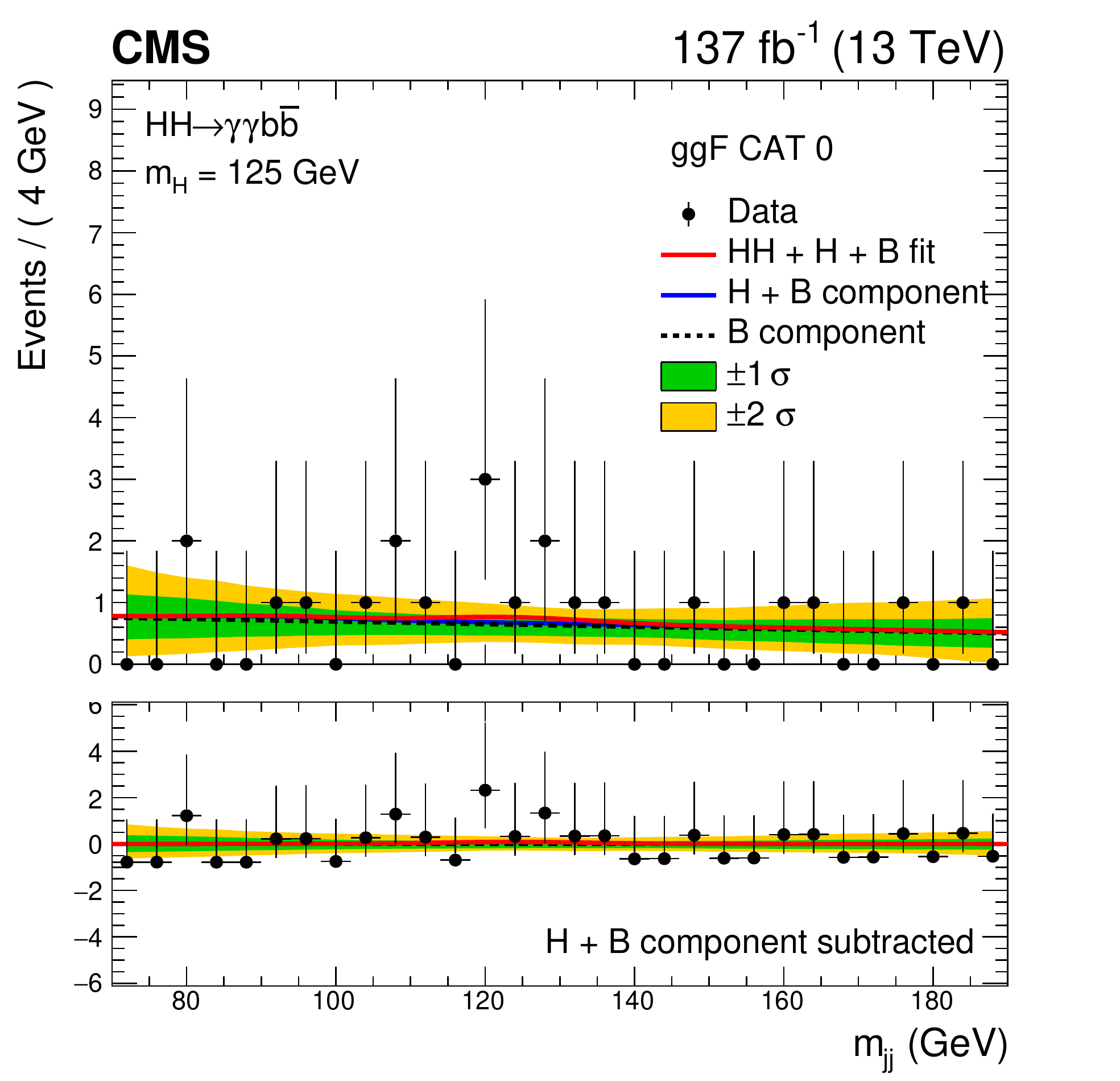}
\includegraphics[width=0.40\textwidth]{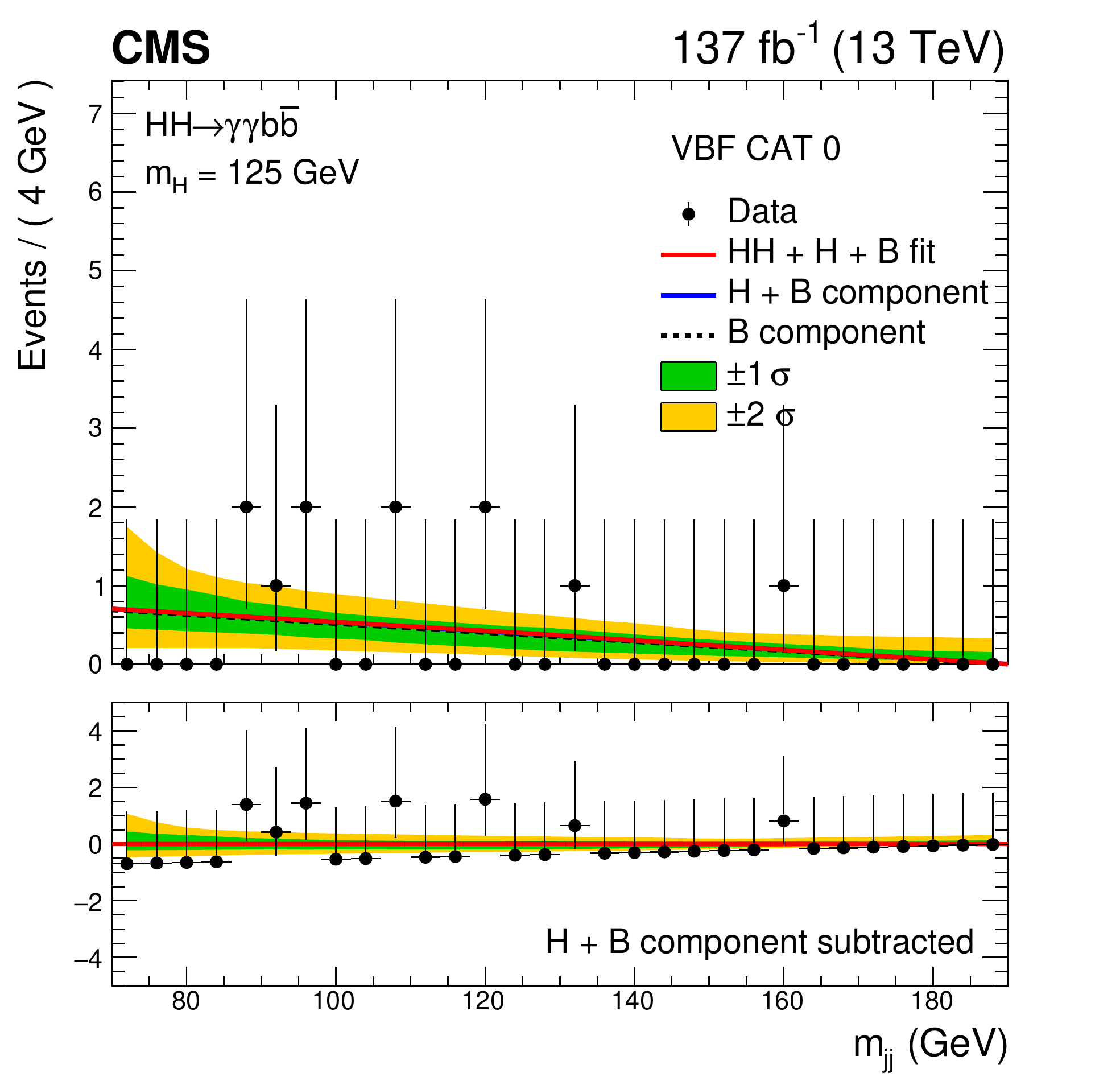}\\
  \caption{
Invariant mass distributions $\Mgg$ (upper) and $\Mjj$ (lower) for the selected events in data (black points) in the best resolution ggF (CAT0) and VBF (CAT0) categories. The solid red line shows the sum of the fitted signal and background (HH+H+B), the solid blue line shows the background component from the single Higgs boson and the nonresonant processes (H+B), and the dashed black line shows the nonresonant background component (B). The normalization of each component (HH, H, B) is extracted from the combined fit to the data in all analysis categories. The one (green) and two (yellow) standard
  deviation bands include the uncertainties in the background component
  of the fit. The lower panel in each plot shows the residual signal yield after the
 background (H+B) subtraction.}
 \label{fig:sigBkgPlotsvbf}
\end{figure}

\begin{figure}[htbp]
 \centering
\includegraphics[width=0.40\textwidth]{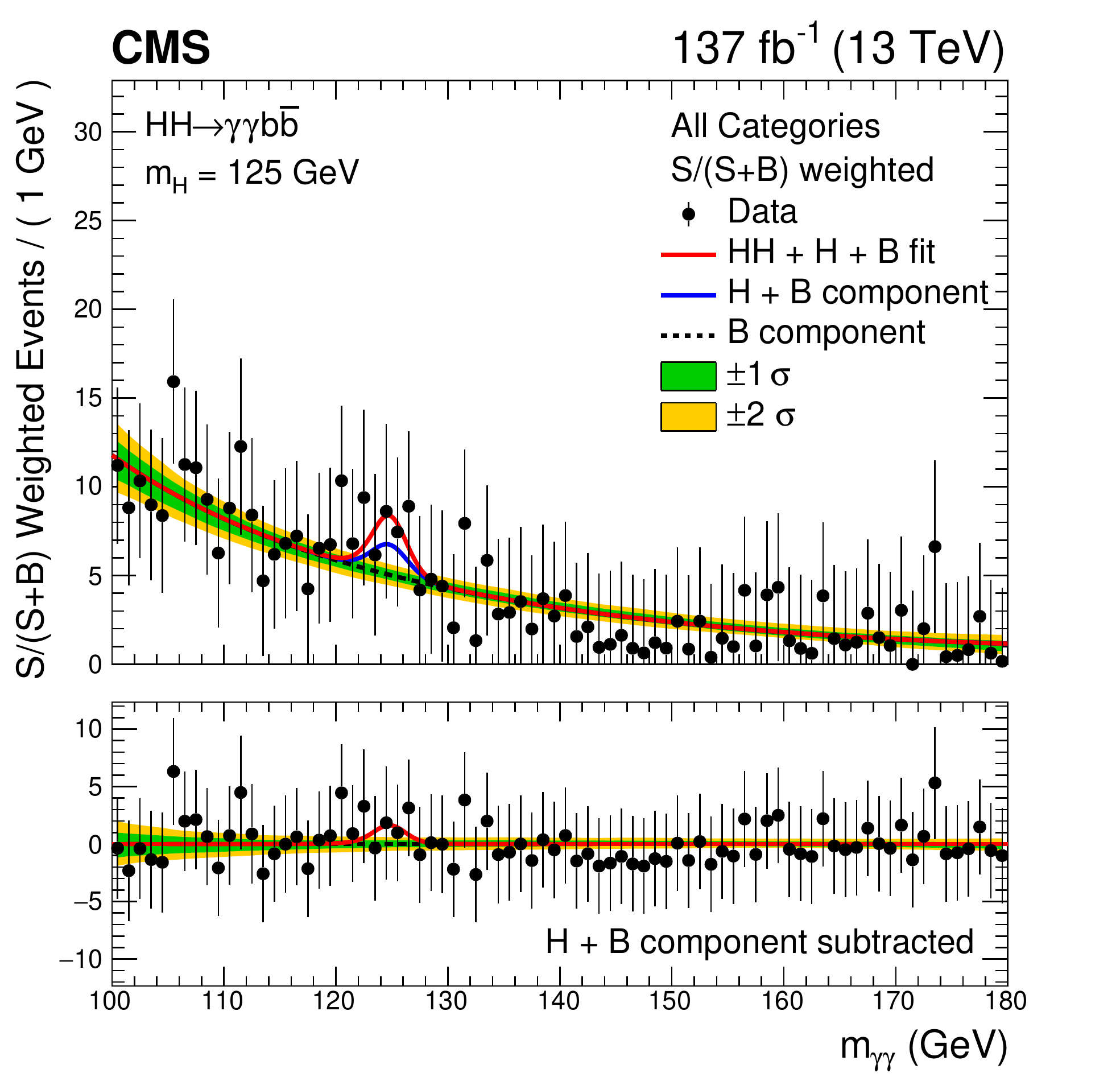}
\includegraphics[width=0.40\textwidth]{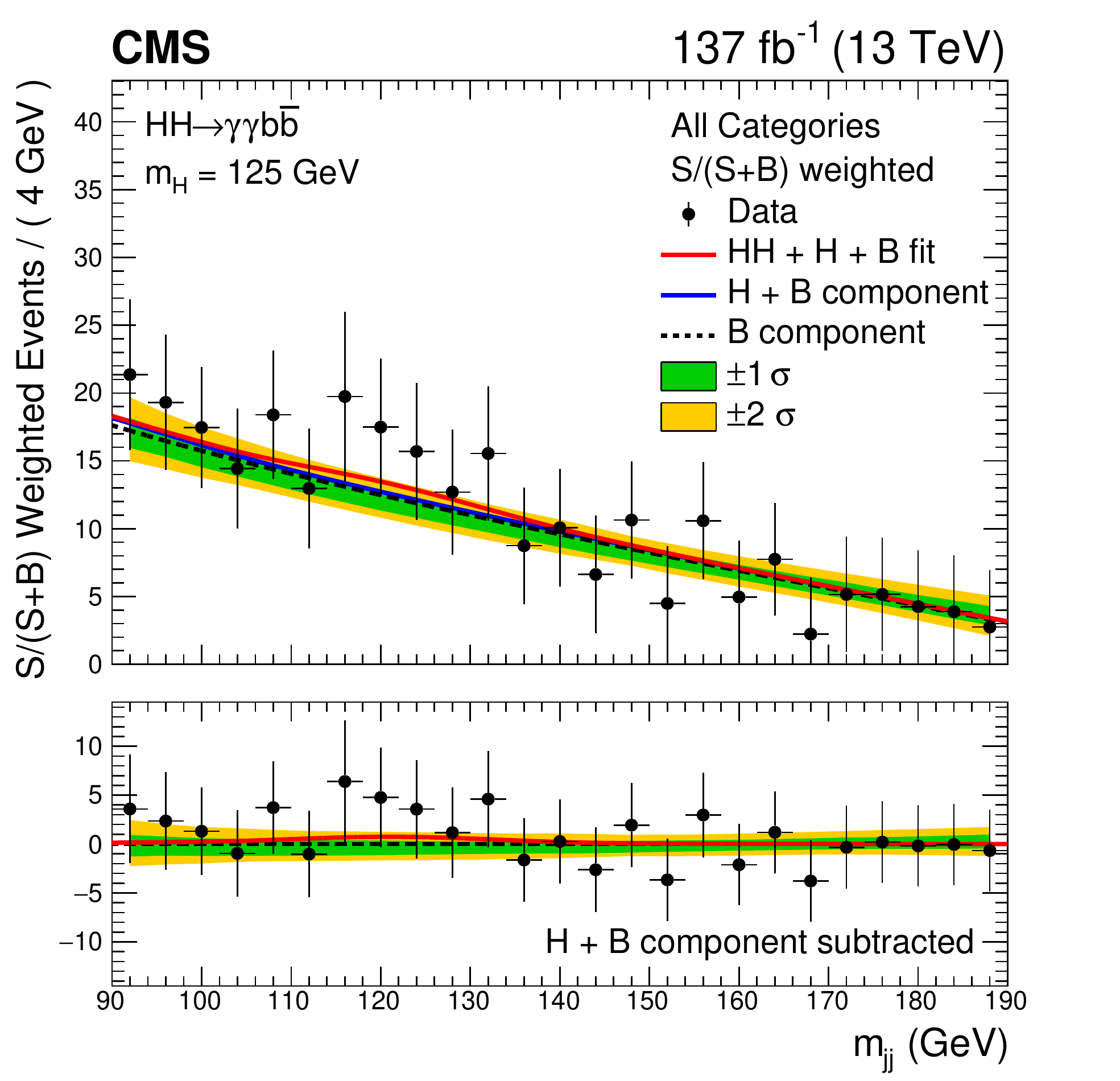}
  \caption{
Invariant mass distributions $\Mgg$ (left) and $\Mjj$ (right) for the selected events in data (black points) weighted by S/(S+B),
        where S (B) is the number of signal (background) events extracted from the signal-plus-background fit.
 The solid red line shows the sum of the fitted signal and background (HH+H+B), the solid blue line shows the background component from the single Higgs boson and the nonresonant processes (H+B), and the dashed black line shows the nonresonant background component (B). The normalization of each component (HH, H, B) is extracted from the combined fit to the data in all analysis categories. The one (green) and two (yellow) standard
  deviation bands include the uncertainties in the background component
  of the fit. The lower panel in each plot shows the residual signal yield after the
 background (H+B) subtraction.} 
 \label{fig:wall_mgg_mjj}
\end{figure}

The data and the signal-plus-background model fit to $\Mgg$ and $\Mjj$ are shown in Fig.~\ref{fig:sigBkgPlotsvbf} for the best resolution ggF and VBF categories. 
The distribution of events weighted by S/(S+B) from all \HH categories is shown in Fig.~\ref{fig:wall_mgg_mjj} for $\Mgg$ and $\Mjj$. In this expression, S (B) is the number of signal (background) events extracted from the signal-plus-background fit.

No significant deviation from the background-only hypothesis is observed. We set upper limits at 95\% \CL on the product of the production cross section of a pair of Higgs bosons and the branching fraction into $\bbgg$, $\sigmaHH \BR$, using the modified frequentist approach for confidence levels (\CLs), taking the LHC profile likelihood ratio 
as a test statistic~\cite{CLS2,CLS1,CLSA,CMS-NOTE-2011-005} in the asymptotic approximation. The observed (expected) 95\% \CL upper limit on $\sigmaHH \BR$
 amounts to 0.67 (0.45)\unit{fb}. The observed (expected) limit corresponds to 7.7 (5.2) times the SM prediction. 
 All results were extracted assuming $\mH = 125 \GeV$. We observe a variation smaller than 1\% in both the expected and observed upper limits when using $\mH = 125.38 \pm 0.14\GeV$, corresponding to the most precise measurement of the
Higgs boson mass to date~\cite{Sirunyan:2020xwk}.

\begin{figure}[!ht]
  \centering
\includegraphics[width=0.7\textwidth]{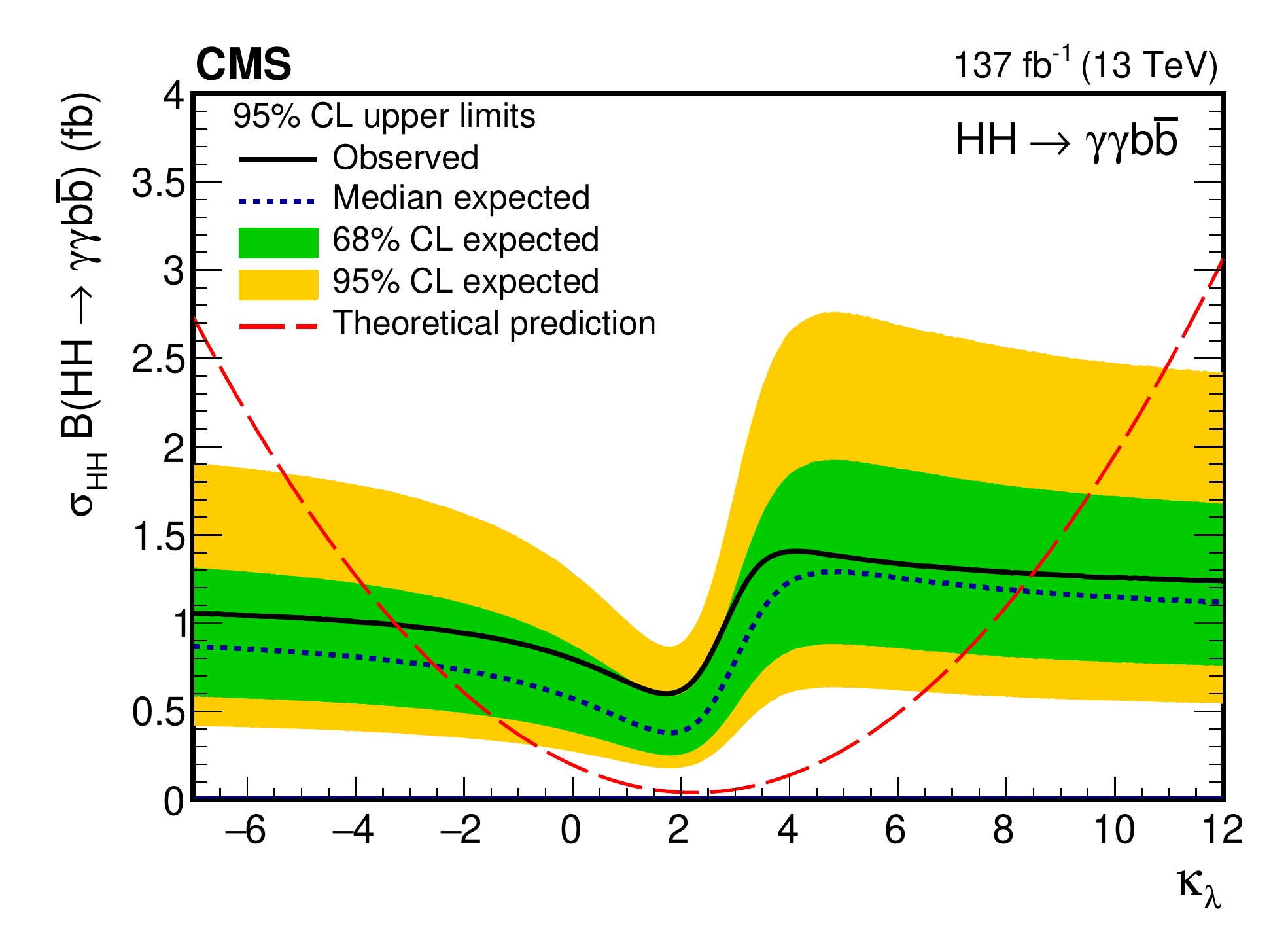}\hfil 

  \caption{Expected and observed 95\% \CL upper limits on the product of the \HH production cross section and $\mathcal{B}(\HH\to\bbgg)$ obtained for different values of $\kapl$ assuming $\kapt$ = 1. The green and yellow bands represent, respectively, the one and two standard deviation extensions beyond the expected limit. The long-dashed red line shows the theoretical prediction.
}
  \label{fig:klambdascan}
\end{figure}

Limits are also derived as a function of $\kapl$, assuming that the top quark Yukawa coupling is SM-like ($\kapt = 1$). The result is shown in Fig.~\ref{fig:klambdascan}. The variation in the excluded cross section as a function of $\kapl$ is directly related to changes in the kinematical properties of \HH production. At 95\% \CL, \kapl is constrained to values in the interval $[-3.3, 8.5]$, while the expected constraint on $\kapl$ is in the interval $[-2.5, 8.2]$. This is the most sensitive search to date.

\begin{figure}[!htb]
  \centering
\includegraphics[width=0.43\textwidth]{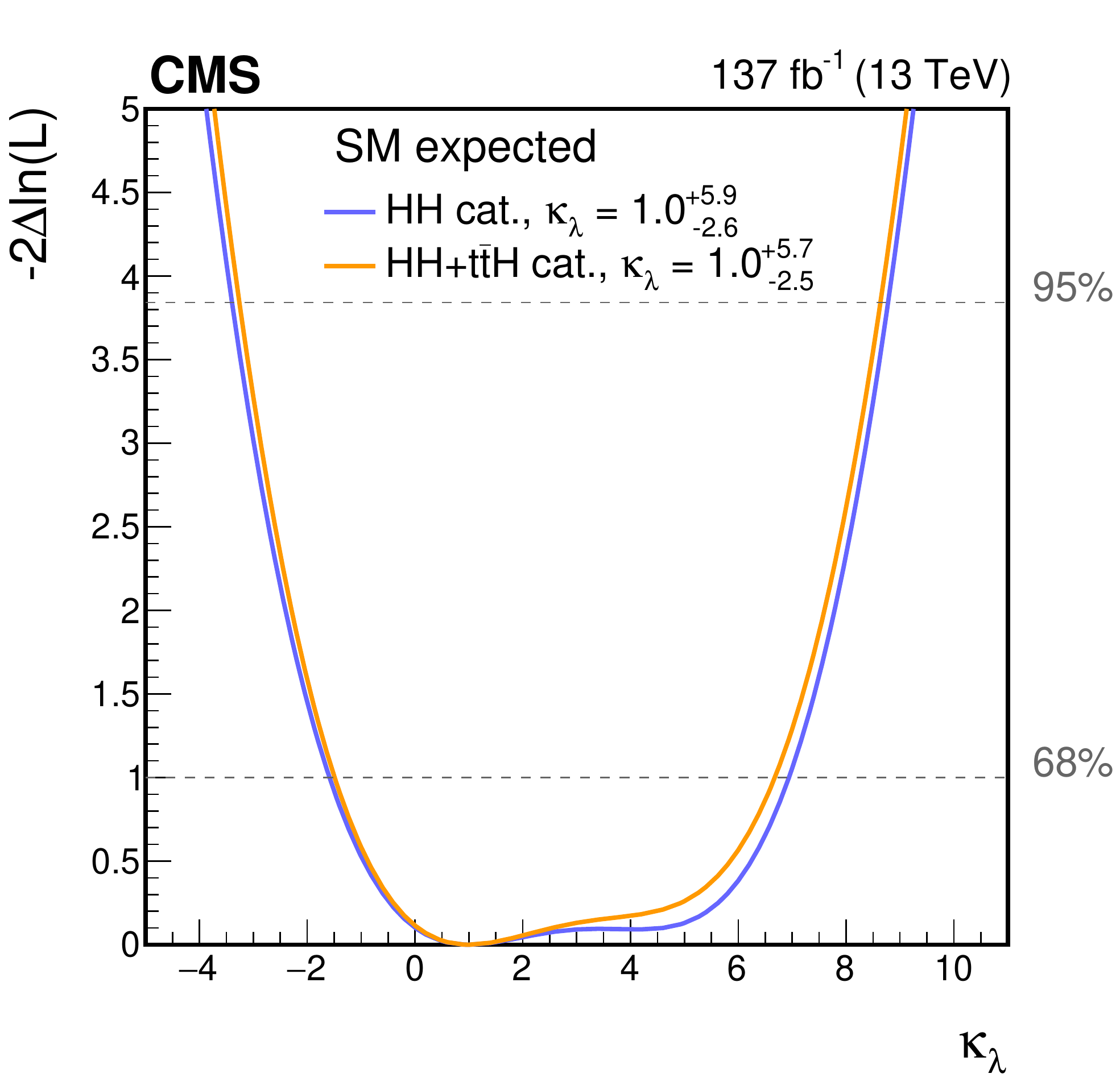} 
\includegraphics[width=0.43\textwidth]{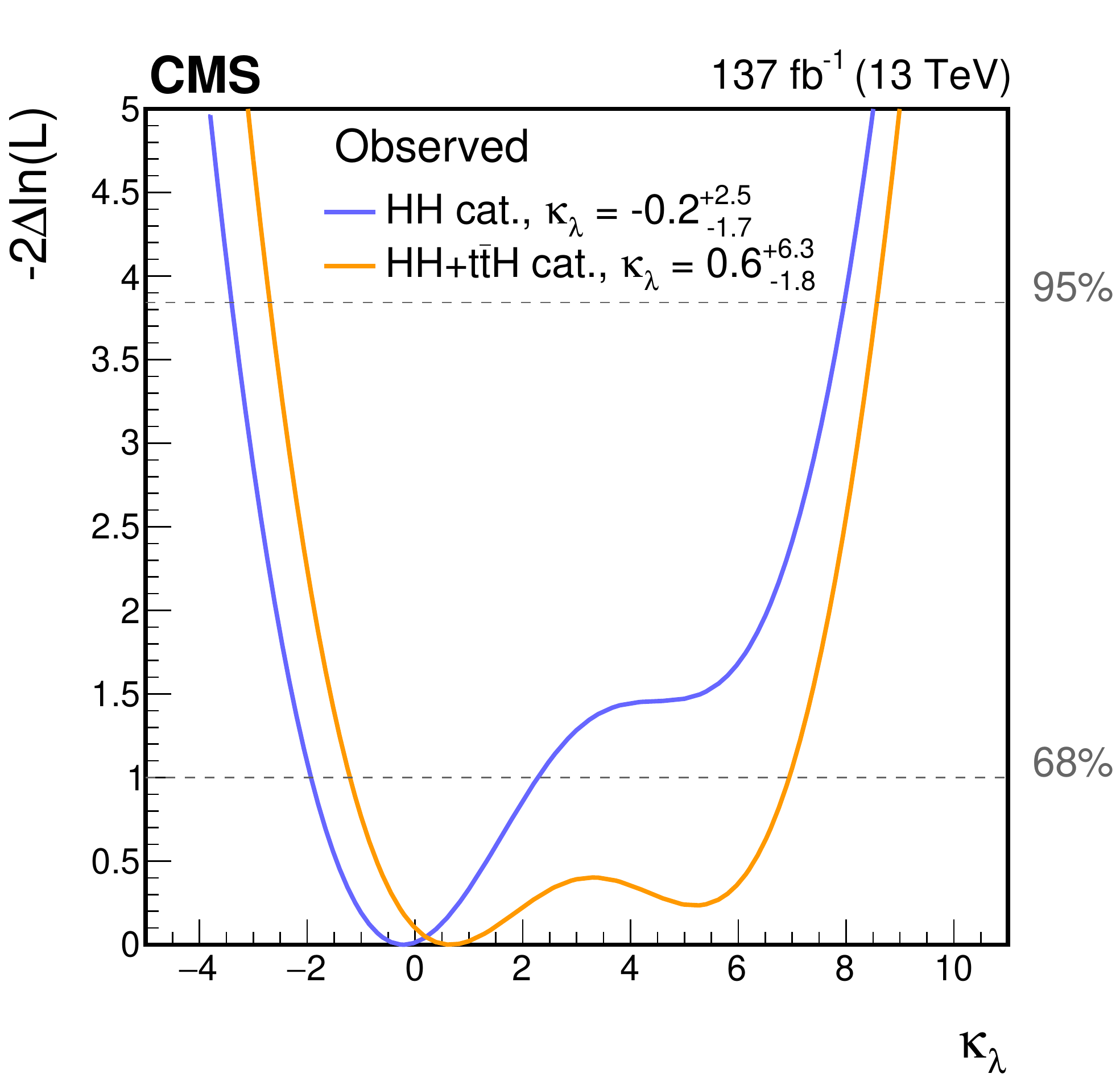} 
  \caption{Negative log-likelihood, as a function of $\kapl$, evaluated with an Asimov data set assuming the SM hypothesis (left) and the observed data (right).
 The 68 and 95\% \CL intervals are shown with the dashed gray lines.
 The two curves are shown for the \HH (blue) and \HH+\ttH (orange) analysis categories. All other couplings are set to their SM values.}
\label{fig:kllikelihood}
\end{figure}

Assuming instead that an \HH signal exists with the properties predicted by the SM, constraints on \lbdHHH can be set. The results are obtained both with the \HH categories only, and with the \HH categories combined with the \ttH categories in a simultaneous maximum likelihood fit. 
The \HH signal is considered together with the single \PH processes (\ttH, \ggH, \VBFH,\VH, and Higgs boson production in association with a single top
quark). The cross sections and branching fractions of the \HH and single \PH processes are scaled as a function of \kapl, while the top quark Yukawa coupling is assumed to be SM-like, $\kapt=1$.
One-dimensional negative log-likelihood scans for $\kapl$ are shown in Fig.~\ref{fig:kllikelihood} for an Asimov data set~\cite{CLSA} generated with the SM signal-plus-background hypothesis, $\kapl=1$, and for the observed data.
When combining the \HH analysis categories with the \ttH categories, we obtain $\kapl = 0.6^{+6.3}_{-1.8}$ ($1.0^{+5.7}_{-2.5}$ expected).
Values of \kapl outside the interval $[-2.7$, $8.6]$ are excluded at 95\% \CL. The expected exclusion at 95\% \CL corresponds to the region outside the interval $[-3.3,8.6]$. The shape of the likelihood as function of \kapl in Fig.~\ref{fig:kllikelihood} is characterized by 2 minima. This is related to an interplay between the cross section dependence on \kapl and differences in acceptance between the analysis categories.

\begin{figure}[!htb]
  \centering
\includegraphics[width=0.43\textwidth]{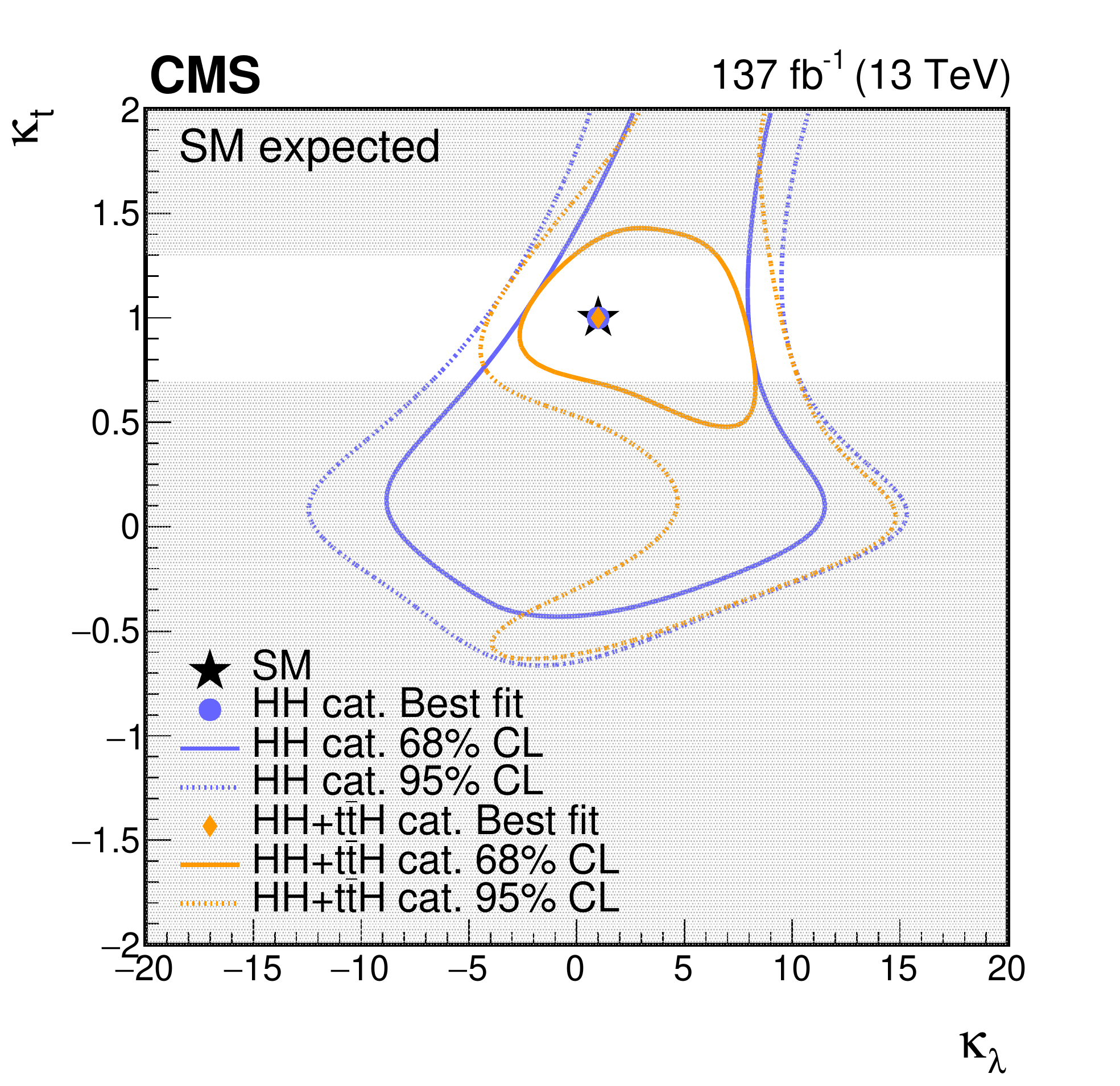} 
\includegraphics[width=0.43\textwidth]{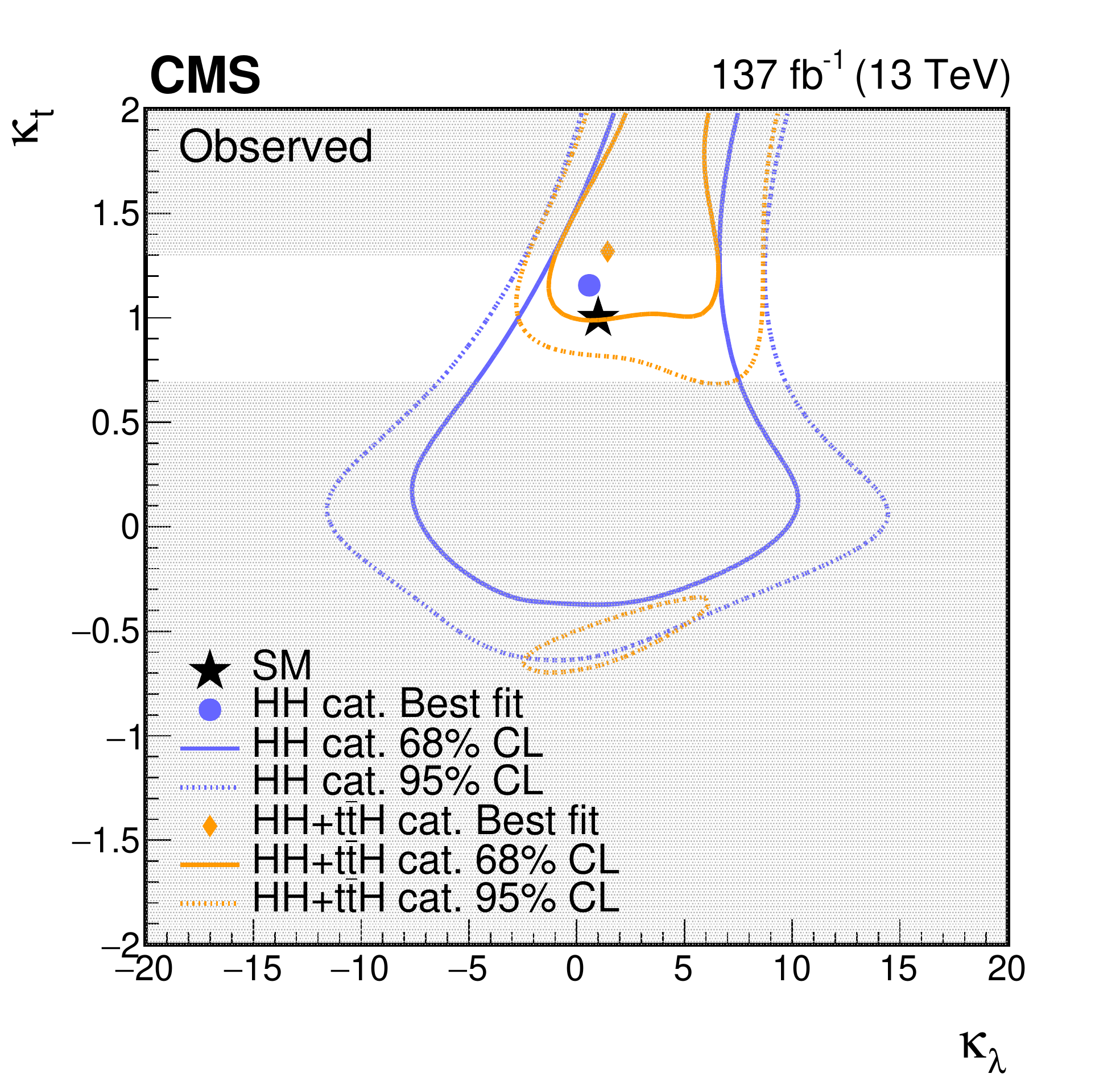} 
  \caption{Negative log-likelihood contours at 68 and 95\% \CL in the ($\kapl$,~$\kapt$) plane evaluated with an Asimov data set assuming the SM hypothesis (left) and the observed data (right). The contours obtained using the \HH analysis categories only are shown in blue, and in orange when combined with the \ttH categories. The best fit value for the \HH categories only ($\kapl$ = 0.6,~$\kapt$ = 1.2) is indicated by a blue circle, for the \HH+~\ttH categories ($\kapl$ = 1.4,~$\kapt$ = 1.3) by an orange diamond, and the SM prediction ($\kapl$ = 1.0,~$\kapt$ = 1.0) by a black star. The regions of the 2D scan where the \kapt parametrization for anomalous values of \kapl at LO is not reliable are shown with a gray band.}
  \label{fig:klkaptlikelihood_tthcombination}
\end{figure}

The \HH and single Higgs boson production cross sections depend not only on \kapl, but also on \kapt. To better constrain the \kapl and \kapt coupling modifiers, a 2D negative log-likelihood scan in the ($\kapl$,~\kapt) plane is performed, taking into account the modification of the production cross sections and $\mathcal{B}(\Hbb)$, $\mathcal{B}(\Hgg)$ for anomalous (\kapl,~\kapt) values. The modification of the single \PH production cross section for anomalous \kapl is modeled at NLO, while the dependence on \kapt is parametrized at LO only, neglecting NLO effects~\cite{Maltoni:2017ims}. This approximation holds as long as the value of $\abs{\kapt}$ is close to unity, roughly in the range $ 0.7<\kapt<1.3 $. The parametric model is not reliable outside of this range. 
Figure~\ref{fig:klkaptlikelihood_tthcombination} shows the 2D likelihood scans of \kapl versus \kapt for an Asimov data set assuming the SM hypothesis and for the observed data.
 The regions of the 2D scan where the \kapt parametrization for anomalous values of \kapl at LO is not reliable are shown with a gray band.

\begin{figure}[!htb]
  \centering
\includegraphics[width=0.43\textwidth]{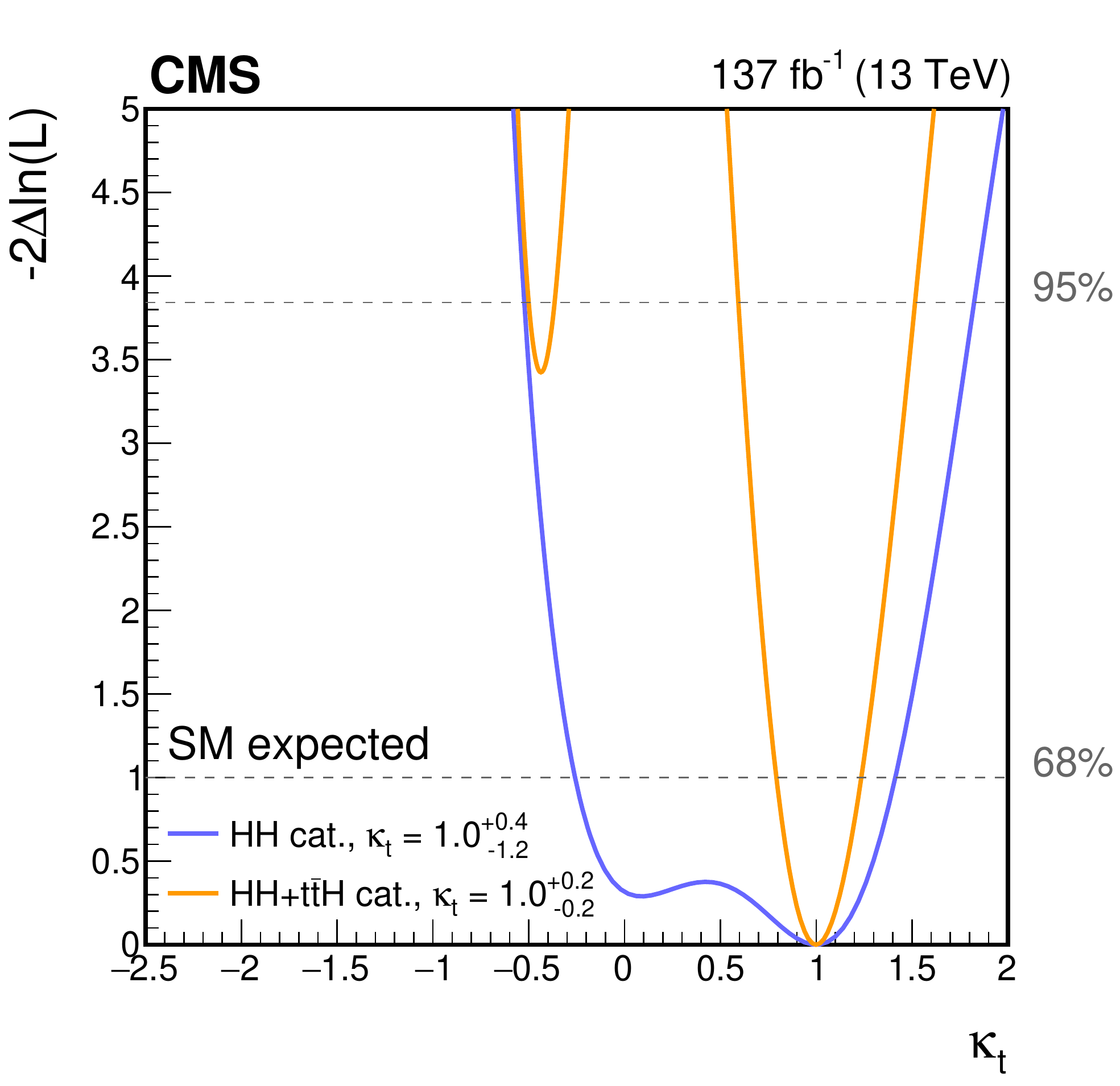}
\includegraphics[width=0.43\textwidth]{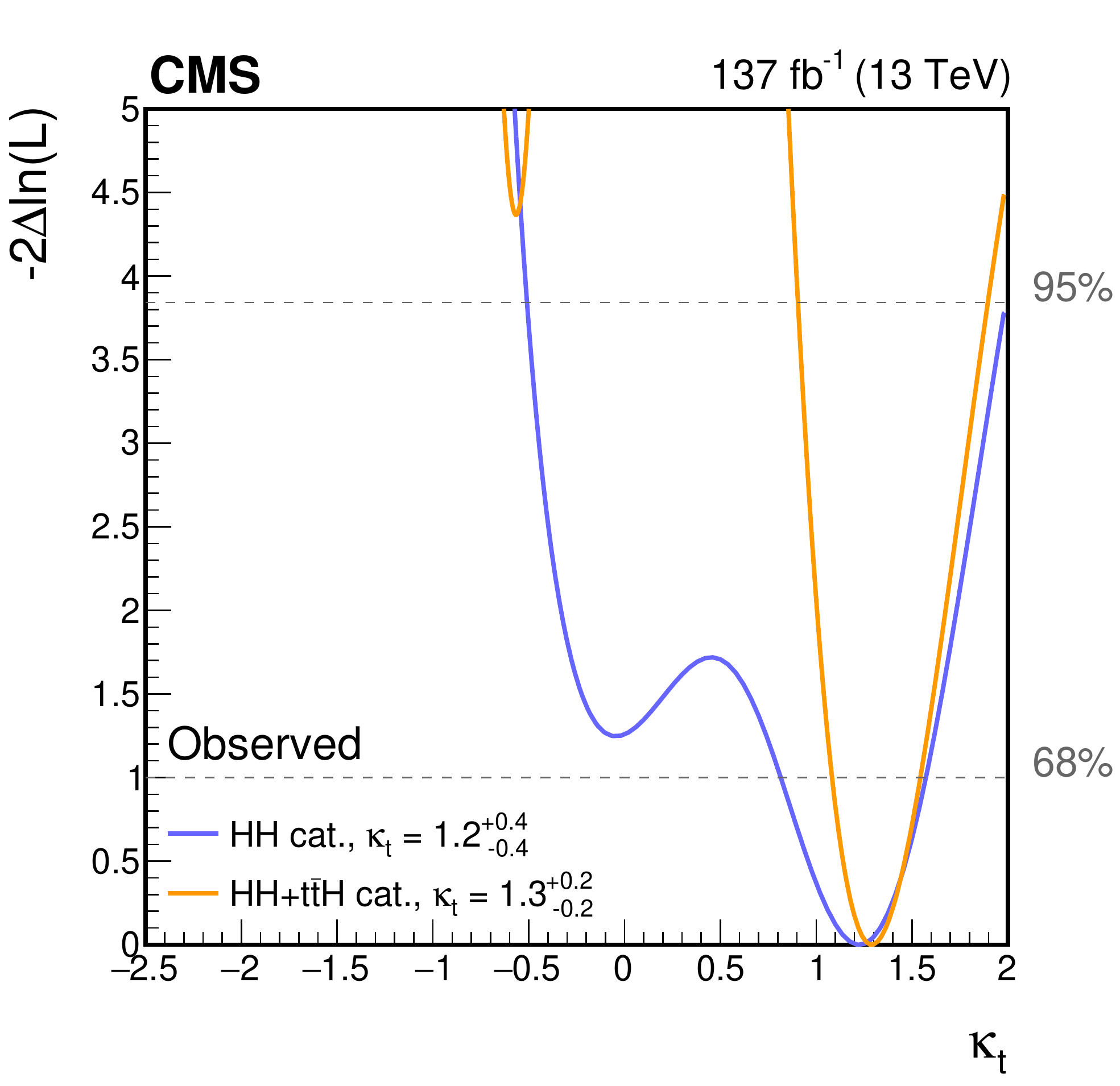}
  \caption{Negative log-likelihood scan, as a function of $\kapt$, evaluated with an Asimov data set assuming the SM hypothesis (left) and the observed data (right).
 The 68 and 95\% \CL intervals are shown with the dashed gray lines.
 The two curves are shown for the \HH (blue) and the \HH+\ttH (orange) analysis categories. All other couplings are fixed to their SM values.
}
\label{fig:kaptlikelihood}
\end{figure}

The inclusion of the \ttH categories significantly improves the constraint on \kapt. The 1D negative log-likelihood scan, as a function of \kapt with \kapl fixed at $\kapl$ = 1, is shown in Fig.~\ref{fig:kaptlikelihood} for an Asimov data set generated assuming the SM hypothesis, $\kapt$ = 1, as well as for the observed data. The measured value of $\kapt$ is $\kapt = 1.3^{+0.2}_{-0.2}$ ($1.0^{+0.2}_{-0.2}$ expected).
Values of \kapt outside the interval $[0.9$, $1.9]$ are excluded at 95\% \CL. The constraint on \kapt is comparable to the one recently set in Ref.~\cite{Sirunyan:2020icl}, where anomalous values of \conev were also considered. 

Upper limits at 95\% \CL are also set on the product of the \HH VBF production cross section and branching fraction, $\sigmaVBF \BR$, with the yield of the ggF \HH signal constrained within uncertainties to the one predicted in the SM.
 The observed (expected) 95\% \CL upper limit on 
 $\sigmaVBF \BR$ 
amounts to 1.02 (0.94)\unit{fb}. The limit corresponds to 225 (208) times the SM prediction. This is the most stringent constraint on
 $\sigmaVBF \BR$ 
to date. 

\begin{figure}[!ht]
  \centering
\includegraphics[width=0.7\textwidth]{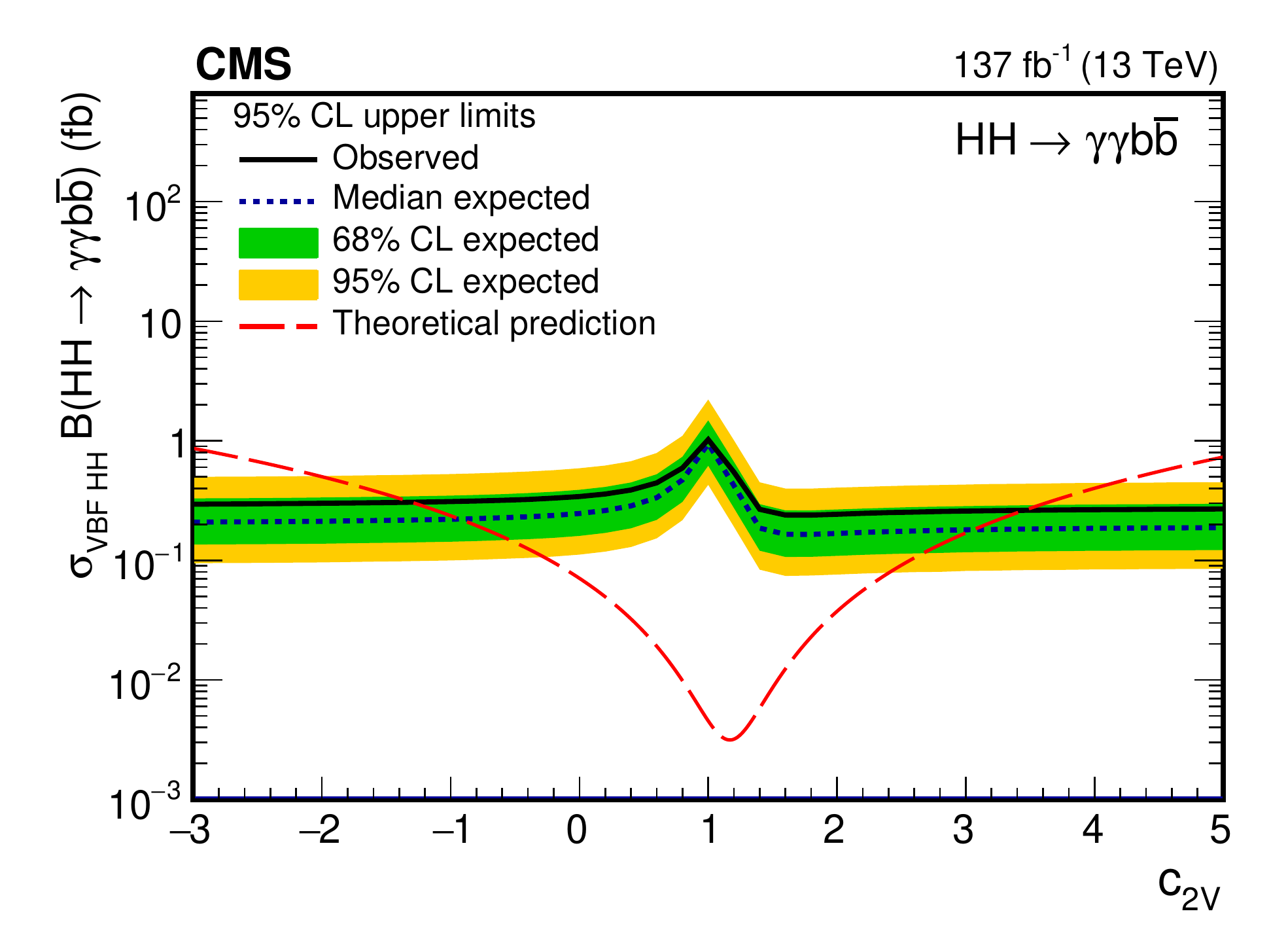}\hfil 

  \caption{Expected and observed 95\% \CL upper limits on the product of the VBF \HH production cross section and $\mathcal{B}(\HH\to\bbgg)$ obtained for different values of $\ctwov$. The green and yellow bands represent, respectively, the one and two standard deviation extensions beyond the expected limit.
The long-dashed red line shows the theoretical prediction.
}
\label{fig:ctwovscan}
\end{figure}

Limits are also set, as a function of $\ctwov$, as presented in Fig.~\ref{fig:ctwovscan}. The observed excluded region
corresponds to $\ctwov < -1.3$ and $\ctwov > 3.5$, while the expected exclusion is $\ctwov < -0.9$ and $\ctwov > 3.1$. It can be seen in Fig.~\ref{fig:ctwovscan} that this analysis is more sensitive to anomalous values of $\ctwov$ than to the region around the SM prediction. This is related to the fact that, for anomalous values of $\ctwov$, the total cross section is enhanced and the $\Mtilde$ spectrum is harder as shown in Fig.~\ref{fig:mx} (right). This leads to an
increase in the product of signal acceptance and efficiency as well as a more distinct signal topology.

\begin{figure}[!htb]
  \centering
\includegraphics[width=0.43\textwidth]{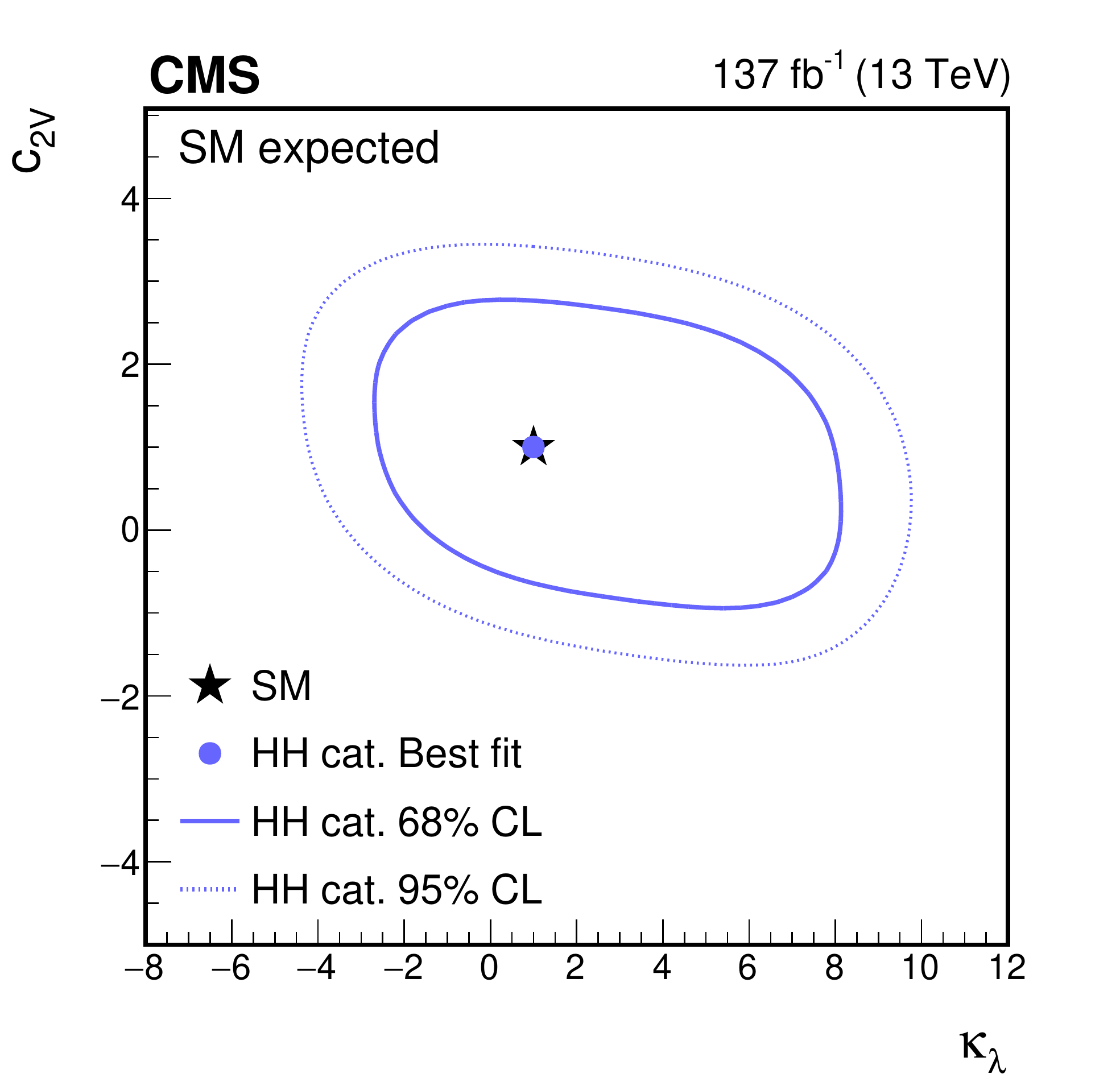}
\includegraphics[width=0.43\textwidth]{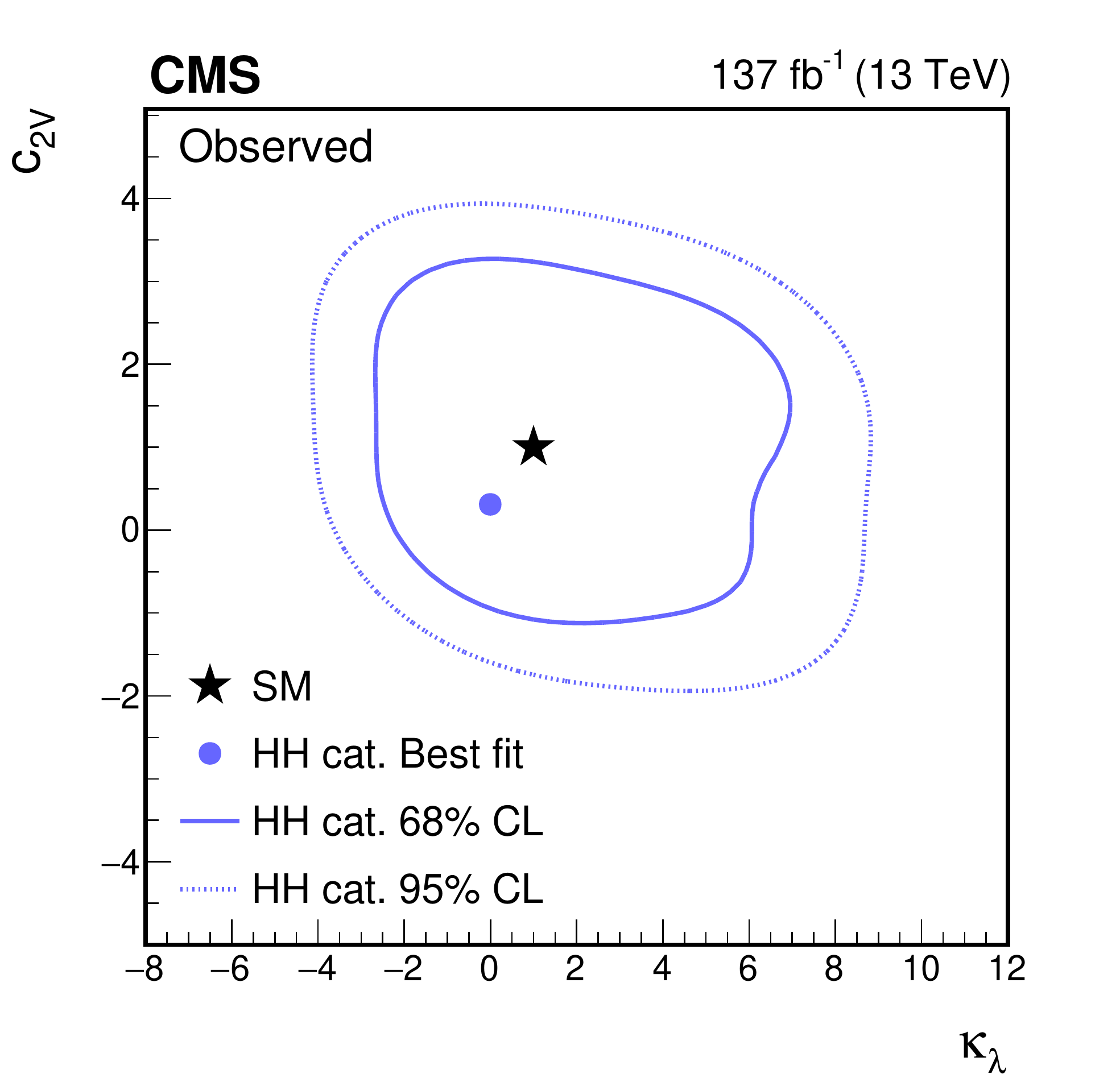}
  \caption{Negative log-likelihood contours at 68 and 95\% \CL in the ($\kapl$,~$\ctwov$) plane evaluated with an Asimov data set assuming the SM hypothesis (left) and with the observed data (right). The contours are obtained using the \HH analysis categories only. The best fit value ($\kapl$ = 0.0,~$\ctwov$ = 0.3) is indicated by a blue circle, and the SM prediction ($\kapl$ = 1.0,~$\ctwov$ = 1.0) by a black star.}
  \label{fig:klctwovlikelihood}
\end{figure}

Assuming \HH production occurs via the VBF and ggF modes, we set constraints on the \kapl and \ctwov coupling modifiers simultaneously. A 2D negative log-likelihood scan in the ($\kapl$,~\ctwov) plane is performed using the 14 \HH analysis categories. 
Figure~\ref{fig:klctwovlikelihood} shows 2D likelihood scans for the observed data and for an Asimov data set assuming all couplings are at their SM values.

\begin{figure}[!ht]
  \centering
\includegraphics[width=0.7\textwidth]{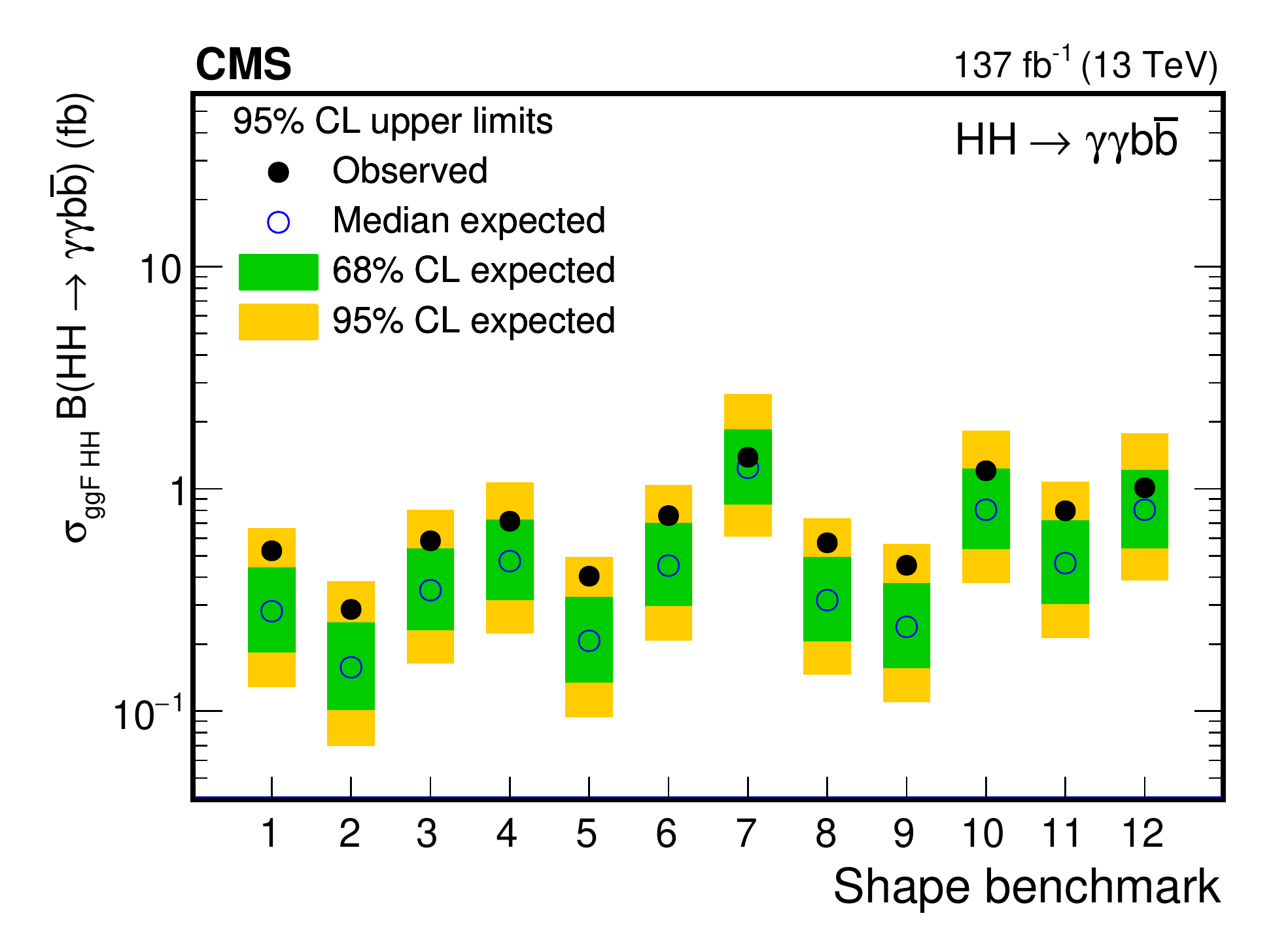}\hfil \\
\includegraphics[width=0.7\textwidth]{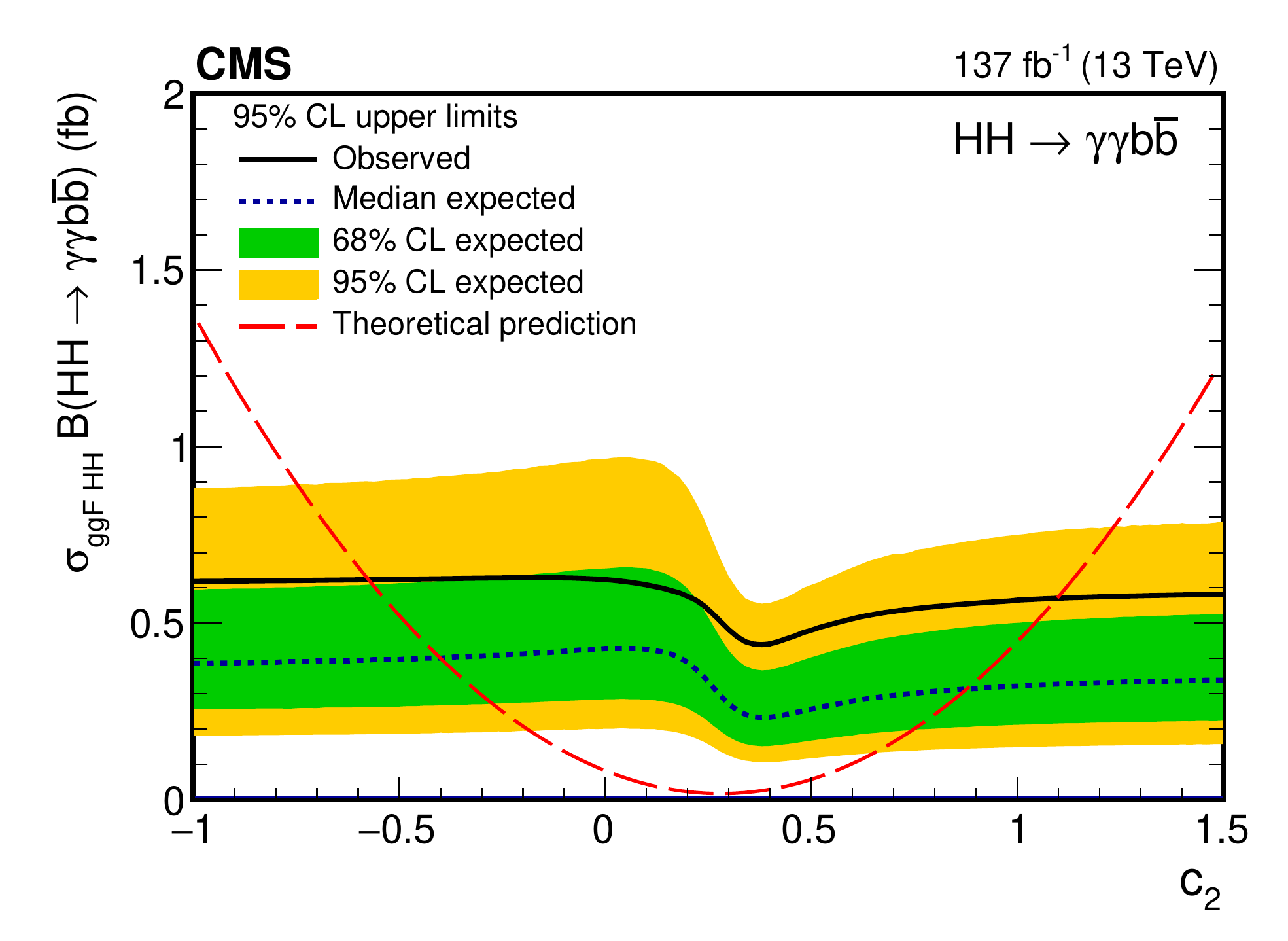}\hfil 
  \caption{Expected and observed 95\% \CL upper limits on the product of the ggF \HH production cross section and $\mathcal{B}(\HH\to\bbgg)$ obtained for different nonresonant benchmark models (defined in Table \ref{tab:bench}) (upper) and BSM coupling $\ctwo$ (lower). In this fit, the yield of the VBF \HH signal is constrained within uncertainties to the one predicted in the SM. The green and yellow bands represent, respectively, the one and two standard deviation extensions beyond the expected limit. On the lower plot the long-dashed red line shows the theoretical prediction.
}
  \label{fig:klambdascan_bench}
\end{figure}

We also set upper limits at 95\% \CL for the twelve BSM benchmark hypotheses defined in Table~\ref{tab:bench}. In this fit, the yield of the VBF \HH signal is constrained within uncertainties to the one predicted in the SM. The limits for different BSM hypotheses are shown in Fig.~\ref{fig:klambdascan_bench} (upper). In addition, limits are also calculated as a function of the BSM coupling between two Higgs bosons and two top quarks, $\ctwo$, as presented in Fig.~\ref{fig:klambdascan_bench} (lower). The observed excluded region corresponds to $\ctwo < -0.6$ and $\ctwo > 1.1$, while the expected exclusion is $\ctwo < -0.4$ and $\ctwo > 0.9$.

\section{Summary}\label{sec:sum}

A search for nonresonant Higgs boson pair production (\HH) has been presented, where one of the Higgs bosons decays to a pair of bottom quarks and the other to a pair of photons. This search uses 
proton-proton collision data collected at $\sqrt{s} = 13\TeV$ by the CMS
experiment at the LHC, corresponding to a total integrated luminosity
of 137\fbinv. No significant deviation from the background-only hypothesis is observed.
Upper limits at 95\% confidence level (\CL) on the product of the \HH production cross section and the branching fraction into $\bbgg$ are extracted for production in the standard model
(SM) and in several scenarios beyond the SM. 
The expected upper limit at 95\% \CL on $\sigmaHH \BR$ is 0.45\unit{fb}, corresponding to about 5.2 times the SM prediction, while the observed upper limit is 0.67\unit{fb}, corresponding to 7.7 times the expected value for the SM process. The presented search has the highest sensitivity to the SM \HH production to date. Upper limits at 95\% \CL on the SM \HH production cross section are also derived as a function of the Higgs boson self-coupling modifier $\kapl \equiv \lbdHHH/\lbdSM$ assuming that the top quark Yukawa coupling is SM-like. The coupling modifier $\kapl$ is constrained within a range $-3.3<\kappa_{\lambda}< 8.5$, while the expected constraint is within a range $-2.5<$ $\kappa_{\lambda}<8.2$ at 95\% \CL.

This search is combined with an analysis that targets top quark-antiquark associated production of a single Higgs boson decaying to a diphoton pair.
In the scenario in which the \HH signal has the properties predicted by the SM, the coupling modifier \kapl has been constrained. 
In addition, a simultaneous constraint on \kapl and the modifier of the coupling between the Higgs boson and the top quark \kapt is presented when both the \HH and single Higgs boson processes are considered as signals.

Limits are also set on the cross section of
nonresonant \HH production via vector boson fusion (VBF). 
The most stringent limit to date is set on the product of the \HH VBF production cross section and the branching fraction into $\bbgg$. The observed (expected) upper limit at 95\% \CL amounts to 1.02 (0.94)\unit{fb}, corresponding to 225 (208) times the SM prediction.
Limits are also set as a function of the modifier of the coupling between two vector bosons and two Higgs bosons, $\ctwov$. 
 The observed excluded region
corresponds to $\ctwov < -1.3$ and $\ctwov > 3.5$, while the expected exclusion is $\ctwov < -0.9$ and $\ctwov > 3.1$.

Numerous hypotheses on coupling modifiers beyond the 
SM have been explored, both
in the context of inclusive Higgs boson pair production and for \HH
production via gluon-gluon fusion and VBF. 
The production of Higgs boson pairs was also combined with the top
quark-antiquark pair associated production of a single Higgs boson. 
Overall, all of the results are consistent with the SM predictions.

\begin{acknowledgments}
  We congratulate our colleagues in the CERN accelerator departments for the excellent performance of the LHC and thank the technical and administrative staffs at CERN and at other CMS institutes for their contributions to the success of the CMS effort. In addition, we gratefully acknowledge the computing centers and personnel of the Worldwide LHC Computing Grid for delivering so effectively the computing infrastructure essential to our analyses. Finally, we acknowledge the enduring support for the construction and operation of the LHC and the CMS detector provided by the following funding agencies: BMBWF and FWF (Austria); FNRS and FWO (Belgium); CNPq, CAPES, FAPERJ, FAPERGS, and FAPESP (Brazil); MES (Bulgaria); CERN; CAS, MoST, and NSFC (China); COLCIENCIAS (Colombia); MSES and CSF (Croatia); RIF (Cyprus); SENESCYT (Ecuador); MoER, ERC PUT and ERDF (Estonia); Academy of Finland, MEC, and HIP (Finland); CEA and CNRS/IN2P3 (France); BMBF, DFG, and HGF (Germany); GSRT (Greece); NKFIA (Hungary); DAE and DST (India); IPM (Iran); SFI (Ireland); INFN (Italy); MSIP and NRF (Republic of Korea); MES (Latvia); LAS (Lithuania); MOE and UM (Malaysia); BUAP, CINVESTAV, CONACYT, LNS, SEP, and UASLP-FAI (Mexico); MOS (Montenegro); MBIE (New Zealand); PAEC (Pakistan); MSHE and NSC (Poland); FCT (Portugal); JINR (Dubna); MON, RosAtom, RAS, RFBR, and NRC KI (Russia); MESTD (Serbia); SEIDI, CPAN, PCTI, and FEDER (Spain); MOSTR (Sri Lanka); Swiss Funding Agencies (Switzerland); MST (Taipei); ThEPCenter, IPST, STAR, and NSTDA (Thailand); TUBITAK and TAEK (Turkey); NASU (Ukraine); STFC (United Kingdom); DOE and NSF (USA).
  
  \hyphenation{Rachada-pisek} Individuals have received support from the Marie-Curie program and the European Research Council and Horizon 2020 Grant, contract Nos.\ 675440, 724704, 752730, and 765710 (European Union); the Leventis Foundation; the A.P.\ Sloan Foundation; the Alexander von Humboldt Foundation; the Belgian Federal Science Policy Office; the Fonds pour la Formation \`a la Recherche dans l'Industrie et dans l'Agriculture (FRIA-Belgium); the Agentschap voor Innovatie door Wetenschap en Technologie (IWT-Belgium); the F.R.S.-FNRS and FWO (Belgium) under the ``Excellence of Science -- EOS" -- be.h project n.\ 30820817; the Beijing Municipal Science \& Technology Commission, No. Z191100007219010; the Ministry of Education, Youth and Sports (MEYS) of the Czech Republic; the Deutsche Forschungsgemeinschaft (DFG) under Germany's Excellence Strategy -- EXC 2121 ``Quantum Universe" -- 390833306; the Lend\"ulet (``Momentum") Program and the J\'anos Bolyai Research Scholarship of the Hungarian Academy of Sciences, the New National Excellence Program \'UNKP, the NKFIA research grants 123842, 123959, 124845, 124850, 125105, 128713, 128786, and 129058 (Hungary); the Council of Science and Industrial Research, India; the HOMING PLUS program of the Foundation for Polish Science, cofinanced from European Union, Regional Development Fund, the Mobility Plus program of the Ministry of Science and Higher Education, the National Science Center (Poland), contracts Harmonia 2014/14/M/ST2/00428, Opus 2014/13/B/ST2/02543, 2014/15/B/ST2/03998, and 2015/19/B/ST2/02861, Sonata-bis 2012/07/E/ST2/01406; the National Priorities Research Program by Qatar National Research Fund; the Ministry of Science and Higher Education, project no. 0723-2020-0041 (Russia); the Tomsk Polytechnic University Competitiveness Enhancement Program; the Programa Estatal de Fomento de la Investigaci{\'o}n Cient{\'i}fica y T{\'e}cnica de Excelencia Mar\'{\i}a de Maeztu, grant MDM-2015-0509 and the Programa Severo Ochoa del Principado de Asturias; the Thalis and Aristeia programs cofinanced by EU-ESF and the Greek NSRF; the Rachadapisek Sompot Fund for Postdoctoral Fellowship, Chulalongkorn University and the Chulalongkorn Academic into Its 2nd Century Project Advancement Project (Thailand); the Kavli Foundation; the Nvidia Corporation; the SuperMicro Corporation; the Welch Foundation, contract C-1845; and the Weston Havens Foundation (USA).
  
\end{acknowledgments}

\bibliography{auto_generated}   
\cleardoublepage \appendix\section{The CMS Collaboration \label{app:collab}}\begin{sloppypar}\hyphenpenalty=5000\widowpenalty=500\clubpenalty=5000\vskip\cmsinstskip
\textbf{Yerevan Physics Institute, Yerevan, Armenia}\\*[0pt]
A.M.~Sirunyan$^{\textrm{\dag}}$, A.~Tumasyan
\vskip\cmsinstskip
\textbf{Institut f\"{u}r Hochenergiephysik, Wien, Austria}\\*[0pt]
W.~Adam, T.~Bergauer, M.~Dragicevic, A.~Escalante~Del~Valle, R.~Fr\"{u}hwirth\cmsAuthorMark{1}, M.~Jeitler\cmsAuthorMark{1}, N.~Krammer, L.~Lechner, D.~Liko, I.~Mikulec, F.M.~Pitters, J.~Schieck\cmsAuthorMark{1}, R.~Sch\"{o}fbeck, M.~Spanring, S.~Templ, W.~Waltenberger, C.-E.~Wulz\cmsAuthorMark{1}, M.~Zarucki
% ...
\vskip\cmsinstskip
\dag: Deceased\\
1:  Also at Vienna University of Technology, Vienna, Austria\\
% ...
\end{sloppypar}
\end{document}